\documentclass[prc,aps,floatfix,showpacs,twocolumn,nofootinbib]{revtex4-1}
\usepackage{dcolumn}
\usepackage{slashed}
\usepackage{bm}
\usepackage{color}
\usepackage{graphicx}
\usepackage{pstricks}
\usepackage{amssymb,amsmath}
\usepackage[utf8]{inputenc} 							% lettere accentate da tastiera
\usepackage{bbm}
\usepackage{simplewick}
\usepackage{pstricks}
\usepackage{float}
\usepackage{simplewick}

\def\tri{{{}^3{\rm H}}}
\def\het{{{}^3{\rm He}}}
\def\heq{{{}^4{\rm He}}}
\def\lis{{{}^7{\rm Li}}}
\def\beo{{{}^8{\rm Be}}}

\def\bmr{{\bm r}}

\def\bmp{{\bm p}}
\def\bmk{{\bm k}}
\def\bmq{{\bm q}}
\def\bmj{{\bm j}}
\def\bme{{\bm e}}

\def\bmP{{\bm P}}

\def\ii{{\text i}}
\def\n{\phantom{0}}

\def\y{{\bm y}}
\newcommand{\bmsi}{{\bm \sigma}}

\begin{document}
%%%%%%%%%%%%%%%%%%%%%%%%%%%%%%%%%%%%%%%%%%%%%%%%%%%%%%%%%%%%%%%%%%%%%%%%%%%%%%%%%%%%
\title{The X17 boson and the $\tri(p,e^+ e^-)\heq$ and $\het(n,e^+ e^-)\heq$ processes: a theoretical analysis}

\author{M. Viviani$^1$, E. Filandri$^{2,3}$, L. Girlanda$^{4,5}$, C. Gustavino$^6$, A. Kievsky$^1$, 
L.E. Marcucci$^{1,7}$, and R. Schiavilla$^{8,9}$}

\affiliation{
$^1$INFN-Pisa, I-56127, Pisa, Italy \\
$^2$University of Trento, 38123 Trento, Italy\\
$^3$INFN-TIFPA, 
%Trento Institute of Fundamental Physics and Applications,
38123 Povo-Trento, Italy\\
$^4$Department of Mathematics and Physics, University of Salento, I-73100 Lecce, Italy \\
$^5$INFN-Lecce, I-73100 Lecce, Italy \\  
$^6$INFN Sezione di Roma, 00185 Rome, Italy\\
$^7$Department of Physics ``E. Fermi'', University of Pisa, I-56127 Pisa, Italy \\
$^8$Department of Physics, Old Dominion University, Norfolk, VA 23529, USA\\
$^9$Theory Center, Jefferson Lab, Newport News, VA 23606, USA}
\date{\today}

\begin{abstract}
The present work deals with $e^+$-$e^-$ pair production in the four-nucleon system.
We first analyze the process as a purely electromagnetic one
in the context of a state-of-the-art approach
to nuclear strong-interaction dynamics and nuclear electromagnetic currents,
derived from chiral effective field theory ($\chi$EFT).  Next, we examine
how the exchange of a hypothetical low-mass boson would impact the cross section
for such a process.
We consider several possibilities, that this boson is either a scalar, pseudoscalar, vector, or
axial particle.  The {\it ab initio} calculations use exact hyperspherical-harmonics methods
to describe the bound state and low-energy spectrum of the $A\,$=$\,4$ continuum,
and fully account for initial state interaction effects in the $3+1$ clusters.
While electromagnetic interactions are treated to
high orders in the chiral expansion, the interactions of the hypothetical boson
with nucleons are modeled in leading-order $\chi$EFT (albeit, in some instances,
selected subleading contributions are also accounted for).  We also provide an
overview of possible future experiments probing pair production in the $A\,$=$\,4$
system at a number of candidate facilities.
\end{abstract}

\pacs{}

\maketitle

\section{Overview}
The present work reports on a comprehensive analysis of low-energy $e^+$-$e^-$ pair production
in the four-nucleon system, both as a purely electromagnetic process and by including
the contribution of a hypothetical low-mass boson. It is organized as follows. The present
section provides an overview of the complete study.  Starting from an up-to-date review of the
current status of experimental searches and theoretical analyses, we delineate next the
chiral-effective-field-theory ($\chi$EFT) framework adopted here to describe nuclear dynamics and
to model the interactions of nucleons with the hypothetical boson.  We then proceed to a
summary of the {\it ab initio} predictions obtained for the differential cross sections of
the $\tri(p,e^+ e^-)\heq$ and $\het(n,e^+ e^-)\heq$ reactions.  We close with some concluding
remarks and a discussion of possible future experiments at a number of candidate facilities.  
The remaining sections are meant to elaborate
more expansively on these various aspects of the calculations.

\subsection{Motivation and current status}
\label{sec:status}
The possible existence of a new kind of low mass particle (at the MeV scale)
is a problem of current and intense theoretical and experimental interest
(see, for example, Ref.~\cite{Battaglieri:2017aum} and references therein).
 This interest is, in fact, part of a broader effort aimed at identifying dark matter
 (DM).  Its existence has been postulated to explain a
 number of gravitational anomalies that have been observed at galactic
 scales and beyond~\cite{Bertone:2018aaa} since the 1930's.  However, no
 conclusive experimental signatures of DM have been reported up until now.

A few years ago, there were claims~\cite{Krasznahorkay:2015iga} that an unknown
particle (denoted as ``X17'') had been observed in the process $^7$Li$(p,e^+ e^-)^8$Be
at the ATOMKI experimental facility situated in Debrecen (Hungary).
These claims were based on a $\approx 7 \sigma$ excess of events in the
angular distribution of leptonic pairs produced in this reaction, which has
a $Q$-value of about $18$ MeV.  The excess---referred to below as the
``anomaly''---could be explained by positing the emission of an unknown
boson with a mass of about $17$ MeV decaying into $e^+ e^-$ pairs.

The search for bosonic DM candidates had already started several
years earlier, by attempting to establish the possible existence of additional forces
(beyond gravity), mediated by these bosons~\cite{Pospelov:2008aaa}, between
DM and visible matter.  To one such class of particles belongs the so-called
``dark-photon'', namely a boson of mass $M_{X}$ having the same
quantum numbers as the photon, and interacting with a Standard Model (SM)
fermion $f$ with a coupling constant given by $\varepsilon \,q_f$, where $q_f$ is the fermion
electric charge.  Following several years of experimental searches for dark photons,
``exclusion plots'' in the $\varepsilon$-$M_{X}$ parameter space were produced,
restricting more and more the allowed region~\cite{Battaglieri:2017aum,Banerjee:2018vgk}.
One of the most stringent limits was provided by the NA48/2 experiment,
which set $\varepsilon < 8 \times 10^{-4}$ at $90$\% confidence level~\cite{Batley:2015lha}.
Similar limits have been produced by assuming that the X17 is a pseudoscalar
particle~\cite{Andreas:2010ms}, that is, an axion-like particle.
More recently, stringent limits were also set by the NA64 experiment at
CERN~\cite{Banerjee:2018vgk}.  Of course, the analysis becomes more complicated
and the ensuing picture far less unambiguous, if the coupling constants are
assumed to be different for each of the SM fermions.

The claim of the $\beo$ anomaly by the ATOMKI group~\cite{Krasznahorkay:2015iga}
soon spurred several theoretical studies.  In Ref.~\cite{Feng:2016jff}, 
a number of alternatives---that the X17 could be a scalar, a pseudoscalar, or a vector
boson---were analyzed.  The first two were quickly dismissed, while
the possibility that the X17 could be a vector boson, that is, a dark photon,
was investigated in detail.  In order to circumvent the NA48/2 limit, 
it was conjectured that the X17 could be ``proto-phobic'', namely that it would couple much more
weakly to the proton than to the neutron~\cite{Feng:2016jff}. 

Soon after, the possibility that the X17 could be a spin-1 particle interacting
via axial couplings to the $u$ and $d$ quarks was explored in
Ref.~\cite{Kozaczuk:2016nma}.  In that work, $\beo$ ground- and
excited-state wave functions obtained in state-of-the-art shell model calculations
were used in combination with the ATOMKI data to constrain the 
range of X17-quark couplings that could explain the observed anomaly.
Limits provided by the bounds determined by a number of other experiments
were also analyzed.  In case of an axial coupling, the NA48/2 constraint does
not apply, but other limits have to be taken into account,
for example, from the study of rare $\eta$ and $\phi$ decays and proton fixed target experiments
(see, for more details, Ref.~\cite{Kozaczuk:2016nma} and references therein).

In 2019, the ATOMKI group reported a similar excess at approximately the
same invariant mass in the $\tri(p,e^+ e^-)\heq$
reaction~\cite{Krasznahorkay:2019lyl}; this excess has more recently been confirmed 
in Ref.~\cite{Krasznahorkay:2021joi}.  The authors of Ref.~\cite{Feng:2016jff} published
a new study~\cite{Feng:2020mbt}, considering both the $\beo$ and $\heq$ anomalies
and allowing for scalar, pseudoscalar, vector, and axial coupling of the X17 to quarks and electrons,
with the intent of verifying whether the two results were consistent.
They concluded that the {\it``$\,7\sigma$ anomalies
reported in {\rm[the]} $\beo$ and $\heq$ nuclear decays are both kinematically
and dynamically consistent with the production of a 17 MeV proto-phobic
gauge boson''}~\cite{Feng:2020mbt}.  However, other studies have
challenged the explanation of a  proto-phobic X17 emission in the $\beo$ experiment, by taking into account
existing $\lis(p,\gamma)\beo$ cross section data~\cite{Zhang:2020ukq,Hayes:2021hin}.

In the following, we consider the more general case of a Yukawa-like interaction
between the X17 and a SM fermion of species $f$ (specifically, quarks and electrons) with the
coupling constant expressed as $\varepsilon_f \,e$, where $e$ ($>0$) is
the unit electric charge.  The X17 boson must decay promptly in $e^+$-$e^-$ pairs for these to be detected
inside the experimental setup.  This observation actually introduces a {\it lower limit} to the possible
values of $\varepsilon_e$.  These limits are also established by various electron
beam-dump experiments (see, for example, Ref.~\cite{PhysRevD.86.095019} and
references therein).  For $M_X\approx17$ MeV, the most stringent lower bound, $|\varepsilon_e| > 2 \times 10^{-4}$, comes
from the SLAC E141 experiment~\cite{PhysRevLett.59.755}, while the upper bound $|\varepsilon_e|< 2 \times 10^{-3}$ has been
set by the KLOE-2 experiment~\cite{Anastasi:2015qla}.
See Ref.~\cite{Kozaczuk:2016nma} for a comprehensive
discussion of these and other constraints regarding $\varepsilon_e$.  

If the X17 were to couple also to muons, then its existence
would have consequences for the well known $(g-2)_\mu$ anomaly~\cite{Abi:2021gix}, namely the
discrepancy between the measured and predicted value of the muon anomalous magnetic
moment.\footnote{In this context we should point
out that a recent LQCD calculation of $(g-2)_\mu$ indicates~\cite{Borsanyi:2021} that
there may not be any significant tension between the experimental value and the prediction based on the Standard Model.}
In this instance, a pure axial coupling to muons would worsen this discrepancy,
while a pure vector coupling would reduce it~\cite{PhysRevD.75.115017}.  Another interesting
result comes from new measurements of the fine structure constant
using atomic recoil of Rubidium~\cite{Morel:2020dww}.  Using this measurement,
the SM prediction for $(g-2)_e$ too is in tension with the experimental
value (at about $1.6\,\sigma$).  Also in this case, a pure vector
coupling of the X17 to the electron with $|\varepsilon_e|= (8 \pm 3) \times 10^{-4}$~\cite{Morel:2020dww}
would resolve this tension. 

The observation of the $\beo$ and $\heq$ anomalies has
triggered a rapidly growing number of theoretical studies~\cite{Zhang:2017zap,Ellwanger:2016wfe,Dror:2017ehi,DelleRose:2017xil,DelleRose:2018eic,DelleRose:2018pgm,Alves:2017avw,Alves:2020xhf,Bordes:2019wcp,Nam:2019osu,Kirpichnikov:2020tcf,Fayet:2020bmb,Hayes:2021hin}, see Ref.~\cite{Fornal:2017msy} for a critical review.  On the experimental side, 
there are several experiments (MEGII~\cite{Baldini:2018nnn}, DarkLight~\cite{Balewski:2014pxa},
SHiP~\cite{SHiP:2020noy}, and others~\cite{Battaglieri:2017aum})
planning specifically to search for such a light boson.  In addition, large
collaborations, such as BelleII~\cite{Kou:2018nap}, NA64~\cite{Banerjee:2020bbb}, and others, 
are dedicating part of their efforts in an attempt to clarify this issue.

To date, the explanation of the $\beo$ and $\heq$ anomalies remains an open problem.
If the existence of such a particle were to be experimentally
confirmed, it would have profound repercussions for the study of DM and for beyond
SM theories.  It is worthwhile pointing out here that most theoretical analyses
so far have assumed that the reactions proceed via a two step process.  First,
a resonant state is formed in the collision of the incident nucleon with the target
nucleus and, second, this resonant state decays to the ground state by
emitting either a photon or the hypothetical X17 boson.

\begin{figure}[bth]
\centering
\includegraphics[scale=0.45,clip]{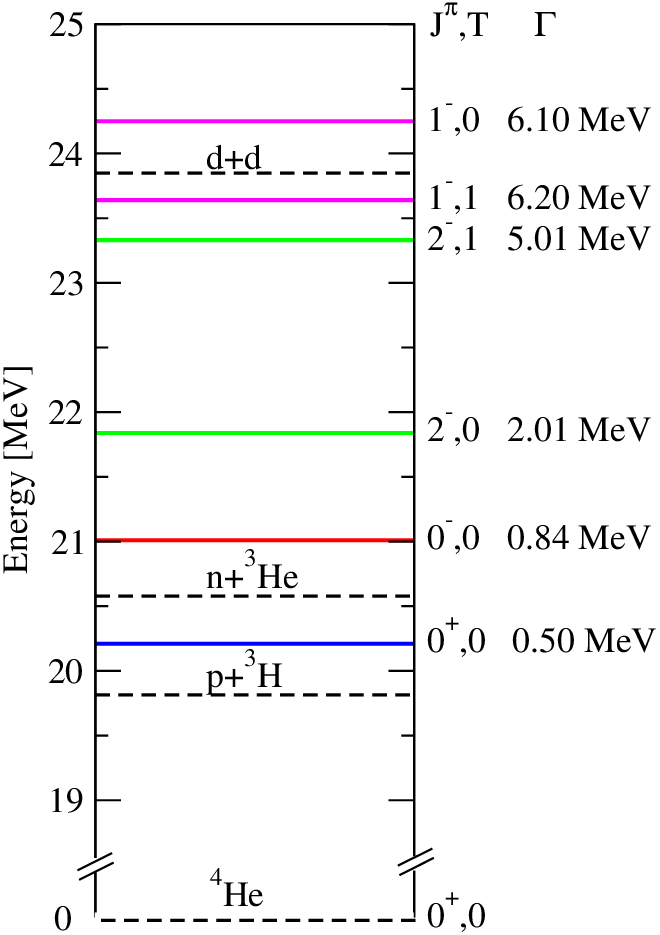}
\caption{The low-energy spectrum of $^4$He~\protect\cite{Tilley:1992zz}.  The dashed
lines indicate the thresholds for the opening of the 1+3 and 2+2 channels,
while the solid lines indicate the energies of the various resonances
with corresponding widths, $J^\pi$, and isospin assignments to the right.}
\label{fig:levels}
\end{figure}

This resonance-saturation
approach permits the properties of the X17 particle to be
inferred~\cite{Feng:2016jff,Kozaczuk:2016nma,Feng:2020mbt} from the quantum numbers of
the decaying and ground states, assuming conservation of parity and other symmetries.
While such a treatment may be justified in the case of $\beo$ where the
initial state $p+\lis$ populates the (relatively narrow isoscalar and isovector) $1^+$ resonances of $\beo$,
it becomes problematic in the case of $\heq$, since the low-energy $p+\tri$ initial states may
populate the fairly narrow $0^+$ and $0^-$ resonances of $\heq$.  Furthermore,
contributions from farther resonances or from direct (non-resonant) capture
cannot be excluded a priori, and  should be estimated. 
As a matter of fact, in Ref.~\cite{Zhang:2020ukq}, the contribution
of the direct $p+\lis$ capture in different waves has been found important,
if not dominant.  For future reference, the low-energy spectrum
of $^4$He is reported in Fig.~\ref{fig:levels}. The
energies and widths of the various resonances
have been determined in R-matrix analyses~\cite{Tilley:1992zz}.

Recently, it was conjectured in Ref.~\cite{Kalman:2020meg} that the effect of populating
higher excited states might cause peaks in the angular distribution of the final pair,
and therefore mimic the resonance-like structure observed in the $^3$H$(p,e^+ e^-)^4$He
ATOMKI experiment.  However, such a conjecture does not seem to be borne out by the present
analysis. Instead, our calculations, which are based on a realistic and fully microscopic
treatment of nuclear dynamics, indicate the absence of any such structure.  In particular, the
$J^\pi\,$=$\,1^-$ scattering state, which plays a dominant role, only yields a smooth behavior
in the angular correlation between the leptons, although this correlation appears to be
rather sensitive to the low-energy structure of the $\heq$ continuum. 

In a more recent paper~\cite{Aleksejevs:2021zjw}---see 
also Ref.~\cite{Koch:2020ouk} for a similar study---the possibility that the peak seen in the
$\beo$ experiment could be caused by higher-order QED effects, beyond the one-photon-exchange
approximation, has been investigated.  It has been found that the contribution of these corrections
increases with the opening angle $\theta_{ee}$ between the lepton momenta, and could explain the
observed behavior of the cross section.  We should point out that for the $\beo$ experiment the
peak structure was observed around $\theta_{ee}\,$=$\,140^\circ$. It is not clear
if such an explanation will remain valid in the case of the $\heq$ experiment, where a sharper
peak is observed at a considerably smaller opening angle, around $110^\circ$.

%
% Here
%
\subsection{Interactions and currents}
 {\it Ab initio} studies of few-nucleon dynamics can be carried out nowadays with great 
accuracy~\cite{Viviani:2020gkm}, not only for bound states but also for the low-energy
portion of the continuum spectrum, including the treatment of resonances.  This capability
and the availability of consistent models of nuclear electroweak currents---that is,
electroweak currents constructed consistently with the underlying strong-interaction
dynamics---make it possible to almost completely remove uncertainties associated with
the nuclear wave functions and/or reaction mechanisms, and therefore to unambiguously
interpret the experimental evidence.  It is within this context that, in the present paper, we
provide fairly complete analyses of the $\tri(p,e^+ e^-)\heq$ and $\het(n,e^+ e^-)\heq$
processes, with and without the inclusion of the hypothetical X17 boson.
Below, we briefly outline the theoretical framework, and refer the reader to the
following sections for more extended discussions of various aspects of this framework.

The nuclear Hamiltonian is taken to consist of non-relativistic kinetic energy,
and two- and three-nucleon interactions.  These interactions are derived from
two different versions of chiral effective field theory ($\chi$EFT): 
one~\cite{Entem:2003ft,Machleidt:2011zz,Epelbaum:2002vt} retains only
pions and nucleons as degrees of freedom, while the
other~\cite{Piarulli:2016vel,Piarulli:2017dwd} also retains $\Delta$-isobars.
Both $\chi$EFT versions account for high orders in the chiral expansion, but
again differ in that the interactions of
Refs.~\cite{Entem:2003ft,Machleidt:2011zz,Epelbaum:2002vt}
are formulated and regularized in momentum space, while those
of Refs.~\cite{Piarulli:2016vel,Piarulli:2017dwd} in coordinate space.
As a consequence, the former are strongly non-local in coordinate
space.

The low-energy constants (LECs) that characterize
the two-nucleon interaction have been determined by fits to the
nucleon-nucleon scattering database (up to the pion production threshold),
while the LECs in the three-nucleon interaction have been constrained
to reproduce selected observables in the three-nucleon sector (see
Sec.~\ref{sec:wf}).  However, the nuclear Hamiltonians resulting
from these two different formulations both lead to an excellent
description of measured bound-state properties and scattering observables
in the three- and four-nucleon systems, including in particular
the $^4$He ground-state energy and 3+1 low-energy
continuum~\cite{Viviani:2020gkm}, germane to the present
endeavor.

Another important aspect of the theoretical framework is the treatment of nuclear
electromagnetic currents.  These currents have been derived
within the two different $\chi$EFT formulations we consider here, namely
without~\cite{Pastore:2008ui,Pastore:2009is,Pastore:2011ip,Kolling:2009iq,Kolling:2011mt,Piarulli:2012bn}
and with~\cite{Schiavilla:2018udt} the inclusion of explicit $\Delta$-isobar degrees of freedom.
They consist of  (i) one-body terms, including relativistic corrections suppressed
by two orders in the power counting relative to the leading order; (ii) two-body
terms associated with one pion exchange (derived from leading and subleading
chiral Lagrangians) as well as pion loops, albeit in the $\Delta$-full $\chi$EFT formulation
the contributions of $\Delta$ intermediate states have been ignored in these loops;
and (iii) two-body contact terms originating from minimal and non-minimal contact couplings. 

The subleading one-pion-exchange and non-minimal contact electromagnetic
currents are characterized by a number of unknown LECs that have
been fixed by reproducing the experimental values of the two- and
three-nucleon magnetic moments and by relying on either resonance
saturation for the case of the $\Delta$-less formulation~\cite{Piarulli:2012bn}
or fits to low-momentum transfer data on the deuteron threshold electrodisintegration
cross section at backward angles for the case of the $\Delta$-full
formulation~\cite{Schiavilla:2018udt}.   

These interactions and the associated electromagnetic currents lead to
predictions for light-nuclei electromagnetic observables, including
ground-state magnetic moments, radiative transition rates between
low-lying states, and elastic form factors, in very satisfactory agreement with
the measured values (for recent reviews, see~\cite{Bacca:2014tla,Carlson:2014vla,Marcucci:2015rca}).
\begin{figure}[bth]
\centering
\includegraphics[scale=0.35,clip]{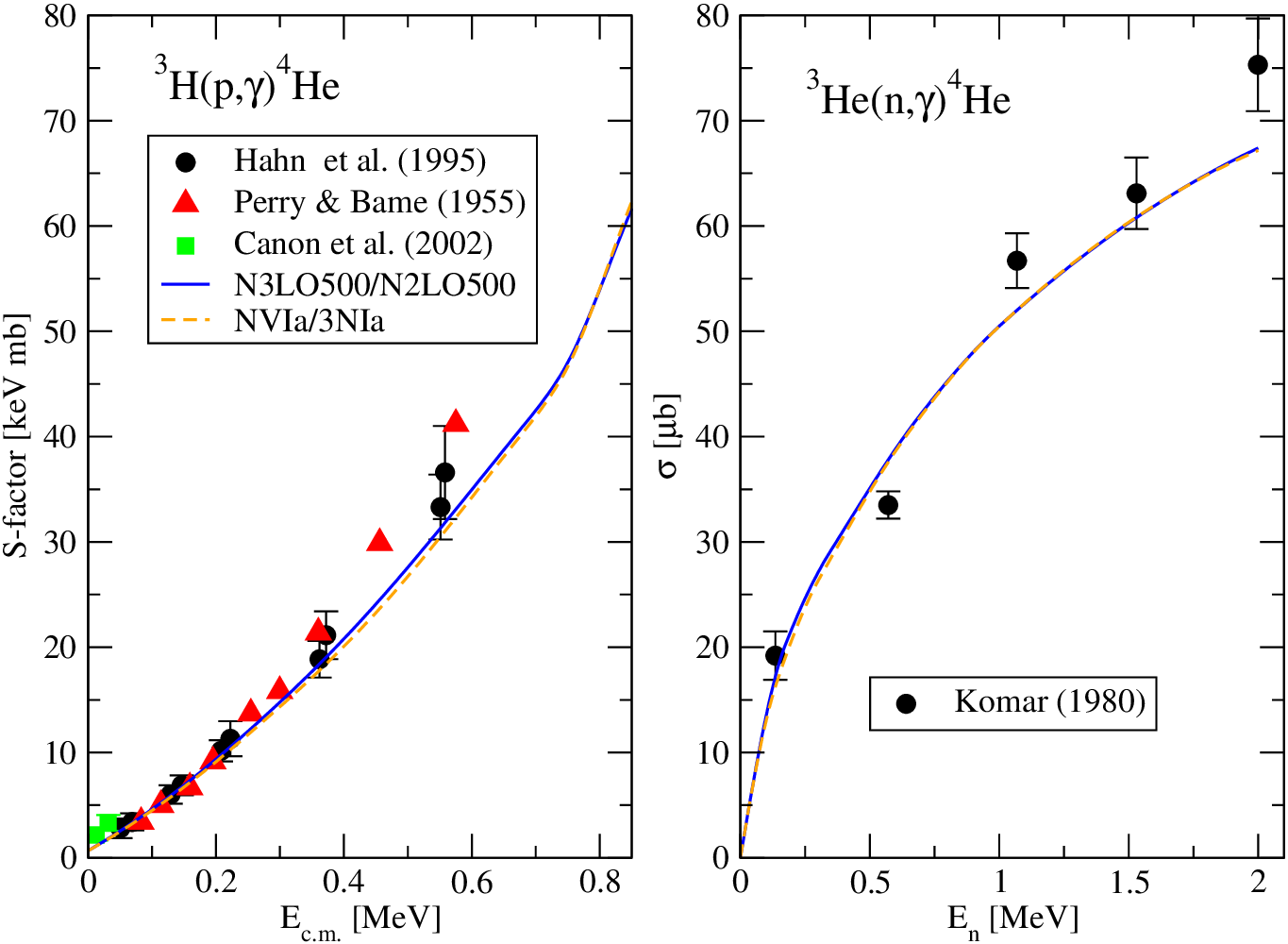}
\caption{Astrophysical factor for the capture $\tri(p,\gamma)\heq$ (left panel) and total
cross section for the capture $\het(n,\gamma)\heq$ (right panel),
calculated in both the $\Delta$-less (N3LO500/N2LO500)
and $\Delta$-full (NVIa/3NIa) $\chi$EFT formulations considered in
the present work. The experimental data are from
Refs.~\protect\cite{Perry:1955ptc,Hahn:1995ptc,Canon:2002ds,Komar:1993nhc}.}
\label{fig:capture}
\end{figure}
As an illustration of the validity of the present theoretical framework
and the level of sophistication achieved in these {\it ab initio}
few-nucleon calculations, we present in Fig.~\ref{fig:capture}
predictions for the low-energy radiative captures of protons
on $^3$H and of neutrons on $^3$He, compared to available
experimental data.  We use bound and continuum wave functions
obtained with hyperspherical-harmonics (HH) methods~\cite{Kievsky:2008es,Marcucci:2020fip},
and fully account in the $n$+$^3$He initial state for its coupling to the
energetically open $p$+$^3$H channel.  Matrix elements of the (complete) current
are calculated with quantum Monte Carlo methods~\cite{Schiavilla:1989zz} without
any approximations---the
statistical errors (not shown in the figure) due to the Monte Carlo
integrations are at the \% level.

In the energy regime of Fig.~\ref{fig:capture}, these radiative captures
proceed primarily via $M_1$- and $E_1$-transitions between, respectively,
the $^3S_1$ and $^1P_1$ incoming states and the
$^4$He ground state.\footnote{We use the spectroscopic notation $^{2S+1}L_{J}$,
where $S$ and $L$ are, respectively, the channel spin and relative orbital angular momentum
between the two clusters, and $J$ is the total angular momentum.}
We should stress here that in the lepton-pair production
processes we consider below, additional $S$- and $P$-wave channels
play a prominent role, in particular the $^1S_0$ channel.  Lastly, Fig.~\ref{fig:capture}
shows that, at least in the low-energy regime of interest in the present work,
the model dependence resulting from the two different $\chi$EFT
formulations is weak. 
\subsection{Including the X17 boson}
\label{sec:s1c}
In the SM and one-photon-exchange approximation, the $e^+$-$e^-$ pair
production on a nucleus is driven by the amplitude (in a schematic notation)
\begin{equation}
\label{eq:e1}
T_{fi}=4\pi\alpha\, \frac{(\overline{u}_-\,\gamma_\mu \,v_+)\, j^{\mu}_{fi}}
{q^\mu\, q_\mu}\ ,
\end{equation} 
where $\alpha$ is the fine structure constant, $q^\mu$ is the four-momentum transfer
defined as the sum of the outgoing-lepton four momenta,
$u_-$ and $v_+$ are, respectively, the electron and positron spinors, and $j^{\mu}_{fi}$
is the matrix element of the nuclear electromagnetic current between the initial and final
nuclear states (here, either the $p$+$^3$H or
$n$+$^3$He scattering state and $^4$He ground-state); a less
cursory description of this as well as the X17-induced
amplitudes to follow is provided in Secs.~\ref{sec:s-em} and \ref{sec:x17curr} below.

We consider four different possibilities for the coupling of the X17 to electrons
and hadrons: scalar ($S$), pseudoscalar ($P$), vector ($V$), and axial ($A$).
The electron-X17 interaction Lagrangian density reads
\begin{equation}
{\cal L}^c_{e,X}(x)=e\, \varepsilon_e\,\overline{e}(x)\, \Gamma^{c}\, e(x)\, X_{c}(x) \ ,\label{eq:LeX}
\end{equation}
where $c$ specifies the nature of the coupling and the associated Lorentz
structure,  
\begin{equation}
\Gamma^{c=S,P,V,A}=1, i\, \gamma^5 , \gamma^\mu, \gamma^\mu \, \gamma^5 \ ,
\end{equation}
$e(x)$ is the electron field, and $X_{c}(x)\,$=$\,X(x)$ for $c\,$=$\,S, P$
and $X_{c}(x)\,$=$\,X_\mu(x)$ for $c\,$=$\,V, A$ represents the X17 field.
The single coupling constant $\varepsilon_e$ is written in units of the electric
charge $e>0$ ($e^2\,$=$\,4\pi\alpha$).

The hadron-X17 interaction Lagrangian densities are derived
in $\chi$EFT by considering only leading-order contributions (and selected subleading ones in the vector and pseudoscalar cases).
A fairly self-contained summary of this derivation is provided in
Sec.~\ref{sec:x17lagr} for completeness.  Here we write them as
\begin{eqnarray}
\label{eq:e4s}
{\cal L}^{S}_{X}(x)\!&=&\!e\, \overline{N}(x)[\eta^{S}_0 +\eta^{S}_z\, \tau_{3}]N(x)\,X(x) \ ,\\
\label{eq:e4p}
{\cal L}^{P}_{X}(x)\!&=&\!e\, \eta_z^{P}\, \pi_3 (x)\, X(x) + e \, \eta^{P}_0 \overline{N}(x)i\,\gamma^5N(x)\,X(x)\ ,\\
\label{eq:e4v}
{\cal L}^{V}_{X}(x)\!&=&\!e\,  \overline{N}(x)[\eta^{V}_0 +\eta^{V}_z\, \tau_{3}]\gamma^\mu\,N(x)\,X_\mu(x)  \\
&&\!+\frac{e}{4\, m_N}\,  \overline{N}(x)[\kappa_0 \eta^{V}_0 +\kappa_z\eta^{V}_z\tau_{3}]\sigma^{\mu\nu}\,N(x) F^X_{\mu\nu}(x)  \ ,\nonumber\\
{\cal L}^{A}_{X}(x)&=&\!e\,  \overline{N}(x)[\eta^{A}_0 +\eta^{A}_z\, \tau_{3}]\gamma^\mu\gamma^5\,N(x)\,X_\mu(x)\ ,
\label{eq:e4a}
\end{eqnarray}
where $m_N$ is the nucleon mass, $N(x)$ is the iso-doublet of nucleon fields, $\pi_3(x)$ is the third component of
the triplet of pion fields, and $F^X_{\mu\nu}(x)\,$=$\,\partial_\mu\, X_\nu(x)-\partial_\nu\, X_\mu(x)$
is the X17 field tensor.
The combinations of LECs (rescaled by $e$) are lumped into
the isoscalar and isovector coupling constants $\eta_0^{c}$ and $\eta_z^{c}$.
In the vector case,
we have included also the subleading term proportional to $F^X_{\mu\nu}(x)$ and where 
\begin{equation}
    \kappa_0\,=\,\kappa_p+\kappa_n\ , \qquad
    \kappa_z\,=\,\kappa_p-\kappa_n\ , \label{eq:kappa}
\end{equation}
$\kappa_p$ and $\kappa_n$ being the anomalous
magnetic moments of the proton and neutron, respectively.
In the pseudoscalar case, the leading-order interaction
originates from the direct coupling of the X17 to the pion.  However, since
the associated coupling constant is expected to be suppressed~\cite{Alves:2017avw,Alves:2020xhf},
we have also considered an isoscalar coupling of the X17 to the nucleon, even though
it is subleading, at least nominally, in the $\chi$EFT power counting relative
to the isovector one.  As per the axial case,
the tree-level ${\cal L}^{A}_{X}(x)$ contains an additional term  
of the form $\partial^\mu \pi_3 (x)\, X_\mu(x)$, which we have dropped.\footnote{This
term leads to a X17-nucleon current proportional to $q^\mu/m_\pi^2$ (for low momentum
transfers) which, when contracted
with the lepton axial current, produces a contribution proportional to $m_e/m_\pi^2$,
and hence negligible when compared to that resulting from the X17 direct coupling to the nucleon.}

The X17-induced amplitude for emission of the $e^+$-$e^-$ pair is then obtained from
\begin{equation}
\label{eq:e8t}
T_{fi}^{cX}=4\pi\alpha \, \frac{\varepsilon_e\,(\overline{u}_-\,\Gamma_{c} \,
v_+)\, j_{fi}^{c X}}{q^\mu\, q_\mu-M_X^2}\ ,
\end{equation} 
where $M_X$ is the mass of the X17 particle, and
$j^{cX}_{fi}$ represents the matrix element of
the nuclear current, including the coupling
constants $\eta_\alpha^c$
associated with the X17 particle.  To account for its width $\Gamma_X$,
we make the replacement
\begin{equation}
\label{eq:e9}
M_X\longrightarrow M_X-i\, \Gamma_X/2 \ .
\end{equation}
At the leading order we are considering here,
the nuclear current $j^{cX}({\bf q})$ consists of the sum
of one-body terms,
\begin{equation}
\label{eq:e10}
j^{cX}({\bf q})=\sum_{i=1}^A  \,O^{cX}_i \, e^{i\bmq\cdot{\bf r}_i} \ ,
\end{equation}
where $\bmq$ is the three-momentum transfer (that is, the sum of
the outgoing-lepton momenta), and $O_i^{cX}$ involves generally
the momentum, spin, and isospin operators of nucleon $i$---the specific
operator structure obviously depending on the nature of the coupling assumed for the
X17---as well as the coupling constants $\eta_\alpha^c$.\footnote{In evaluating the matrix elements
$j^{\mu}_{fi}$ and $j^{cX}_{fi}$, we find it convenient to have the
current operators act on the final bound state, namely to the left.}
As shown in Fig.~\ref{fig:levels}, the low-energy spectrum of $^4$He consists
of fairly narrow resonant states.
By varying the energy of the incident beam, it might be possible
to predominantly populate a specific resonant state by exploiting
the selectivity of the transition operator, and therefore infer
the nature---scalar, pseudoscalar, vector, or axial---of the (hypothetical )
X17 particle.

\subsection{$\tri(p,e^+ e^-)\heq$ and $\het(n,e^+ e^-)\heq$ cross sections}
\label{sec:s1d}
The amplitudes $T_{fi}$ and $T^{cX}_{fi}$ generally interfere (except
when the X17 is a pseudoscalar particle) and
the resulting cross section has
a complicated structure; it is derived in Sec.~\ref{sec:x17cross}.
In the laboratory frame, the initial state consists of an incoming proton 
or neutron of momentum ${\bmp}$ and spin projection $m_1$, and a
bound $^3$H or $^3$He cluster in spin state $m_3$ at rest.  Its wave function $ \Psi_{m_3,m_1}^{(\gamma)}(\bmp)$ is
such that, in the asymptotic region of large separation ${\bm y}_\ell$ between the isolated nucleon (particle $\ell$) and  the trinucleon cluster (particles $ijk$), it reduces to
\begin{equation}
\label{eq:e11}
\Psi_{m_3,m_1}^{(\gamma)}(\bmp) \longrightarrow 
{1\over\sqrt{4}} \sum_{\ell=1}^4 \phi^{m_3}_\gamma(ijk) \chi^{m_1}_\gamma(\ell) \, \Phi_{\bmp}^{(\gamma)}(\y_\ell) \ ,
\end{equation}
where $\Phi_{\bmp}^{(\gamma)}(\y_\ell)$ is either a Coulomb distorted wave
or simply the plane wave $e^{i{\bmp}\cdot{\y_\ell}}$ depending on whether
we are dealing with the $p+^3$H ($\gamma\,$=$\,1$) or $n+^3$He ($\gamma\,$=$\,2$) state.
The final state consists of the lepton pair---with the $e^-$ having momentum (energy) ${\bm k}$ ($\epsilon$)
and spin $s$, and the $e^+$ having momentum (energy) ${\bm k}^\prime$ ($\epsilon^\prime$)
and spin $s^\prime$---and the $^4$He ground state recoiling
with momentum $\bmp-\bmk-\bmk^\prime$.  Energy conservation
requires
\begin{equation}
\label{eq:e12}
\epsilon+\epsilon^\prime+ \frac{(\bmp-\bmk-\bmk^\prime)^2}{2\,M}= E_0\ ,
\end{equation}
where $M$ is the rest mass of the $^4$He ground state.
Further, we have defined $E_0=T_p+B_4-B_3\approx 20\,$ MeV, where $T_p$
is the kinetic energy of the incident nucleon, and
$B_3$ and $B_4$ are the binding energies
of, respectively, the bound three-nucleon cluster and $^4$He.

After integrating out the energy and momentum $\delta$-functions,
the five-fold differential cross section, averaged over the azimuthal
quantum numbers $m_1$ and $m_3$ of the nuclear clusters, and
summed over the spins of the final leptons, has generally the form
\begin{equation}
\label{eq:e13}
\frac{d^5\sigma^{(\gamma)}}{d\epsilon\,d\hat{\bmk}\, d\hat{\bmk}^\prime}\!=\!
\sigma^{(\gamma)}(\epsilon,\hat{\bmk}, \hat{\bmk}^\prime)+
\sigma^{(\gamma)}_X(\epsilon,\hat{\bmk}, \hat{\bmk}^\prime)+
\sigma_{XX}^{(\gamma)}(\epsilon,\hat{\bmk}, \hat{\bmk}^\prime) \ ,
\end{equation}
where $\sigma^{(\gamma)}$, $\sigma_X^{(\gamma)}$, and $\sigma_{XX}^{(\gamma)}$
denote the contributions coming solely from electromagnetic currents,
the interference between electromagnetic and X17-induced currents, and
purely X17-induced currents, respectively, and $\hat{\bmk}$ and $\hat{\bmk}^\prime$
specify the directions of the electron and positron momenta.  The positron
energy $\epsilon^\prime$ is fixed by energy conservation, while the electron
energy $\epsilon$ is restricted to the range $m_e \le \epsilon \le \epsilon_{max}$,
where the maximum allowed energy $\epsilon_{max}$ is obtained from the
solution of Eq.~(\ref{eq:e12}) corresponding to a vanishing positron
momentum; in fact, since the kinetic energy of the recoiling $^4$He
is tiny, $\epsilon_{max}$ is close to $E_0-m_e$.  In the 
laboratory coordinate system, where the $z$-axis is oriented
along the incident beam momentum $\bmp$, the spherical angles 
specifying the $\hat\bmk$ ($\hat\bmk'$) direction are denoted
as $\theta$ and $\phi$ ($\theta'$ and $\phi'$), and
\begin{equation}
 \hat\bmk \cdot \hat\bmk'  \equiv \cos\theta_{ee}=\cos\theta\cos\theta'+\sin\theta\sin\theta'\cos(\phi'-\phi)\ .
    \label{eq:thetaee}
\end{equation}

Of course, the numerical results presented below for the various cross sections
depend on the mass $M_X$ and width $\Gamma_X$ as well as on the
values of the coupling constants $\varepsilon_e$ and $\eta_\alpha^{c}$ of the X17
particle to electrons and hadrons, respectively.  We report these and, in particular, the values we
have adopted for the LECs entering the combinations $\eta_\alpha^{c}$ in
Sec.~\ref{sec:resX}.  It is important to stress, though, that in this
first exploratory study, we are not interested in determining precisely these various
parameters (as well as their associated uncertainties), also
in view of the fact that the experimental evidence for the existence of the X17
boson is yet to be confirmed unambiguously.  Rather, our intent here is
(i) to setup the theoretical framework, and (ii) to investigate
possible experimental signatures of the X17, in particular, by establishing how
its nature affects the behavior of the cross section as function of
the energy and lepton angles.

\subsubsection{Numerical results: energy dependence}
We assume the width of the X17 to come from its decay into $e^+$-$e^-$ pairs
(the branching ratio for decay in a channel other than $e^+$-$e^-$, such as $\gamma\gamma$ or neutrinos,
is estimated to be negligible~\cite{Feng:2016jff,DelleRose:2018pgm}).
This width is seen to scale as $\Gamma^c_X\sim x_c\, \alpha\, \varepsilon_e^2\, M_X$
with a numerical factor $x_c$ of order unity---its precise value depending on the assumed
coupling between the X17 and the electron.  Current bounds indicate
$|\varepsilon_e| \lesssim 10^{-3}$, and therefore the expected width is of the
order of the eV or less, tiny relative to the typical energies of the emitted
electrons in the pair production process.

\begin{figure}[bth]
\centering
\includegraphics[scale=0.35,clip]{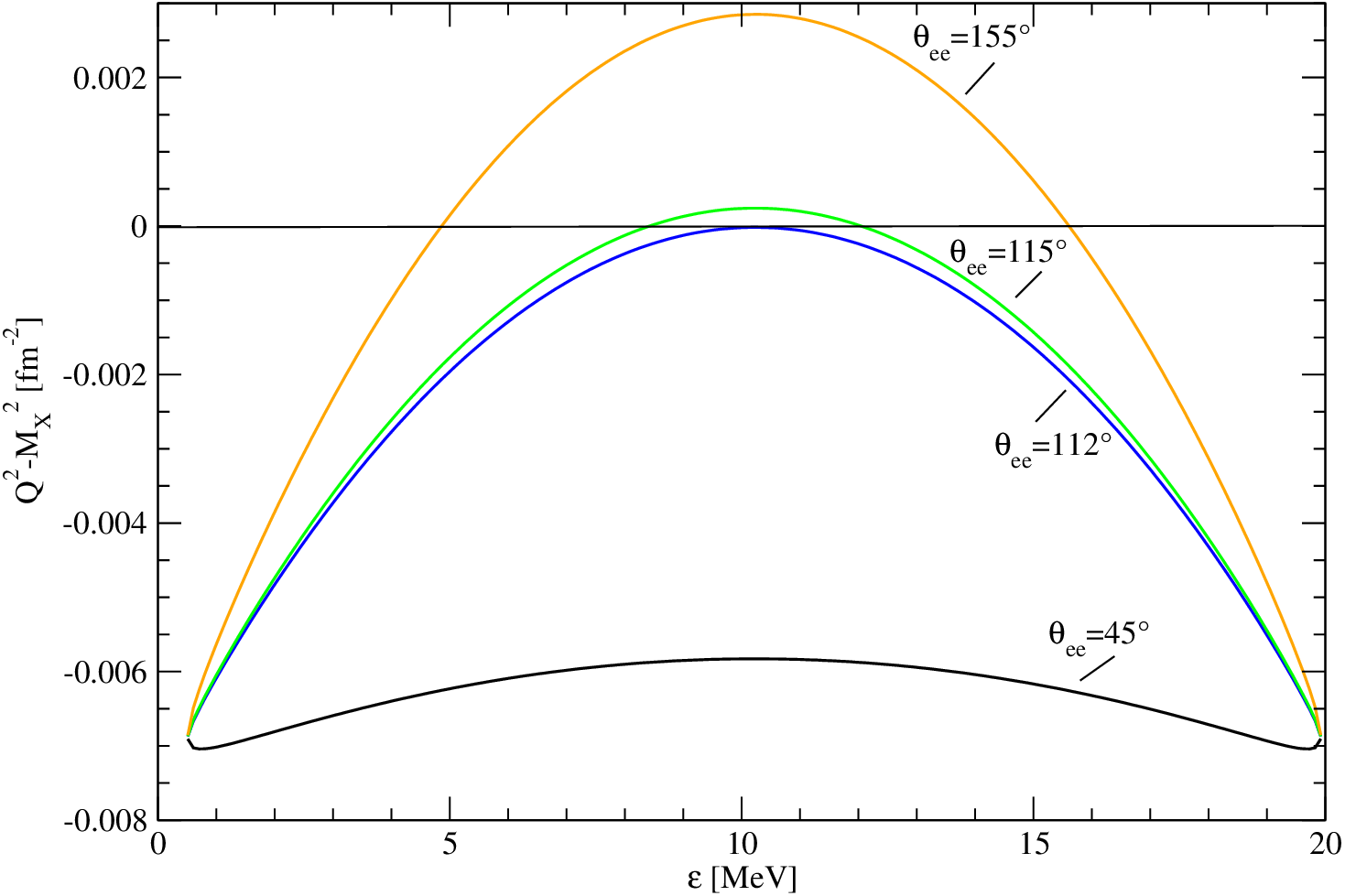}
\caption{Quantity $Q^2-M_X^2$ as function of the electron energy $\epsilon$ for
the configuration in which the lepton pair is produced in a plane
orthogonal to the initial proton momentum of energy 0.90 MeV, and
for various values of $\theta_{ee}$; here, $M_X\,$=$\,17$ MeV.
Note that for $\theta_{ee}<112^\circ$, the quantity $Q^2-M_X^2$ is always negative.}
\label{fig:qqmm}
\end{figure}
As is apparent from Eq.~(\ref{eq:e8t}), the cross section is most sensitive to
the X17 presence for momentum transfers $Q^2=q^\mu q_\mu$ such that
$Q^2\approx M_X^2$.  For the kinematical configuration corresponding to
the electron and positron being emitted in the plane perpendicular to the
proton momentum (to be specific, we are considering the $^3$H$(p,e^+ e^-)^4$He process 
in the setup of the ATOMKI experiment here),
there is a critical angle $\theta^*_{ee} \approx 112^\circ$, such that for
$\theta_{ee}> \theta^*_{ee}$ there are two electron energies $\epsilon_1$ and
$\epsilon_2$ that satisfy the condition above, see Fig.~\ref{fig:qqmm}.
For energies close to $\epsilon_i$, we can approximate
$Q^2-M_X^2\approx \alpha_i\, (\epsilon-\epsilon_i)$, and
the X17 propagator---rather, its magnitude square---as
\begin{equation}
\label{eq:e16aa}
{1\over |Q^2-(M_X-i\Gamma_X/2)^2|^2} 
 \approx  {\alpha_i^{-2}\over (\epsilon-\epsilon_i)^2 + (M_X\Gamma_X/\alpha_i)^2}\ ,
\end{equation}
namely a Lorentzian with a width given by $M_X\Gamma_X/|\alpha_i|$.
While this width is magnified by the factor 
$M_X |\alpha_i|^{-1}$, it is still found to be
tiny in comparison to typical $\epsilon_i$ values.

Such a narrow width makes it very difficult to reveal the interplay between
the electromagnetic and X17-induced amplitudes, in particular to
disentangle their interference.  However, the experiment has
a finite energy resolution, and folding this resolution with
the calculated cross sections corresponding to the X17 theoretical
width results in a broadening of the peaks.  
\begin{figure}[bth]
\centering
\includegraphics[scale=0.35]{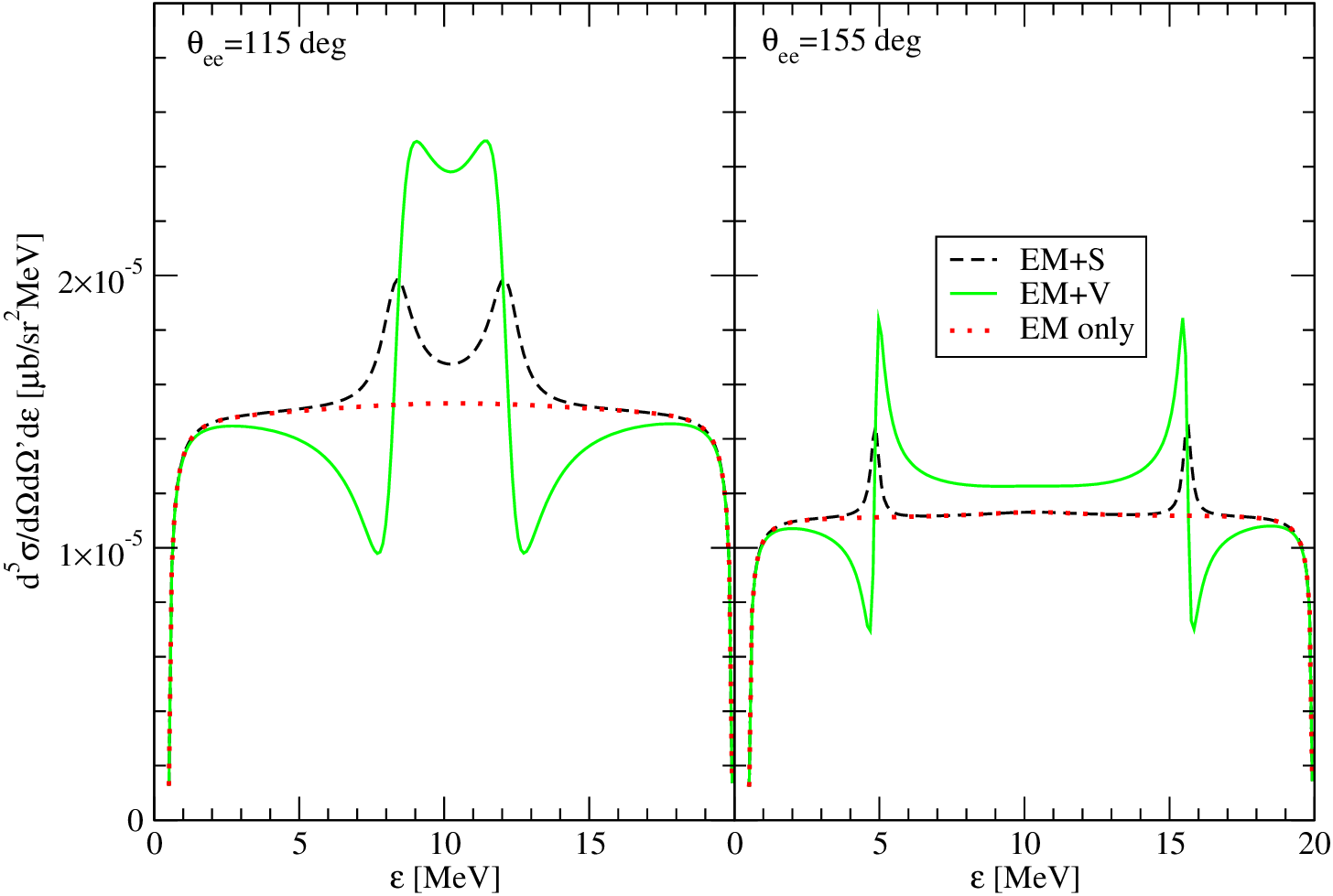}
\caption{The five-fold differential cross section for the process $\tri(p,e^-e^+)\heq$
in the ATOMKI setup
as function of the electron energy, for the configuration in which the $e^+$ and $e^-$ momenta
are at angles $\theta_{ee}$ of 115$^\circ$ (left panel)
and 155$^\circ$ (right panel) between them.  The (red) dotted curve represents the results
obtained by including the electromagnetic
contribution only; the (green) solid curve and (black) dashed curves
represent the results obtained by including a X17 boson interacting via a vector and scalar coupling, respectively. Here $M_X\,$=$\,17$ MeV and $\Gamma^{\rm eff}\,$=$\,0.4$ MeV, see
text for further explanations. The calculations are based on the N3LO500/N2LO500 interactions
and accompanying electromagnetic currents.
}
\label{fig:sig5_155}
\end{figure}
For the sake of illustration, in Fig.~\ref{fig:sig5_155} we show the five-fold differential cross section for the process $\tri(p,e^+e^-)\heq$ as function of the electron energy in the ATOMKI setup
(with proton energy of 0.90 MeV) for an effective, albeit perhaps unrealistic,
width $\Gamma^{\rm eff}_X\,$=$\,0.4$ MeV.
In the figure, we have reported the results obtained at angles $\theta_{ee}$ of 115$^\circ$ (left panel) and 155$^\circ$ (right panel), where the condition $Q^2$=$\,M^2_X$ is verified for two energies.
The (red) dotted curve represents the results obtained by including only electromagnetic contributions;
the (green) solid and (black) dashed curves include
the contribution of the X17 particle interacting with nucleons either via a proto-phobic vector or a scalar coupling. 
We have taken $M_X\,$=$\,17$ MeV with the remaining coupling constants chosen arbitrarily.

By inspecting the figure, we see that there is a strong interference between the
electromagnetic and vector X17-induced amplitudes.
This interference is weaker for the case of a scalar X17.
The energy dependence of the differential cross section
for pseudoscalar and axial couplings is similar to that
obtained for the scalar coupling, and the corresponding results are
not shown in Fig.~\ref{fig:sig5_155}; in particular, for a pseudoscalar
exchange, the interference (between electromagnetic and X17-induced amplitudes)
vanishes identically.  Unfortunately, at angles $\theta_{ee} \gtrsim110^\circ$ the
differential cross section is rather small (of the order of pb): its accurate
measurement would be experimentally very challenging.

\subsubsection{Numerical results: angular correlations}
In order to compare with the ATOMKI data for the
$\tri(p,e^-e^+)\heq$ reaction reported in Ref.~\cite{Krasznahorkay:2019lyl},
we consider the four-fold differential cross sections obtained by integrating over
the electron energy, with the remaining kinematical variables in the same
configuration above (that is, the momenta of the lepton pair in the plane
orthogonal to the incident proton momentum).  Since the ATOMKI 
cross section measurements are unnormalized, we rescale them
to match the calculated values for $\theta_{ee} \lesssim 90^\circ$, where the cross
section is dominated by the purely electromagnetic amplitude.  In the more recent
Ref.~\cite{Krasznahorkay:2021joi}, ``background-free'' $\tri(p,e^-e^+)\heq$ reaction data
are also reported, obtained by subtracting the counting rate due to  ``external''
pair conversion (EPC) processes. This EPC rate is estimated on the basis of a GEANT
simulation of the processes where real photons emitted in the
$\tri(p,\gamma)\heq$ radiative capture convert in lepton pairs
by interacting with the experimental apparatus~\cite{Krasznahorkay:2021joi}.
However, these data at angles $\lesssim 90^\circ$  have large errors,
making the matching between theory and experiment in this angular range rather problematic.
It is for this reason that, in this section, we compare with the (un-subtracted) data of
Ref.~\cite{Krasznahorkay:2019lyl} (qualitatively similar to the un-subtracted
data reported in the 2021 study), where errors at angles $\lesssim 90^\circ$ are much smaller.

\begin{figure*}[bth]
\centering
\includegraphics[scale=0.65]{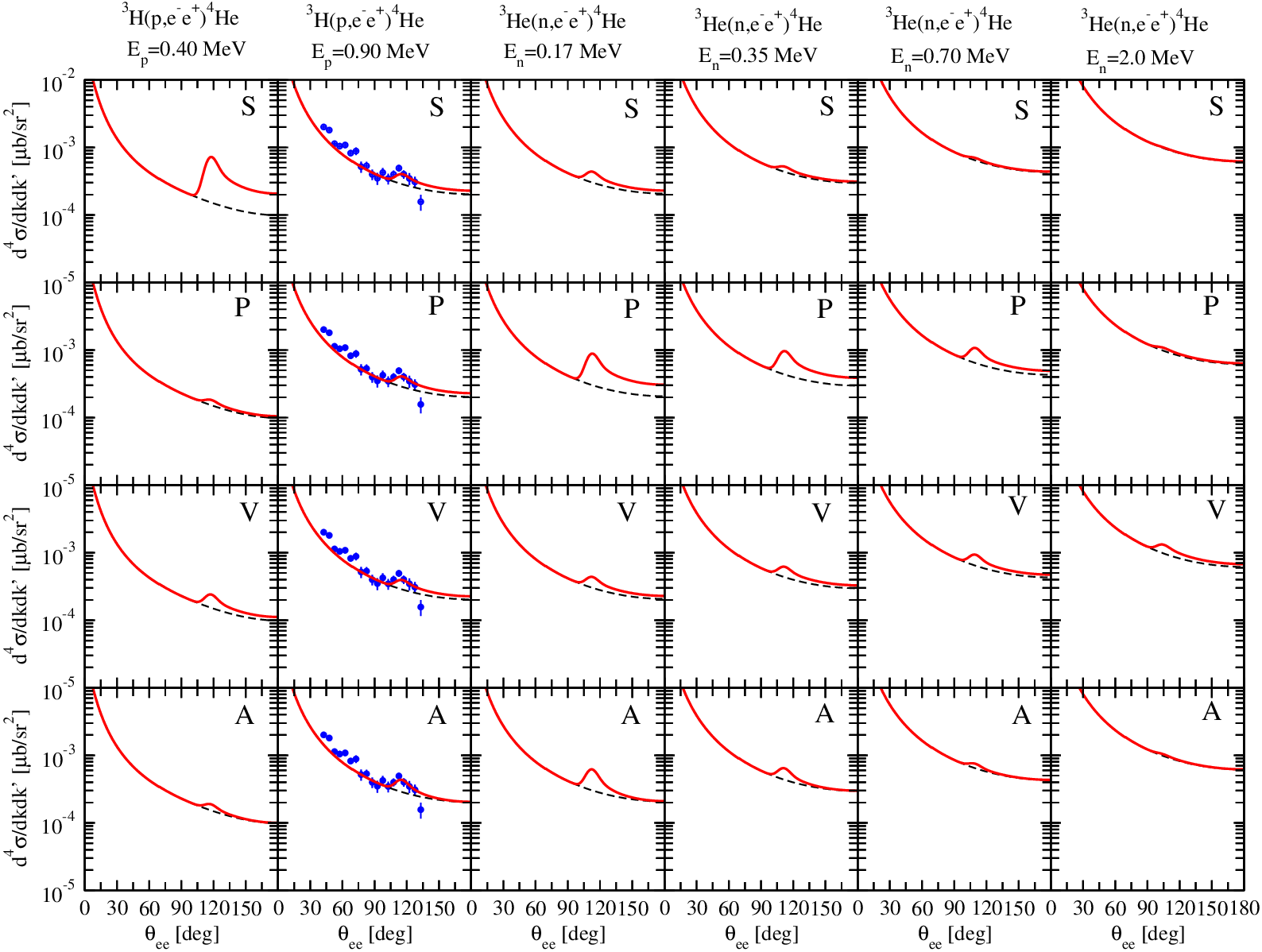}
\caption{The four-fold differential cross section for the $\tri(p,e^-e^+)\heq$
and $^3$He$(n,e^-e^+)\heq$ processes at six different incident nucleon energies
for the configuration in which the $e^+$ and $e^-$ momenta are
in the plane orthogonal to the incident nucleon momentum and as function
of the angle $\theta_{ee}$ between them.  The panels labeled S, P, V, and A show
the results obtained by including the exchange of a scalar, pseudoscalar, vector,
and axial X17, respectively.  In all cases, we have
taken $M_X\,$=$\,17$ MeV and $\Gamma_X\,$ as given from the X17 decay in $e^-e^+$,
and have adjusted the coupling constants so as to reproduce the
ATOMKI $\tri(p,e^-e^+)\heq$ cross section data of
Ref.~\protect\cite{Krasznahorkay:2019lyl} at the incident proton energy of 0.90 MeV
(see text for further explanations).  The dashed (black) and solid (red) curves show the results
obtained by including the electromagnetic only or both the electromagnetic and X17 amplitudes.
The calculations are based on the N3LO500/N2LO500 interactions
and accompanying electromagnetic currents.}
\label{fig:sige_X17_6ene_v6}
\end{figure*}

We report the calculated cross sections for both
$\tri(p,e^+e^-)\heq$ and $^3$He$(n,e^+e^-)\heq$ reactions at a number of
incident proton and neutron energies in Fig.~\ref{fig:sige_X17_6ene_v6}.
In computing the cross sections, we have taken the width $\Gamma_X$
from the X17 decay into $e^+$-$e^-$ pairs; however, we have folded the
resulting calculated values with a Gaussian, in order to account for the
finite angular resolution (see Sec.~\ref{sec:resX}).  For each of the assumed
couplings, we constrain the combinations $\eta^c_\alpha$ by fitting the ATOMKI
data (solid points with the error bars) obtained at an incident proton energy of 0.9 MeV and in the range
$\theta_{ee}> 90^\circ$, where the (purported) X17 signal has been observed
(we take as before $M_X\,$=$\,17$ MeV). The extracted values of the coupling
constants are reported in Sec.~\ref{sec:s7b}: they depend on the parameters
chosen to perform the angular smearing of the theoretical cross sections, and on
the factor used to rescale the Atomki data.  We
anticipate here that both the $\beo$ and $\heq$ anomalies can be consistently 
explained by the hypothesis of a vector X17, while for an axial X17
the coupling constants are found to be seemingly inconsistent with each other.
For the scalar and pseudoscalar case it is more difficult to draw any firm
conclusion. This issue is discussed in more detail in Sec.~\ref{sec:s7b}.

In Fig.~\ref{fig:sige_X17_6ene_v6}, the incident proton energies of 0.40
and 0.90 MeV correspond to energies $E_0$ in Eq.~(\ref{eq:e12}) of 20.12
and 20.50 MeV, respectively.  Referring to Fig.~\ref{fig:levels}, we see that
the lower 20.12 MeV corresponds to the energy of the first $0^+$ excited state,
while the higher 20.50 MeV (the energy selected in the ATOMKI experiment) is just
below the $n+\het$ threshold.  The incident neutron energies of 
0.17, 0.35, 0.70, and 2.0 MeV correspond to $E_0$ values of
20.69, 20.82, 21.08, and 22.08 MeV, and the first three
are close to the energy of the $0^-$ excited state, while
the last is on top of the $2^-$ excited state.

This structure of the $^4$He low-energy spectrum and the selectivity
of the X17-induced transition operator are reflected in the results of
Fig.~\ref{fig:sige_X17_6ene_v6}. Referring to Eq.~(\ref{eq:e10}) and
setting aside the isospin dependence generally of the form
$\eta^c_0+\eta^c_z\, \tau_{i,z}$, the operator structure $O^{cX}_i$ is:
1 (proportional to $\bmq\cdot{\bm \sigma}_i$) for a $S$ ($P$) boson
exchange and 1 (${\bm p}_i\cdot{\bm \sigma}_i$) for the time component
and ${\bm p}_i$ or $\bmq\times{\bm \sigma}_i$ (${\bm \sigma}_i$) for
a $V$ ($A$) boson exchange. The $S$ operator
(specifically, its isoscalar component) connects
the (predominantly isoscalar) $J^\pi\,$=$\, 0^+$ resonance
to the $^4$He ground state.  This transition is
responsible for the prominent low-energy peak seen in Fig.~\ref{fig:sige_X17_6ene_v6};
this peak rapidly fades away as the energy of the incident proton
increases in the $\tri(p,e^-e^+)\heq$ process.  It is barely visible
in the $^3$He$(n,e^-e^+)\heq$ process, since the $n\,+^3$He
threshold is already relatively far away from the $0^+$ resonance
energy, see Fig.~\ref{fig:levels}.

Since the purely isovector (pion-mediated)
$P$ transition operator has a small matrix element between
the (predominantly $T\,$=$\,0$) $0^-$ resonance and $^4$He ($0^+$) ground state,
we have ignored it altogether in the panels of Fig.~\ref{fig:sige_X17_6ene_v6}
by setting $\eta^P_z\,$=$\,0$---that is, by assuming a ``piophobic'' X17---and have
instead considered only the contribution from the isoscalar $P$ transition operator,
which has a large matrix element between these $0^-$ and $0^+$ states.
As a consequence, a pronounced peak structure is seen when the energy
approaches that of the $0^-$ resonance.
As this energy increases well beyond the $n+\het$ threshold, additional
significant contributions also come from the (relatively broad)
$2^-$ resonances.

The time component of the $V$ operator has the same structure
as the $S$ operator.  Its space component, however, produces large
electric dipole ($E_1$) matrix elements between the $1^-$ scattering state
and $\heq$ ground state (note in Fig.~\ref{fig:levels} the two
wide $1^-$ resonances located at energies close to the $d+d$ threshold).
These $E_1$ matrix elements are found to be much larger than the matrix elements
due to the $0^+\longrightarrow 0^+$ transition, and are the same for both photon and
X17-induced amplitudes (modulo coupling constants, of course), see also Ref.~\cite{Zhang:2020ukq} for a similar
finding in connection with the $\beo$ experiment.
Once the values of $\eta^V_0$ and $\eta^V_z$ have been constrained 
to reproduce the ATOMKI data, the height of the peak in the
full cross section is nearly constant relative to that in the purely
electromagnetic cross section for all energies
considered in this work.  By contrast, the $1^+$ scattering-state
contributions are suppressed since they have magnetic-dipole ($M_1$) character.
Moreover and  most importantly, in this $^3S_1$ wave the Pauli principle prevents the four
nucleons from coming close together.   Even though there are well defined and
fairly narrow $2^-$ resonances, the associated transitions, having $M_2$ character,
are suppressed.

The time component of the $A$ operator, which now has both
isoscalar and isovector terms, induces big matrix elements
between the $0^-$ scattering state and $\heq$ ground state.
In fact, this transition yields the dominant contribution, since
now the $1^-\longrightarrow 0^+$ transition is $M_1$ and therefore
suppressed by $q$ relative to the $0^-\longrightarrow 0^+$ one above.
The $1^+ \longrightarrow 0^+$ transition is $E_1$, but
again inhibited by the Pauli repulsion.  As a consequence, in the axial case
the X17 peak is more pronounced for energies at which the $0^-$ state is populated,
as for the $P$ case.  However, it should be noted that for large angles $\theta_{ee}$
(close to back-to-back configurations), the cross-section enhancement from
$A$ couplings is much reduced, due to the vanishing of the X17 $\rightarrow e^-e^+$ 
amplitude.
 
Next, we explore the dependence of the four-fold differential cross sections as 
function of the polar angles $\theta$ and $\theta'$ formed by the directions of the 
lepton momenta with respect to the incident beam momentum $\bmp$.  We only
consider configurations where $\theta\,$=$\,\theta'$.  The condition 
$Q^2=M_X^2$ is found to be satisfied for $55^\circ<\theta<125^\circ$
($M_X=17$ MeV). Therefore, outside this range
the X17 peaks do not appear in the cross section.  Furthermore, 
moving away from $\theta\,$=$\,90^\circ$, the difference $\Delta\phi=\phi'-\phi$ for which the 
condition $Q^2-M_X^2=0$ is verified, tends to increase, while the values
of the parameters $|\alpha_i|$ in the expansion of the X17 propagator
tend to decrease.  One would therefore expect the X17 peak to be located 
at larger and larger values of $\Delta\phi$, and its height to increase as $|\alpha_i|^{-1}$,
see Eq.~(\ref{eq:e16aa}).
\begin{figure}[bth]
\centering
\includegraphics[scale=0.45,clip]{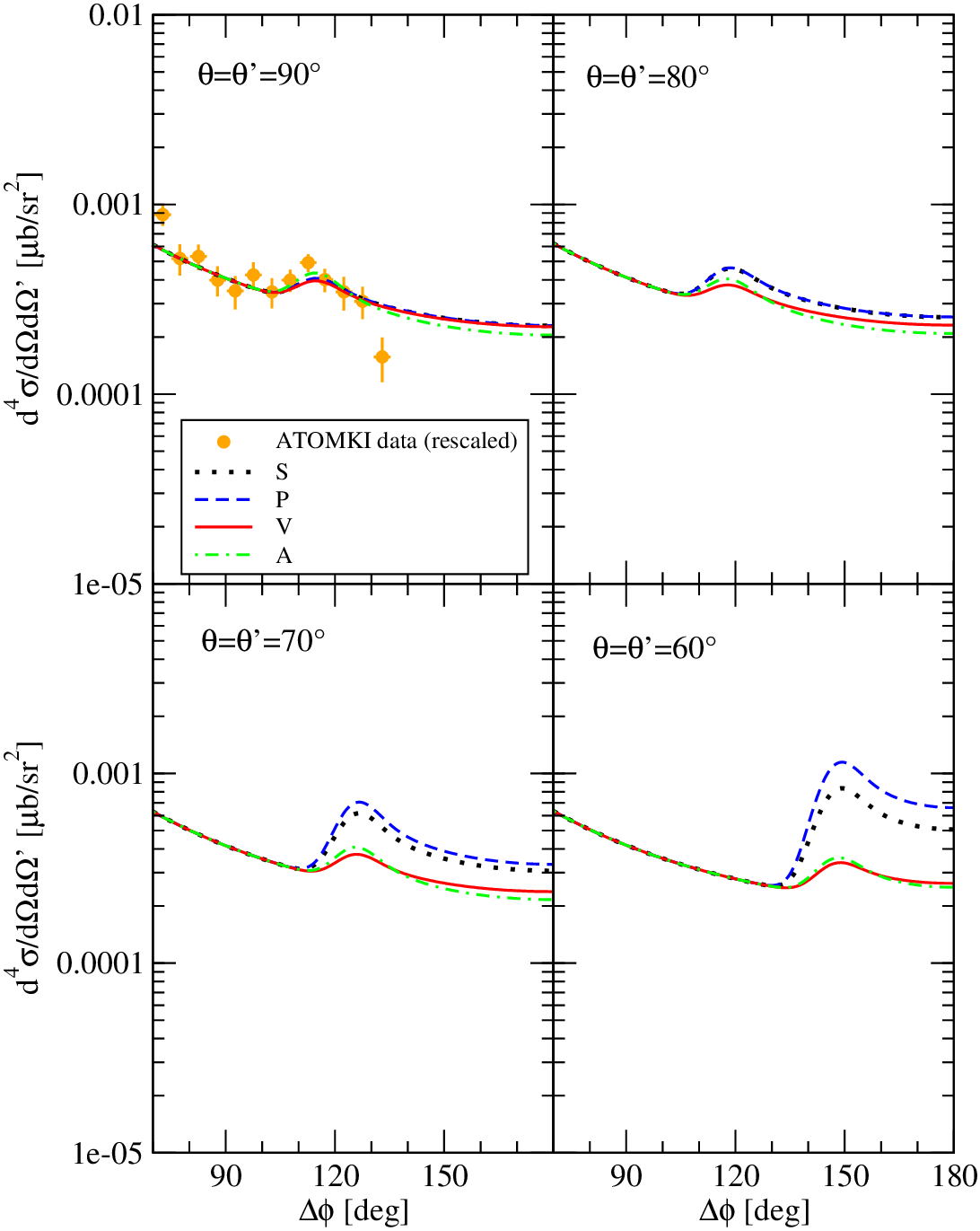}
\caption{The four-fold differential cross section for the $\tri(p,e^-e^+)\heq$
process at 0.90 MeV incident proton energy for the
configuration in which the $e^+$ and $e^-$ momenta are
emitted at angles $\theta\,$=$\,\theta'$ with respect to the incident proton momentum,
and as function of the difference $\Delta\phi=\phi'-\phi$.  
The curves labeled S, P, V, and A show the results obtained by including the exchange
of a scalar, pseudoscalar, vector, and axial X17, respectively.  In all cases, we have
taken $M_X\,$=$\,17$ MeV and $\Gamma_X$ from the decay in $e^-e^+$, and
have adjusted the coupling constants so as to reproduce the ATOMKI $\tri(p,e^-e^+)\heq$
data~\protect\cite{Krasznahorkay:2019lyl} at $\theta\,$=$\,\theta'\,$=$\,90^\circ$, rescaled as discussed in the main text.}
The calculations are based on the N3LO500/N2LO500 interactions
and accompanying electromagnetic currents.
\label{fig:sige_X17_theta}
\end{figure}

These expectations are generally borne out by the actual calculations,
as shown in Fig.~\ref{fig:sige_X17_theta}.  There is a clear dependence
on the assumed nature of the X17 boson.  It is worthwhile 
pointing out that for the pseudoscalar case, the larger increase observed in the
cross section for $\theta\,$=$\,\theta'\,$=$\,60^\circ$ comes from a kinematically enhanced contribution
of the charge multipole connecting the $2^-$ and
$0^+$ ($^4$He) states.\footnote{In the notation of Sec.~\ref{sec:x17cross}, the corresponding
term in the cross section is proportional to  $(1+3\,\cos^2\theta_q)\,|C_2^{112}(q)|^2$,
where $C_2^{112}(q)$ is the reduced matrix
element associated with the
transition between the $2^-$ and 
$0^+$ states and $\theta_q$ is the angle between the incoming nucleon momentum ${\bm p}$
and the momentum transfer
${\bm q}\,$=$\,{\bm k}+{\bm k}^\prime$; for $\theta\,$=$\,\theta^\prime$ we have
$\cos\theta_q\,$=$\,( k+k^\prime)\cos\theta/q$.  At the
X17 peak, the condition $Q^2\,$--$\,M_X^2\,$=$\,0$ and energy conservation in Eq.~(\ref{eq:e12}) lead to
\[
(k+k')/q \approx [1-(M_X/E_0)^2]^{-1/2}\approx 2
\]
for $E_0\approx 20$ MeV, and hence $\cos\theta_q \approx 2\,\cos\theta$.  The resulting cross
section behaves as $1+12\,\cos^2\theta$, rapidly increasing as $\theta\,$=$\,\theta^\prime$ move away from $90^\circ$.}
For the scalar case, the similar enhancement is due to the contribution
of the charge multipole connecting the $1^-$ and
$0^+$ states.\footnote{Again in the notation of Sec.~\ref{sec:x17cross}, this
multipole mainly contributes to the cross section with a term proportional to $\cos\theta_q\,{\rm Re}[C_0^{000}(q)^* C_1^{101}(q)]$, where
as discussed in the previous footnote, $\cos\theta_q \approx 2\,\cos\theta$. In the perpendicular 
plane (where $\theta_q=90^\circ$) this term vanishes, but gives a positive contribution for $\theta<90^\circ$,
enhancing the cross section.} Furthermore, we note that in the pseudoscalar (scalar) 
case the cross section increases (decreases) when $\theta=\theta'>90^\circ$. 
This amplifies the differences between the two cases.

\subsection{Experimental perspective}
\label{sec:s1e}
The experimental study of the $\tri(p,e^+ e^-)\heq$ reaction performed by the ATOMKI group seems
to indicate the existence of a X17 boson. However, it is difficult to establish its quantum numbers,
since the data were limited to a few proton energies and only leptons emitted in the plane orthogonal
to the beam line were detected. Furthermore, under certain conditions 
the data may be consistent with standard electromagnetic processes alone~\cite{Aleksejevs:2021zjw}, without 
the need for invoking the creation of a new particle.

In order to clarify the current ambiguous state of affairs, our calculations suggest to perform
an experimental study that covers a wide range in angle and energy, to fully scan the $0^+$, $0^-$, $2^-$, $1^-$ 
excited levels shown in Fig.~\ref{fig:levels}.  Such a study would allow us to either confirm or
exclude the existence of the X17, and ultimately study its properties, if its existence were to be corroborated. 
Although the excess of pair-production events as a function
of the energy depends on the X17 quantum numbers (i.e., on the nature of its coupling to electrons and nucleons),
an experimental setup in which only particles orthogonal 
to the beam axis are detected might be hindering our ability to discriminate among these different quantum numbers,
and hence uniquely identify the X17 properties. This limitation can be appreciated 
by inspecting Fig.~\ref{fig:sige_X17_6ene_v6}, where the predicted trend of the excess is found to 
be quite similar for the pseudoscalar and axial cases.
%By contrast, 
However, as shown in Fig.~\ref{fig:sige_X17_theta}, the use of a detector with a large angular acceptance
would make it possible to discriminate among different options since the angular distribution
of the emitted pair depends appreciably on the X17 quantum numbers (a comprehensive analysis 
of different kinematical configurations will be reported in a future publication). A dedicated
detector could also provide a measurement of the pair four-momenta as well as particle
identification, to ascertain that the pair is truly an $e^+e^-$ one. 

A prerequisite to realize such a program is the availability of high intensity proton and neutron beams.
Concerning the $\tri(p,e^+ e^-)\heq$ reaction, a promising facility is the LUNA-MV accelerator that 
will soon be operative at the underground Gran Sasso Laboratory (LNGS).  At the LNGS the cosmic ray 
induced background is many orders of magnitude 
lower than at overground facilities, and the proton beam intensity is a factor hundred higher than at the ATOMKI facility.
Thus, LUNA-MV is well suited to perform accurate measurements in the proton energy range approximately ($0.2$--$1.0$) MeV, 
and in a relatively short time.
In this energy range lies the $0^+$ resonance located $0.50$ MeV above the $p+^3$H threshold. The maximal proton
energy is determined by the onset of the huge production of neutrons due to the charge exchange reaction
$^3$H$(p,n)^3$He for proton energy $\ge 1.02$ MeV.
The experimental setup could be based on the use of a novel RICH (Ring Imaging Cherenkov) detector
with large angular acceptance, surrounding the tritium target~\cite{Gustavino:2021abcd}. The
RICH detector, currently under study, consists of aerogel radiators producing rings of Cherenkov
light when crossed by a relativistic particle, which is collected 
by an array of Silicon Photomultiplier (SiPM). Such a detector is blind to non-relativistic particles 
(e.g., the scattered protons of the beam) and is almost insensitive to high energy gammas (e.g., 
the $\sim 20$ MeV photons produced by the $^3$H$(p,\gamma)^4$He radiative capture) because of the
large radiation length of aerogels.  In practice, only positrons and electrons exceeding MeV kinetic 
energies are detected.
The RICH detector mentioned above is especially well suited to measure the $\het(n,e^+ e^-)\heq$ cross section.
A good site for this measurement is the CERN $n\_ {\rm TOF}$ facility, which provides a pulsed neutron beam
in a wide energy range ($E_n = 1-10^8$ eV). However, the energy of each interacting 
neutrons can be accurately derived with the Time-of-Flight (TOF) technique~\cite{Sabate-Gilarte:2017biu}. 
Even though the dominant channel induced by neutrons is the $\het(n,p)\tri$ charge-exchange one 
(with a $Q$ value of 764 keV), the RICH is completely blind to the non-relativistic protons produced 
by this reaction, in the whole $n\_ {\rm TOF}$ range.  This neutron-induced experiment would allow 
us to extend the $^4$He de-excitation study up by several MeVs, including the energies reported 
in Fig.~\ref{fig:levels}. Finally, the cross section measurement of both the $\tri(p,e^+ e^-)\heq$ 
and (for the first time) $\het(n,e^+ e^-)\heq$ conjugate reactions could reveal possible
peculiarities of the hypothetical proto-phobic fifth force mediated by the X17 boson.

\subsection{Concluding remarks}
\label{sec:conc}
A major objective of the present work has been to provide an accurate treatment of
the $e^+$-$e^-$ pair production process in the four-nucleon system, based on the
one-photon-exchange approximation and
a state-of-the-art $\chi$EFT description of nuclear interactions and
electromagnetic currents.  The initial $3+1$ scattering-state and $^4$He
bound-state wave functions have been obtained from HH solutions of the Schr\"odinger
equation.  In particular, these solutions fully account for the coupling among
different (energetically open) channels and for the presence of resonances
observed in the $A\,$=$\,4$ low-energy spectrum.

In the kinematics of the ATOMKI experiment where the $e^+$-$e^-$
pair is detected in the plane orthogonal to the incident nucleon momentum, the
predicted cross sections for the $^3$H$(p,e^+ e^-)^4$He and $^3$He$(n,e^+ e^-)^4$He
have been found to be monotonically decreasing as function of the opening angle between
the electron and positron momenta, albeit flattening as these momenta approach the
back-to-back configuration, and to increase as function of the incident nucleon
energy, see dashed (black) lines in Fig.~\ref{fig:sige_X17_6ene_v6}.

In the low-energy regime of the ATOMKI experiment, these cross sections are
dominated by the transitions from the $0^+$ ($^1S_0$) and $1^-$ ($^1P_1$)
scattering states to the $0^+$ ground state via $C_0$, and $C_1$ and $E_1$ multipole
operators.  The model dependence from the two different $\chi$EFT implementations
we have adopted (without and with explicit $\Delta$-isobar degrees of freedom, and
formulated in either momentum or configuration space) appears to be negligible.
These results should provide a reliable and accurate baseline for the analysis of current
(and possibly future) experiments in the four-nucleon system.

Next, we have considered how the X17 boson with a mass of 17 MeV might affect
the pair-production process.  Figures~\ref{fig:sige_X17_6ene_v6} and~\ref{fig:sige_X17_theta}
suggest that a systematic study of the cross section
as function of both the opening angle and electron energy might allow
us to discriminate among the different hypotheses regarding the nature of X17.
Such a study is planned for the $^3$H$(p,e^+ e^-)^4$He and
$^3$He$(n,e^+ e^-)^4$He experiments, currently in the feasibility
and development phase at, respectively, the LNGS LUNA-MV and CERN $n\_ {\rm TOF}$
facilities.\footnote{We are of course available to provide
theoretical support in the analysis and interpretation of these as well as the
new ATOMKI experiment~\cite{Krasznahorkay:2021joi}.} 
 
Lastly, in order to further investigate the robustness of the currently predicted
electromagnetic cross sections, we also plan to go beyond the one-photon-exchange
approximation and include higher-order QED corrections in our treatment of
pair production in the $A\,$=$\,4$ system.

\section{Hamiltonians and wave functions}
\label{sec:wf}
The nuclear Hamiltonians consist of non-relativistic
kinetic energy, two-nucleon ($2N$) and three-nucleon ($3N$)
interactions.  In order to estimate the model dependence of the
various predictions, we have considered interactions derived from
two different $\chi$EFT formulations.  In the first~\cite{Entem:2003ft,Machleidt:2011zz},
nucleons and pions are retained as explicit degrees of freedom and
terms up to next-to-next-to-next-to-leading order (N3LO), in the
standard Weinberg counting, are accounted for.  The
resulting interaction is formulated in momentum space
and is regularized with a cutoff $\Lambda\,$ set equal to 500 MeV; it will be referred to
as N3LO500.  In particular, as a consequence of this regularization procedure,
this interaction is strongly non-local in configuration space.
 
The second formulation, developed in Refs.~\cite{Piarulli:2016vel,Piarulli:2017dwd},
retains nucleons, pions, and $\Delta$-isobars as
degrees of freedom, but utilizes a ``hybrid'' counting rule according to which
terms from contact interactions are promoted relative to those
resulting from pion exchange.  The $2N$ interactions that have been
constructed in this formulation, consist of a long-range component
from one- and two-pion exchange, including $\Delta$-isobar intermediate
states, up to next-to-next-to-leading order (N2LO), and a short-range
component from contact terms up to N3LO.
A distinctive feature is that they are formulated and regularized
in configuration space so as to make them local in this space.
In this paper we use the version NVIa with cutoffs for the short- and long-range
components of the interaction given, respectively, by
$R_{\rm S}\,$=$\,0.8$ fm and $R_{\rm L}\,$=$\,1.2$ fm.

Along with the N3LO500 $2N$ interaction, we
include the $3N$ interaction 
that has been derived up to N2LO~\cite{Epelbaum:2002vt} in the
$\chi$EFT formulation based on pions and nucleons only. It
consists of two-pion exchange and contact terms, the former proportional to
(known) LECs that enter the subleading $\pi N$ chiral Lagrangian
${\cal L}^{(2)}_{\pi N}$, and the latter proportional to the (unknown) LECs,
in standard notation, $c_D$ and $c_E$.  This interaction
is regularized (in momentum space) with a cutoff $\Lambda\,$=$\,500$
MeV.  The LECs $c_D$ and $c_E$ have been
determined by reproducing the experimental values of the
$^3$H binding energy and Gamow-Teller (GT) matrix element in
the $\beta$ decay of tritium.  The original determination used the erroneous relation
between $c_D$ and the LEC that characterizes the contact
axial current.  Consequently, $c_D$ and $c_E$ have been
refitted in Ref.~\cite{Baroni:2018fdn}, and their values
are listed in Table~\ref{tab:par}.

In the $\chi$EFT formulation which also includes $\Delta$ isobars,
the $3N$ interaction at N2LO receives an additional contribution from a
two-pion exchange term with a single $\Delta$-isobar intermediate
state~\cite{Piarulli:2017dwd}; it is characterized by known LECs.
This as well as the two-pion-exchange term from ${\cal L}^{(2)}_{\pi N}$
and the contact terms proportional to $c_D$ and $c_E$ are regularized
in configuration space in a way that is consistent with the NVIa
interaction; in particular, $c_D$ and $c_E$ have been fixed by reproducing
the same observables above~\cite{Baroni:2018fdn} and their values too are
reported in Table~\ref{tab:par}.

Results obtained with these two Hamiltonians,
referred to hereafter as N3LO500/N2LO500 and NVIa/3NIa, should
provide an indication of the sensitivity of the low-energy
matrix elements and cross sections of interest in the
present work to different dynamical inputs.  A more systematic
study of this sensitivity---for example, to interactions and currents
constructed using different cutoffs and chiral orders---is deferred
to a subsequent publication.

HH methods, as described 
in considerable detail in Refs.~\cite{Kievsky:2008es,Viviani:2020gkm,Marcucci:2020fip},
are used to calculate the $A\,$=$\,3$--4 bound- and scattering-state
wave functions.  In this and following subsection, we provide a summary
for completeness.  We begin by discussing bound-state wave functions very briefly.

\begin{table}[bth]
\caption{\label{tab:par}
The combinations of $2N$ and $3N$ interactions used in
the present work and the (adimensional) fitted values for the LECs $c_D$ and
$c_E$ corresponding to each of these combinations, along with
the $\tri$, $\het$, and $\heq$ calculated binding energies in MeV (experimental
values are reported in the last line).  The values for $1/m_N$ used in the calculations
corresponding to the N3LO500/N2LO500 and NVIa/3NIa Hamiltonians are 41.47 and
41.47107 in units of MeV-fm$^2$, respectively.}
\begin{center}
\begin{tabular}{l||c|c||c|c|c}
\hline 
\hline 
Model &  $c_D$ & $c_E$ & $\tri$ & $\het$ & $\heq$ \\
\hline
N3LO500/N2LO500  & $+0.945$ & $-0.041$ & $8.471$ & $7.729$ & $28.34$ \\
%N3LO600/N2LO600  & $+1.145$ & $-0.6095$ & $8.467$ & $7.733$ & $28.59$ \\
%\hline
NVIa/3NIa  & $-0.635$ & $-0.090$ & $8.482$ & $7.714$ & $28.53$ \\
%NVIb/3NIb  & $-4.71$ & $+0.550$ & $8.473$ & $7.728$ & $28.15$ \\
\hline
Exp  & & & $8.480$ & $7.718$ & $28.30$ \\
\hline
\end{tabular}
\end{center}
\end{table}

The $\tri$, $\het$, and $\heq$ wave functions
are written as an expansion over spin-isospin-HH states times
hyperradial functions, which are themselves expanded on a basis
of Laguerre polynomials.  The Laguerre-expansion coefficients are
then taken as variational parameters (see Refs.~\cite{Kievsky:2008es,Marcucci:2020fip} for a more comprehensive discussion),
\begin{equation}
  \Psi= \sum_{\mu} c_{\mu}\; \Phi_{\mu}   \ , \label{eq:bswf}
\end{equation}
where $\mu$ denotes collectively the quantum numbers specifying
the combination $\Phi_\mu$ of spin-isospin-HH states.  The Rayleigh-Ritz
variational principle,
\begin{equation}
\langle \delta_c\Psi|H-E|\Psi \rangle=0 \ ,
\label{eq:rrvar}
\end{equation}
is used to determine the expansion coefficients $c_{\mu}$ in
Eq.~(\ref{eq:bswf}) and bound state energy $E$.  The variational
energies obtained for $\tri$, $\het$, and $\heq$ with the nuclear
Hamiltonians considered here are reported in Table~\ref{tab:par}.
We note that the predicted energies for the helium isotopes
(the $^3$H energy is fitted) are close to the experimental values.

\subsection{The $p\,$+$\,\tri$ and $n\,$+$\,\het$ wave functions}
\label{sec:cs}

We use the index $\gamma$ to specify
the asymptotic clusterization under consideration: $\gamma\,$=$\,1$ for $p+\tri$
and $\gamma\,$=$\,2$ for $n+\het$.  Of course, depending on the relative
energy between the 1+3 clusters, the states $p+\tri$
and $n+\het$ may be coupled; however, coupling of these
to 2+2 states will not play a role for the energies we will be
considering below.  We also find it convenient to
introduce the following asymptotic wave functions with relative
orbital angular momentum $L$, channel spin $S$, and total
angular momentum $J$,
\begin{eqnarray}
  \Omega_{\gamma LS,JJ_z}^F &=& {1\over\sqrt{4}} \sum_{\ell=1}^4 
  \Bigl [ Y_{L}(\hat{\bm y}_\ell) \otimes  [ \phi_\gamma(ijk) \otimes \chi_\gamma(\ell)]_{S} 
    \Bigr ]_{JJ_z} \nonumber \\
        && \times {\frac{F_L(\eta_\gamma,p_\gamma y_\ell)}{p_\gamma y_\ell}}  \ ,  \label{eq:fpt}\\
  \Omega_{\gamma LS,JJ_z}^G &=& {1\over\sqrt{4}} \sum_{\ell=1}^4 
  \Bigl [ Y_{L}(\hat{\bm y}_\ell) \otimes  [ \phi_\gamma(ijk) \otimes \chi_\gamma(\ell)]_{S} 
    \Bigr ]_{JJ_z}\nonumber\\
        && \times {\frac{G_L(\eta_\gamma,p_\gamma  y_\ell)}{p_\gamma y_\ell}} g(y_\ell) \ ,  \label{eq:gpt}
\end{eqnarray}
where $y_\ell$ is the 1-3 separation with $ijk$ and $\ell$ denoting the
particles in the bound cluster and the
isolated nucleon, $p_\gamma$ is the magnitude of the relative momentum
between the two clusters, $\phi_{1}$ ($\phi_{2}$)  is the $\tri$ ($\het$) bound-state wave
function, $\chi_1$ ($\chi_2$) the proton (neutron) spin state,
and $F_L$ and $G_L$ are the regular and irregular Coulomb functions, respectively.
The function $g(y_\ell)$ modifies the $G_L(p  y_\ell)$ at small $y_\ell$ by
regularizing it at the origin, and $g(y_\ell) \longrightarrow 1$
as $y_\ell \gtrsim 10$ fm, thus
not affecting the asymptotic behavior of $\Omega_{\gamma LS,JJ_z}^G$
(see Ref.~\cite{Viviani:2020gkm}).  The parameter $\eta_\gamma$ is defined as
\begin{equation}
\eta_\gamma={e^2\,\mu_\gamma \over p_\gamma}\ ,\label{eq:eta}
\end{equation}    
and $e^2$ is taken as 1.43997 MeV-fm.  For $\gamma=2$, there is no Coulomb
interaction, and the functions $F_L$ and $G_L$ reduce to
\begin{equation}
 {F_L(\eta,py)\over py } \longrightarrow j_L(py)\ , \qquad
 {G_L(\eta,py)\over py } \longrightarrow - y_L(py)\ , 
 \label{eq:eta0}
\end{equation}
where $j_L$ and $y_L$ are the regular and irregular spherical Bessel functions,
respectively. 
Lastly, the total energy of the scattering state in the center-of-mass frame is 
\begin{equation}
  E=-B_3^\gamma+T_{\gamma}\, \label{eq:energy} \ ,
\end{equation}
where $B_3^{\gamma=1}$ ($B_3^{\gamma=2}$) specifies the $\tri$ ($\het$) binding energy,
$T_{\gamma}\,$=$\,p_\gamma^2/(2\mu_\gamma)$ the relative kinetic energy, and $\mu_\gamma$ is
the 3+1 reduced mass.

The scattering wave function of total angular momentum $J$ with an incoming  cluster $\gamma$ 
having orbital angular momentum $L$ and channel spin $S$ (with $S\,$=$\,0$ or  1)
is written as 
\begin{eqnarray}
\label{eq:psica} 
  \Psi^{(\gamma)}_{LSJJ_z}&\!=\!&\Psi^C_{\gamma LS, JJ_z}+\Omega_{\gamma LS,JJ_z}^F \\
  &\!+\!& \sum_{\gamma' L'S'}\!\!
  T^{J}_{\gamma LS,\gamma' L'S'} \Bigl( \Omega_{\gamma' L'S',JJ_z}^G \!+\! i\, \Omega_{\gamma' L' S',JJ_z}^F\Bigr)\ ,\nonumber
\end{eqnarray}
where the term $\Psi^C$ vanishes in the limit of large inter-cluster separations, and
hence describes the system in the region where the particles are close
to each other and their mutual interactions are strong.
The other terms describe the system in the asymptotic region,
where inter-cluster interactions are negligible (except for the long-range
Coulomb interaction in the $\tri+p$ case). This asymptotic
wave function is expressed in terms of $T$-matrix elements, that is,
it consists of a (distorted) plane wave plus an outgoing wave.
%In fact, the combination
%$ \Omega_{\gamma LS,JJ_z}^G+ \ii \,\Omega_{\gamma LS,JJ_z}^F$ has a term proportional
%to $G+\ii \,F\longrightarrow e^{\ii p y}$ for very large $y$. 
The core wave function $\Psi^C$ is expanded as in Eq.~(\ref{eq:bswf}),
and the expansion coefficients along with the $T$-matrix elements
$T^{J}_{\gamma LS,\gamma' L'S'}$ are determined by making use
of the Kohn variational principle~\cite{Viviani:2020gkm}.

The more technical aspects in the application of this technique and, in particular,
the issues relating to convergence and numerical stability are discussed
thoroughly in Ref.~\cite{Viviani:2020gkm}. 
The convergence of the HH expansion is generally not a problem
for chiral interactions, except for the $p+\tri$ $J^\pi=0^+$
state below the $n+\het$ threshold.
As a matter of fact, in the present study we have been
able to improve significantly the convergence rate in this channel by including the
$\gamma=2$ asymptotic states $\Omega_{2 L'S',JJ_z}^{F,G}$ also for energies
below the $n+\het$ threshold, that is, for $E<B_3^{\gamma=2}$.
We do so by setting in this regime
\begin{equation}
 {F_L(\eta,py)\over py } \longrightarrow 0\ , \qquad
 {G_L(\eta,py)\over py } \longrightarrow {e^{-\beta y}\over \beta y}\ , 
 \label{eq:etam0}
\end{equation}
where $\beta$ is the imaginary part of $p$.

Accurate benchmarks between the results otained with the HH method and those
calculated by means of the Faddeev-Yakubovsky equations
(solved in momentum and configuration space) were reported
in Ref.~\cite{Viviani:2011ax} for $n+\tri$ and
$p+\het$ elastic scattering, and in Ref.~\cite{Viviani:2016cww} for $p+\tri$ and
$n+\het$ elastic and charge-exchange reactions.
These calculations were limited to energies below the threshold
for three-body breakup. The good agreement found among these
drastically different methods attests to the high accuracy
achieved in solving the $A\,$=$\,4$ scattering problem.

In reference to the model dependence of the nuclear Hamiltonian,
it is weak for scattering observables at energies above
the $n+\het$ threshold. By contrast, the model dependence---especially,
that originating from the cutoff used to regularize the $2N$ and $3N$
chiral interactions---becomes strong at energies below this threshold,
in particular for the $0^+$ state~\cite{Viviani:2020gkm}. 
We have speculated that this effect might be related to a critical dependence
of the position and width of the resonance representing the first excited
state of $\heq$ upon the $3N$ interaction. Experimental studies of this
resonance are currently in progress using electron scattering on
$\heq$~\cite{Kegel:2021abcd}.

Finally, in calculating transition matrix elements we utilize
the wave functions $\Psi^{(\gamma)}_{m_3,m_1}(\bmp_\gamma)$, defined as
  \begin{eqnarray}
  \Psi_{m_3,m_1}^{(\gamma)}(\bmp_\gamma)&=&
    \sum_{SS_z LM JJ_z} \langle {1\over2} m_3 {1\over2} m_1 | SS_z\rangle \langle LM SS_z| JJ_z\rangle \nonumber\\
    &\times&
    4\pi i^L e^{i\phi^{(\gamma)}_L} Y_{LM}(\hat\bmp_\gamma) \,\Psi^{(\gamma)}_{LSJJ_z} \ ,\label{eq:psi1}
  \end{eqnarray}  
where $\bmp_\gamma$ is the relative momentum
between the two clusters, $m_3$ ($m_1$) is the spin projection
of the trinucleon bound state (isolated nucleon),
and $\phi^{(\gamma)}_L$ is the Coulomb phase shift $\sigma_L$ for
$\gamma\,$=$\,1$ or simply vanishes for $\gamma\,$=$\,2$.  These wave functions
are normalized so that in the absence of inter-cluster interactions
they reduces to Eq.~(\ref{eq:e11}).

\section{Electromagnetic cross sections}
\label{sec:s-em}
In this section, we report on the calculation of the $\tri(p,e^+e^-)\heq$
and $\het(n,e^+e^-)\heq$ cross sections in the one-photon exchange
approximation. 

\subsection{The cross section}
\label{sec:xs_em}
The relevant transition matrix element in the lab frame reads
\begin{equation}
 T^{(\gamma)}_{fi} = \frac{4\pi\alpha}{q^\mu\, q_\mu}\,
    \ell^\mu_{\bmk s,\bmk' s'}\,
    \langle \Psi({\bm p}-\bmq) | j^\dagger_\mu(\bmq) | \Psi^{(\gamma)}_{m_3,m_1}({\bm p})\rangle \ ,
    \label{eq:ampli1}
\end{equation}
where $\ell^\mu$ denotes the matrix element of the
leptonic current, $\Psi^{(\gamma)}({\bm p})$ is the initial 1+3 state with
the incident nucleon having momentum ${\bm p}$, $\Psi({\bm p}-\bmq)$ is
the final $^4$He ground state recoiling with momentum ${\bm p}-\bmq$,
and $j^\mu(\bmq)$ is the
nuclear electromagnetic current operator.\footnote{The relation between
the nucleon lab momentum ${\bm p}$ and the relative momentum ${\bm p}_\gamma$
of the previous section is ${\bm p}_\gamma=(\mu_\gamma/m_N)\,{\bm p}$.}  We have defined
the four-momentum transfer $q^\mu\,$=$\,k^\mu+k^{\prime\mu}\,$$\equiv$$\,(\omega,\bmq)$,
where $k^\mu\,$=$(\epsilon,{\bm k})$ and $k^{\prime\mu}\,$=$\,(\epsilon^\prime,{\bm k}^\prime)$
are the outgoing electron and positron four
momenta with corresponding spin projections $s$ and $s^\prime$, and
the leptonic current matrix element as
\begin{equation}
\ell^\mu_{\bmk s,\bmk' s'} = \overline{u}(\bmk,s) \,\gamma^\mu \,v(\bmk',s') \ ,
\end{equation}
where we have chosen to normalize the spinors as $u^\dagger u\,$=$\, v^\dagger v\,$=$\,1$.

After enforcing momentum conservation, the unpolarized cross section follows as
\begin{equation}
 d\sigma^{(\gamma)}={1\over 4}\sum_{m_3 m_1}\sum_{s s'} 
 \frac{|T^{(\gamma)}_{fi}|^2}{v_r}\, d\phi \ ,
\end{equation}
where $v_r\,$=$\,p/m_N$ is the relative velocity ($m_N$ being the nucleon mass),
the phase-space factor $d\phi$ is
\begin{equation}
d\phi=2\pi\delta(E_i-E_f)\,
 {d^3\bmk\over (2\pi)^3} \,{d^3\bmk'\over (2\pi)^3}\ ,
 \label{eq:e28a}
\end{equation}
and $E_i$ and $E_f$ are initial and final energies, respectively.
Carrying out the sum over the lepton-pair spins yields
\begin{equation}
\label{eq:e26}
  {d^6\sigma^{(\gamma)}\over d^3\bmk \, d^3\bmk^\prime}=\frac{1}{ (2\pi)^3}\,
  {\alpha^2\over Q^4} {k\,k'\over  v_r } \delta(E_i-E_f) \,R_{fi} \ ,
 \end{equation}
 where we have defined $Q^2=q^\mu\, q_\mu$, and have introduced the nuclear electromagnetic
 response $R_{fi}$ (the superscript $(\gamma)$ is understood).
 In terms of matrix elements of the
 current, denoted schematically below as $j^\mu_{fi}=(\rho_{fi},{\bm j}_{fi})$, this response reads
 \begin{equation} 
 R_{fi}=\sum_{m_3 m_1} \left[ -(m_e^2+k\cdot k') j^{*}_{fi} \cdot j_{fi} -
 |P\cdot j_{fi}|^2/2 \right ] \ ,
\end{equation}
where $m_e$ is the lepton mass and $P^\mu\,$=$\,k^\mu-k^{\prime\mu}$.  Of course, in
the expression above, we made use of current conservation, that is,
$q\cdot j_{fi}\,$=$\,0$.  Lastly, conservation of energy leads to
Eq.~(\ref{eq:e12}).  

In order to make the dependence on the lepton pair kinematics
explicit, we introduce the basis of unit vectors
\begin{equation}
\hat{\bm e}_{z}=\hat\bmq\ ,\qquad \hat{\bm e}_{y}=\frac{{\bm p}\times\bmq}{|{\bm p}\times\bmq|}
\ ,\qquad \hat{\bm e}_{x}=\hat{\bm e}_{y}\times\hat{\bm e}_{z} \ ,
\end{equation}
where the incident nucleon momentum ${\bm p}$ defines the quantization axis of
the nuclear spins.  Then, the nuclear electromagnetic response can be written as
\begin{equation}
R_{fi}=\sum_{n=1}^6v_n\, R_n \ ,
\end{equation}
where the $v_n$ only involve the lepton kinematical variables and the
reduced response functions $R_n$ denote appropriate combinations
of the nuclear current matrix elements, as specified below.  We find
\begin{eqnarray}
\label{eq:e30}
  v_1 &=& (Q^4/q^4)(\epsilon\epsilon^\prime+ \bmk\cdot\bmk^\prime-m_e^2)\ , \nonumber\\
  v_2&=& -P_x\,[ \epsilon-\epsilon^\prime-(\omega/q)\, P_z ]/\sqrt{2} \ , \nonumber\\
  v_3&=& -P_y\,[ \epsilon-\epsilon'-(\omega/q) \,P_z]/\sqrt{2} \ ,\label{eq:vi}\\
  v_4&=& -(P_x^2+P_y^2)/4 + m_e^2 + \epsilon\epsilon' - \bmk\cdot\bmk'\ , \nonumber\\
  v_5&=& (P_x^2-P_y^2)/2\ , \nonumber\\
  v_6&=& -P_x P_y \ , \nonumber
\end{eqnarray}
and
\begin{eqnarray}
\label{eq:e31}
  R_1 &=& \sum_{m_3,m_1} |\rho_{fi}|^2\ , \nonumber\\
  R_2 &=& \sum_{m_3,m_1} {\rm Re}\, [ \,\rho_{fi}^*\,  (\, j^+_{fi}-j^-_{fi}\,) \,]\ , \nonumber\\
  R_3 &=& \sum_{m_3,m_1} {\rm Im}\,[ \,\rho_{fi}^*\,  (\, j^+_{fi}+j^-_{fi}\,) \,]\ ,\label{eq:Ri}\\
  R_4 &=& \sum_{m_3,m_1} (\, |j^+_{fi}|^2 + |j^-_{fi}|^2 \,)\ ,\nonumber \\
  R_5 &=& \sum_{m_3,m_1} {\rm Re}\, (\,j_{fi}^{+\,*} \,j^-_{fi}\, )\ , \nonumber\\
  R_6 &=& \sum_{m_3,m_1} {\rm Im}\,(\,j_{fi}^{+\,*} \,j^-_{fi}\, ) \ ,\nonumber
\end{eqnarray}
where $P_a$ denotes the component of
${\bm P}={\bmk}-{\bmk}^\prime$ along ${\bm e}_a$, and
the matrix elements are defined (schematically) as
\begin{equation}
\label{eq:e34a}
\rho_{fi}=\langle \Psi| \rho^\dagger|\Psi^{(\gamma)}\rangle \ ,
\qquad {j}^\pm_{fi}=\langle \Psi| \hat{\bm e}^*_{\pm} \cdot{\bm j}^\dagger|\Psi^{(\gamma)}\rangle\ ,
\end{equation}
where ${\bm e}_\pm=\mp (\hat{\bm e}_x\pm i\, \hat{\bm e}_y)/\sqrt{2}$.
Integrating out the energy-conserving $\delta$- function in Eq.~(\ref{eq:e26})
relative to the positron energy yields the five-fold differential cross
section (in the lab frame)
\begin{equation}
\label{eq:e26a}
 \sigma^{(\gamma)}(\epsilon,\hat{\bmk},\hat{\bmk}^\prime)\equiv {d^5\sigma^{(\gamma)}\over d\epsilon\, d\hat{\bmk} \, d\hat{\bmk}^\prime}=\frac{\alpha^2}{ (2\pi)^3}\,
   {k\,k'\over Q^4 } \, \frac{f_{\rm rec}}{v_r} \,R_{fi} \ ,
 \end{equation}
where have defined the recoil factor as
\begin{equation}
  f_{\rm rec}^{-1}=\left | 1 +{1\over M}(k'-p \cos\theta' +k \cos\theta_{ee}) {\epsilon'\over k'}\right|\ .
\end{equation}
Here $\theta'$ is the angle between the directions of the positron and incident nucleon momenta, and $\theta_{ee}$
is the angle between the momenta of the two leptons defined in Eq.~(\ref{eq:thetaee}).
The electron energy is in the range $m_e\le \epsilon\le\epsilon_{max}$, where
$\epsilon_{max}$ would be simply given by $E_0-m_e\approx 20$ MeV---see Eq.~(\ref{eq:e12})---for the
energies under consideration here, were it not for a small correction due to
the $^4$He recoil energy, which we account for explicitly.  Given $\epsilon$,
the positron energy $\epsilon^\prime$ is fixed by energy conservation.

The four-fold differential cross section integrated
over the electron energy is given by
\begin{equation}
  {d^4\sigma^{(\gamma)}\over d\hat \bmk \,d\hat\bmk'}=
  \int_{m_e}^{\epsilon_{max}} d\epsilon\, \sigma^{(\gamma)}(\epsilon,\hat{\bmk},\hat{\bmk}^\prime)
  \ .\label{eq:xs4}
\end{equation}
Finally, the total cross section follows from
\begin{equation}
  \sigma^{(\gamma)}=
   \int d\hat\bmk \int d\hat\bmk' \, \int_{m_e}^{\epsilon_{max}} d\epsilon\, 
  \sigma^{(\gamma)}(\epsilon,\hat{\bmk},\hat{\bmk}^\prime)
  \ .\label{eq:xstot}
\end{equation}
These integrations can be accurately carried out numerically by
standard techniques.

\subsection{Nuclear electromagnetic current}
\label{sec:emj}

Nuclear electromagnetic charge ($\rho$) and current (${\bm j}$) operators have been
constructed up to next-to-next-to-next-to-next-to-leading order (N4LO)
within the two different $\chi$EFT formulations we have adopted here,
without~\cite{Pastore:2008ui,Pastore:2009is,Pastore:2011ip,Kolling:2009iq,Kolling:2011mt,Piarulli:2012bn}
and with~\cite{Schiavilla:2018udt} the inclusion of explicit $\Delta$-isobar degrees of
freedom.  They consist of one-body terms, including relativistic corrections, and two-body
terms associated with one- and two-pion exchange (OPE and TPE, respectively)
as well from minimal and non-minimal couplings.  From a power counting
perspective, one-body charge and current operators come in at LO and NLO,
respectively; one-body relativistic corrections to the charge operator and two-body
OPE current operators from the leading $\pi N$ chiral Lagrangian both enter N2LO;
two-body OPE charge operators and one-body relativistic corrections to the current
operator in the $\Delta$-less formulation contribute at N3LO; in the $\Delta$-full
formulation, however, there is an additional N3LO contribution to the current operator
associated with a OPE term involving a $\Delta$-isobar intermediate state; lastly, two-body
OPE (from subleading $\pi N$ Lagrangians), TPE, and contact terms in both
the charge and current operators come in at N4LO.  We should stress that
the present $\Delta$-full formulation ignores the contributions of $\Delta$
intermediate states in the pion loops.

The TPE charge and current operators contain loop integrals that
are ultraviolet divergent and are regularized in dimensional
regularization~\cite{Pastore:2009is,Pastore:2011ip,Kolling:2009iq,Kolling:2011mt}.
In the current the divergent parts of these loop integrals are
reabsorbed in the LECs of a set of contact
currents~\cite{Pastore:2009is,Kolling:2011mt}, while those in the charge
cancel out, in line with the fact that there are no counterterms at 
N4LO~\cite{Pastore:2011ip,Kolling:2009iq,Kolling:2011mt}. 
Even after renormalization, these operators
have power law behavior for large momenta, and need to be
regularized before they can be sandwiched between nuclear
wave functions.  This (further) regularization is made
in momentum space~\cite{Piarulli:2012bn} or in configuration
space~\cite{Schiavilla:2018udt} depending on whether the charge
and current operators are used
in combination with the N3LO500 or NVIa interactions.

An important requirement is that of current conservation
$\bmq\cdot{\bm j}(\bmq)\,$=$\,\left[\, H\, ,\, \rho(\bmq)\,\right]$ with
the two-nucleon Hamiltonian given by
\begin{equation}
H=[T^{(-1)} +\cdots ]+[ v^{(0)}+v^{(2)}+v^{(3)}+v^{(4)}+\cdots]\ ,
\end{equation}
with $T$ denoting here the kinetic energy operator,
and the charge and current operators having the expansions
\begin{eqnarray}
\rho&=&\rho^{(-3)}+\rho^{(-1)}+\rho^{(0)}+\rho^{(1)}+\cdots \ ,\\
{\bm j}&=&{\bm j}^{(-2)}+{\bm j}^{(-1)}+{\bm j}^{(0)}+{\bm j}^{(1)}+\cdots
\end{eqnarray}
where the superscript $(n)$ specifies the order ${\cal P}^n$ in the
power counting with ${\cal P}$ denoting generically a
low-momentum scale, and the $\cdots$ indicate higher-order
terms that have been neglected here.  Current conservation
then implies~\cite{Pastore:2009is}, order
by order in the power counting, a set of non-trivial relations
between the ${\bm j}^{(n)}$ and the $T^{(n)}$, $v^{(n)}$, and
$\rho^{(n)}$ (note that commutators implicitly bring in factors
of ${\cal P}^{3}$).
These relations couple different orders in the power counting
of the operators, making it impossible to carry out a calculation,
which at a given $n$ for ${\bm j}^{(n)}$, $T^{(n)}$, $v^{(n)}$, and $\rho^{(n)}$
(and hence ``consistent'' from a power-counting perspective)
also leads to a conserved current. 

Another aspect is the treatment of hadronic electromagnetic form factors.
In the $e^-$-$e^+$ processes, these form factors enter at momentum
transfers $Q^2$ =$\,q^\mu q_\mu >0$, that is, in the time-like region.
While they could be consistently calculated in chiral perturbation theory,
here we extrapolate available parametrizations obtained from fits to
electron scattering data, as detailed in Refs.~\cite{Piarulli:2012bn,Piarulli:2014fta,Schiavilla:2018udt}, in
the time-like region.  In fact, the relevant $Q^2$ in the processes we
are considering are close to the photon point $Q^2\,$=$\,0$.

\subsection{Reduced matrix elements}
Because of the low energy and momentum transfers of interest here,
in the multipole expansion of the charge and current operators only
a few terms give significant contributions. In the case of interest here,
these expansions read~\cite{Walecka1995,Marcucci:2000xy,Schiavilla:2002}
\begin{eqnarray}
\label{eq:c}
\lefteqn{ \langle\Psi\,|\,\rho^{\dag}(\bmq)\,|\,
\Psi^{(\gamma)}_ {LSJJ_{z}}\rangle =\qquad\qquad} &&  \nonumber \\
  &&  \sqrt{4\pi} (-{i})^{J} 
 (-)^{J-J_z} D_{-J_{z},0}^{J}(-\phi_q,-\theta_q,0)\> 
  C_{J}^{LSJ}(q), \\
\lefteqn{ \langle\Psi \,|\,{\hat{\bme}}^{*}_{\lambda}\cdot
  \bmj^{\dag}(\bmq)\,|\, \Psi^{(\gamma)}_{LSJJ_{z}}\rangle
  =\qquad\qquad} &&  \nonumber \\
&& -\sqrt{2\pi} (-i)^{J} (-)^{J-J_z} D_{-J_{z},-\lambda}^{J}(-\phi_q,-\theta_q,0) \nonumber \\
    &&\quad \times\left[ \lambda \,M_{J}^{LSJ}(q) +E_{J}^{LSJ}(q)\right] \ ,    \label{eq:me}
\end{eqnarray}
where $\lambda\,$=$\, \pm 1$, and $C_{J}^{LSJ}$,
$E_{J}^{LSJ}$, and $M_{J}^{LSJ}$ denote the reduced matrix 
elements (RMEs) of the charge $(C)$, transverse electric $(E)$,
and transverse magnetic $(M)$ multipole operators, defined as in
Ref.~\cite{Walecka1995}; the (additional) superscript $\gamma$ is understood. 
Since the spin quantization axis of the nuclear states is taken along
the incident nucleon momentum ${\bm p}\,$=$\,p\, \hat{\bm z}$ rather
than the three-momentum transfer $\bmq\,$=$\,q\, \hat{\bm e}_z$,
in order to carry out the multipole expansion,
these states need to be expressed as linear combinations of those with
spins quantized along $\bmq\,$.  This is accomplished by the rotation
matrices ${D}_{J_{z}^{\prime}J_{z}}^{J}$~\cite{Edmond1957}, where the angles
$\theta_q$ and $\phi_q$ specify the direction of $\bmq$
in the lab frame (with ${\bm p}$ along $\hat{\bm z}$).\footnote{The notation used for the
  rotation matrix is the following
\[
    {D}_{M',M}^{J}(\gamma,\beta,\alpha)=e^{\ii M'\gamma}\,d^{J}_{M',M}(\beta)\,e^{\ii M\alpha}\ .
\]
}

We report in Table~\ref{tab:em-sr}
the RMEs contributing to the transition
from an initial $^{2S+1}L_J$ $3+1$ state to the final $^4$He ground state
with $J^\pi\,$=$\,0^+$. 
\begin{table}[bth]
  \caption{\label{tab:em-sr}
   The RMEs $C^{LSJ}_J$, $E^{LSJ}_J$, and $M^{LSJ}_J$ contributing
   to the electromagnetic transition 
   from an initial $3+1$ $^{2S+1}L_J$ scattering state to the final
   $^4$He ground state.}
    \begin{center}
      \begin{tabular}{l|ccc}
        \hline\hline
        state & ${}^{2S+1}L_J$ & charge multipoles & current multipoles  \\
        \hline
        $0^+$  &  ${}^1S_0$ & $C_0^{000}$ & $-$ \\
        $0^-$  &  ${}^3P_0$ & $-$  &  $-$ \\
        \hline
        $1^+$  &  ${}^3S_1, {}^3D_1$  &      $-$     &  $M_1^{L11}$ \\
        $1^-$  &  ${}^1P_1, {}^3P_1$  &  $C_1^{1S1}$ & $E_1^{1S1}$ \\
        \hline
        $2^+$  &  ${}^1D_2, {}^3D_2$  &  $C_2^{2S2}$ & $E_2^{2S2}$ \\
        $2^-$  &  ${}^3P_2, {}^3F_2$  &  $-$         &  $M_2^{L12}$ \\
        \hline
      \end{tabular}    
    \end{center}
  \end{table}
In the long-wavelength approximation of relevance here, Siegert's
theorem~\cite{Siegert1937} relates the electric and Coulomb multipole
operators, respectively $E_{JM}(q)$ and $C_{JM}(q)$, via~\cite{Walecka1995}
\begin{equation}
  E_{JM} (q) \approx \sqrt{J+1\over J} {\Delta E\over q}\, C_{JM}(q)\ ,
  \label{eq:siegert}
\end{equation}
where $\Delta E=E_i-E_f$ is the difference between the initial $1+3$ scattering state
and $\heq$ ground state energies. This relation implies a relationship between
the corresponding RMEs $E^{LSJ}_J(q)$ and
$C^{LSJ}_J(q)$ of Table~\ref{tab:em-sr}. It is worthwhile stressing here
that Siegert's theorem assumes (i) a conserved current and (ii) that
the initial and final states are exact eigenstates of the nuclear
Hamiltonian.  Equation~(\ref{eq:siegert}) provides a
test---indeed, a rather stringent one---of these two assumptions, see Ref.~\cite{Schiavilla:2018udt}
for a discussion of this issue in the context of the chiral interaction
NVIa and accompanying electromagnetic currents. 
  
Finally, we note that the total cross sections for the $\tri(p,\gamma)\heq$
and $\het(n,\gamma)\heq$ radiative captures in Fig.~\ref{fig:capture} follow from
\begin{equation}
  \sigma^{(\gamma)}_C \!=\! {8 \pi^2\alpha\over v_r}  {q\over 1+q/M} \!\!\sum_{LS,J\geq1}
  \Bigl[ |E_J^{LSJ}(q)|^2 + |M_J^{LSJ}(q)|^2\Bigr] ,\label{eq:capture}
\end{equation}
where $q$ is the momentum of the outgoing photon and the sum only includes
transverse RMEs. 
\subsection{Results for the electromagnetic RMEs}
\label{sec:res_em_rmeaa}

We report in Table~\ref{tab:rme-em}
the absolute values of the RMEs contributing to the
$^3$H$(p,e^+ e^-)^4$He process for an incident
proton energy of $\,0.90$ MeV.  They have been calculated
with the N3LO500/N2LO500 chiral interactions and
accompanying electromagnetic current operator
for $3+1$ states with $J\le 2$.  The results in the columns labeled
LO-wf4 and LO-wf5 are obtained with the LO charge and NLO current
operators (in the notation of Sec.~\ref{sec:emj}) and, respectively, a
smaller and a larger number of HH states
included in the ``core'' part $\Psi^C_{LS,JJ_z}$ of the $^3$H$+p$
scattering wave function for each channel $LSJ$ 
(see Ref.~\cite{Viviani:2020gkm} for a comprehensive
discussion of various technical issues relating to the calculation
of these wave functions, the basis sets wf4 and wf5 being
defined in Sec.~IIIa of that paper).  These results demonstrate the high
degree of convergence achieved in the calculation of
the RMEs, even for a ``delicate'' channel like $^1S_0$ in which
the $0^+$ resonant state plays a dominant role.

The columns labeled LO-wf5 and N4LO-wf5 in Table~\ref{tab:rme-em}
show the effect of including the complete set of N4LO charge and current
operators.  While terms beyond LO in the charge give tiny contributions,
those beyond NLO in the current generally lead to a significant increase
in the magnetic and electric RMEs. The relation 
$E_1^{1S1}=\sqrt{2}\,(\Delta E/q) \,C_1^{1S1}\approx\sqrt{2}\,C_1^{1S1}$ (since here $q=0.1$ fm$^{-1}$
$\approx\Delta E$) is reasonably well verified given that the calculated ratio 
$|E_1^{101}|/|C_1^{101}|\approx 1.457$
(of course, this value corresponds to including the full transition operator at N4LO). 

\begin{table}[bth]
  \caption{\label{tab:rme-em}
RMEs in absolute value (in fm$^{3/2}$) corresponding to channels with $J\le 2$ in the $^3$H$+p$
scattering wave function, obtained with the N3LO500/N2LO500 chiral interactions
and accompanying electromagnetic charge and current operators.  The incident
proton energy is 0.9 MeV and the three-momentum transfer $q$ is 0.1 fm$^{-1}$. 
See text for further explanations.
 }  
  \begin{center}
    \begin{tabular}{l|l|lll}
            \hline
            \hline
      RMEs$\times 10^3$ & channel & LO-wf4 &  LO-wf5 & N4LO-wf5 \\      
      \hline
      $|C_0^{000}|$ & ${}^1S_0$ & $14.07$ & $13.81$   & $13.74$ \\
      \hline
      $|M_1^{011}|$ & ${}^3S_1$ & $\n0.52$ & $\n0.52$ & $\n 0.66$ \\
      $|M_1^{211}|$ & ${}^3D_1$ & $\n0.01$ & $\n0.01$ & $\n 0.02$ \\
      \hline
      $|C_1^{101}|$ & ${}^1P_1$ & $25.67$  & $25.84$   & $26.01$ \\
      $|C_1^{111}|$ & ${}^3P_1$ & $\n2.66$ &$\n2.67$ & $\n5.14$ \\
      $|E_1^{101}|$ & ${}^1P_1$ & $29.16$  & $29.34$   & $37.89$ \\
      $|E_1^{111}|$ & ${}^3P_1$ & $\n2.03$ & $\n2.05$ & $\n3.90$\\
      \hline
      $|C_2^{202}|$ & ${}^1D_2$ & $\n0.53$ & $\n0.53$ & $\n0.53$ \\
      $|C_2^{212}|$ & ${}^3D_2$ & $\n0.01$ & $\n0.01$ & $\n0.01$ \\
      $|E_2^{202}|$ & ${}^1D_2$ & $\n0.87$ & $\n0.87$ & $\n0.89$\\
      $|E_2^{212}|$ & ${}^3D_2$ & $\n0.03$ & $\n0.03$ & $\n0.03$\\
      \hline
      $|M_2^{112}|$ & ${}^3P_2$ & $\n3.19$ & $\n3.19$ & $\n3.38$\\
      $|M_2^{312}|$ & ${}^3F_2$ & $\n0.00$ & $\n0.00$ & $\n0.00$\\
      \hline
      \hline
    \end{tabular}
    \end{center}
  \end{table}
 
From Table~\ref{tab:rme-em} we see that
the largest RMEs are $C_0^{000}$, $C_1^{101}$, and $E_1^{101}$, the first
involving the $0^+\longrightarrow 0^+$ transition, and
the last two the $1^-\longrightarrow 0^+$ transition.
The importance of the $C_0^{000}$ RME simply reflects the fact that
at an incident energy of 0.9 MeV the process proceeds via the
formation of the first excited state of $\heq$---the $0^+$
isoscalar resonance, mentioned above---and its subsequent decay to the $\heq$ ground state via the $C_0$ multipole.
 
At the higher end of the $\heq$ spectrum there are a couple of fairly
wide $J^\pi\,$=$\,1^-$ resonances (one isoscalar and the other isovector)
associated with the (coupled) channels $^1P_1$-$^3P_1$. The $^1P_1$ channel
gives, in particular, a large contribution. As a matter
of fact, the RMEs $C_1^{101}$ and
$E_1^{101}$ are even larger than the $C_0$ RME, discussed above.

Also, the corresponding $C^{111}_1$ and
$E_1^{111}$ RMEs are not negligible, as Table~\ref{tab:rme-em} indicates.
These RMEs are significantly smaller than those from the ${}^1P_1$ channel. While the
$E_1$ operator can connect the large $S$-wave component having
total spin $S\,$=$\,0$ in $^4$He to the $^1P_1$ scattering state, it cannot do
so to the $^3P_1$ scattering state because of orthogonality between the spin states
(the $C_1$ and $E_1$ operators are spin independent at LO).  Consequently,
this transition proceeds only through the small components of the $^4$He ground
state (these components account for roughly 15\% of the $^4$He normalization).

By contrast, the $M_1$ transition from the $^3S_1$ channel
is suppressed since the Pauli principle forbids identical nucleons
with parallel spins to come close to each other.
Higher-order transitions with $J\,$=$\, 2$ are even more suppressed
by powers of the three-momentum transfer $q$ which is close to $\lesssim 0.1$ fm$^{-1}$,
the only exception being the $M_2$ transition involving the $^3P_2$ channel,
whose importance (in relative terms) is somewhat magnified owing to
the presence of a couple of
$2^-$ resonant states in the $\heq$ spectrum.

\begin{figure}[bth]
\centering
\includegraphics[scale=0.40,clip]{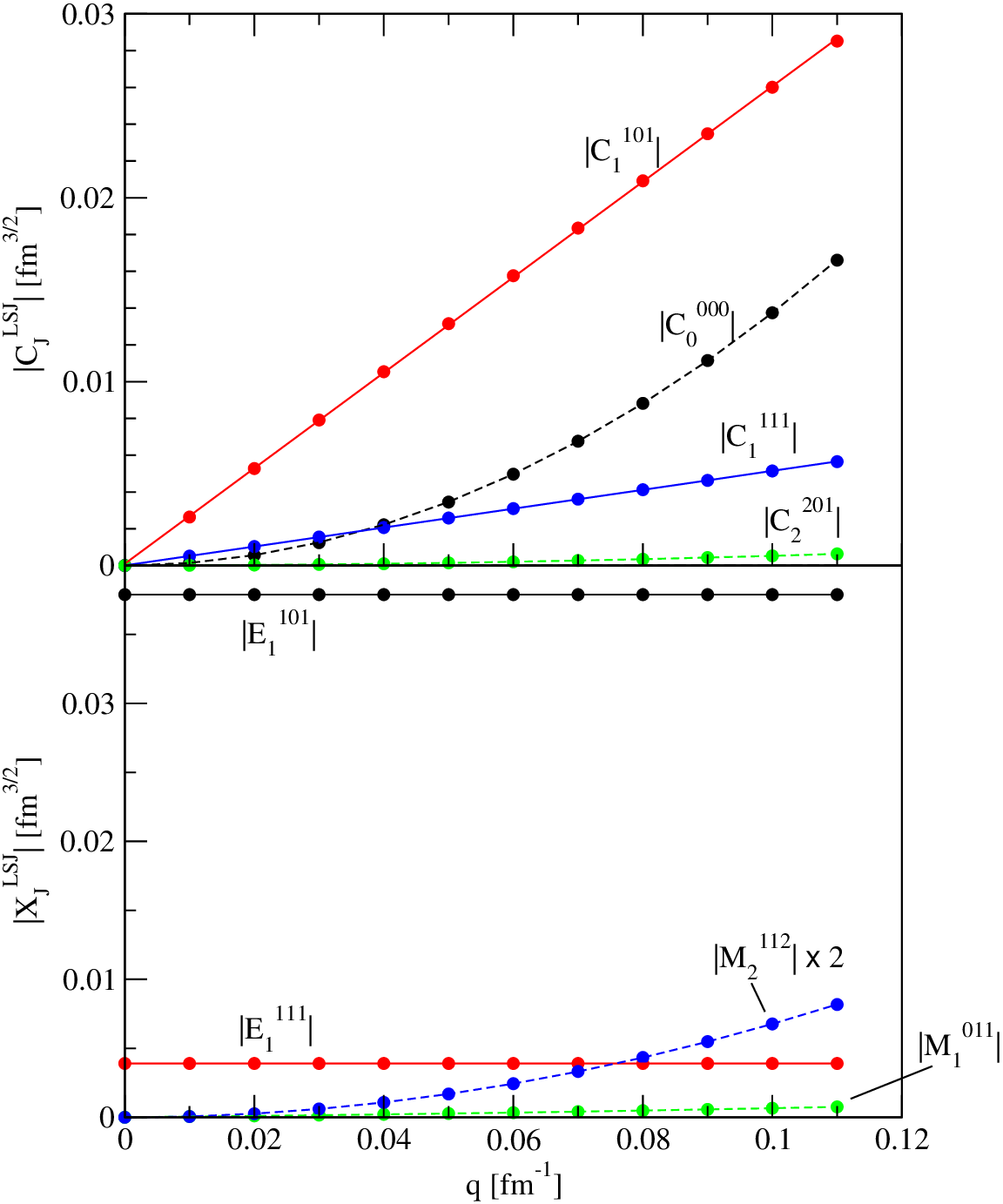}
\caption{The dependence on the three-momentum transfer $q$ of some electromagnetic RMEs (solid circles); the calculations are at incident proton
energy of 0.9 MeV and use the N3LO500/N2LO500 chiral
interactions.  The solid (dashed) lines show fits of the calculated values using linear (quadratic) parametrizations. }
  \label{fig:qdep}
\end{figure}

In order to understand the relative magnitude of the dominant RMEs $C_0^{000}$
and $E_1^{101}$, it is helpful to consider in more detail their dependence on $q$.  At LO $C_0^{000}$
involves the matrix element of the isoscalar $C_0$
multipole operator proportional to $\sum_i j_0(qr_i)$ between the dominant (isoscalar)
$^3$H$+p$ $0^+$ resonance and the (isoscalar) $^4$He ground state.  In the $q$-expansion
of the spherical Bessel function, the leading term gives a vanishing contribution
to the matrix element because of orthogonality, and hence 
the $C_0^{000}$ RME is proportional to $q^2$.  By contrast, the $C_1^{101}$ RME is linear in $q$. Using the relation
given in Eq.~(\ref{eq:siegert}), we observe that $E_1^{1S1}\sim C_1^{1S1}/q$, and therefore the $E_1$ RMEs are independent on $q$. 
This expected $q$-scaling is well verified by the calculated RMEs, as shown 
in Fig.~\ref{fig:qdep}. We note that in the limit $q\,$=$\,0$ the only non-vanishing RMEs are the two $E_1$'s.  This fact
impacts the behavior of the pair-production cross section at backward angles, see below.

One would naively have expected $|C_0^{000}| > |C_1^{101}|, |E_1^{101}|$, since the
energies involved are closer to the $0^+$  than to the $1^-$ resonance.  However,
the further suppression with $q$ ($q^2$) of $C^{000}_0$ relative to $C_1^{101}$ ($E_1^{101}$) 
is responsible for inverting the expected trend.  As a matter of fact, the $1^-$
scattering state plays a very important role in the $\tri(p,e^-e^+)\heq$ and
$\het(n,e^-e^+)\heq$ processes.

\begin{figure}[t]
\centering
\includegraphics[scale=0.35,clip]{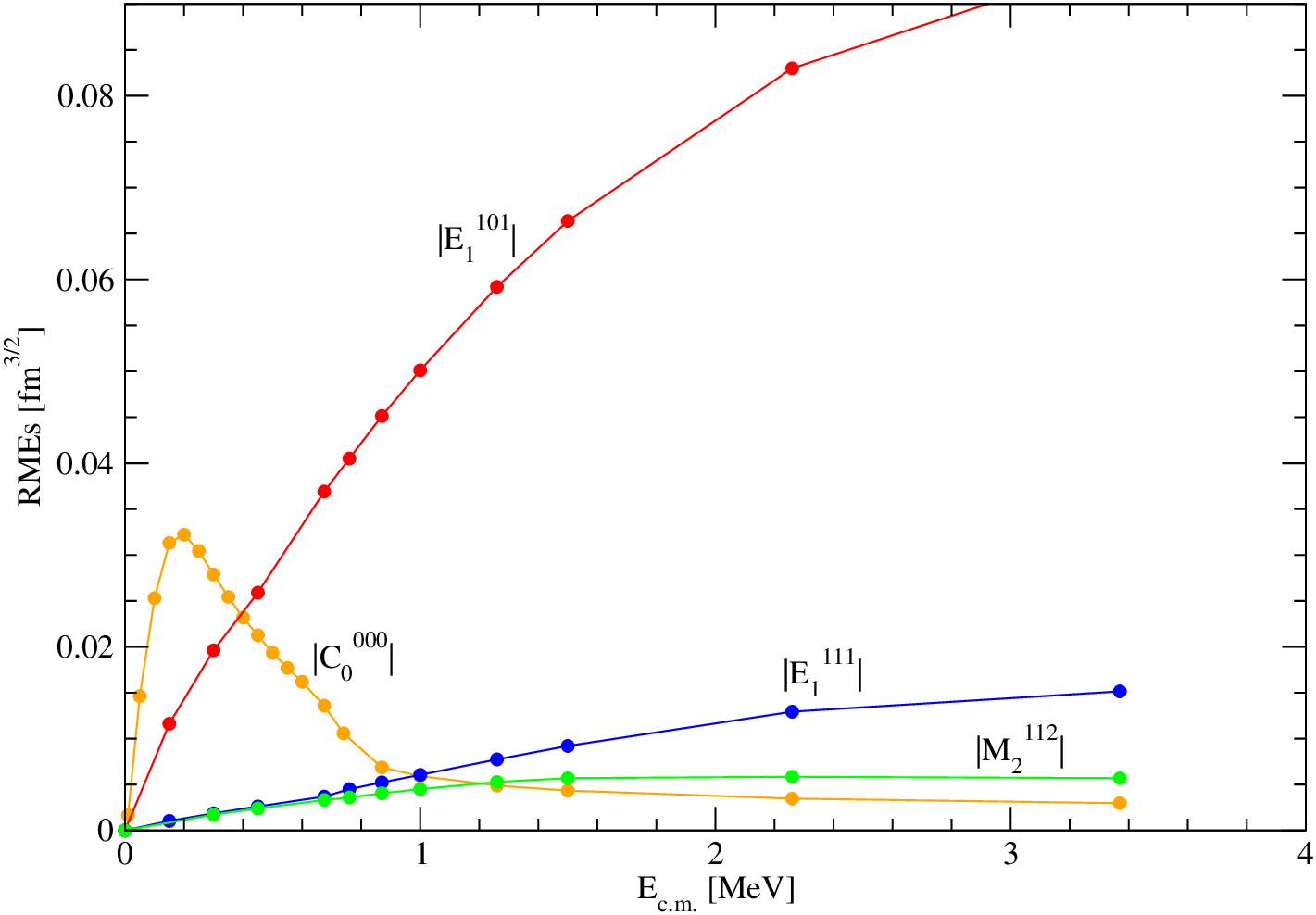}
\caption{The dependence on the proton incident energy of some RMEs (solid circles); the calculations
are at a fixed three-momentum transfer of 0.1 fm$^{-1}$ and use 
the N3LO500/N2LO500 chiral interactions.  The lines are to guide the eyes only.}
  \label{fig:Edep}
\end{figure}

Figure~\ref{fig:Edep} shows the behavior of selected RMEs as function
of the incident proton energy.  At the lower end ($\lesssim 0.4$ MeV), the
$0^+$ resonance is very prominent, but quickly fades away with increasing
energy (see $C^{000}_0$); by contrast, as the energy increases (exceeding
the $n+\het$ threshold) the $1^-$ resonance becomes progressively more
and more dominant (see $E_1^{101}$). The RMEs $E^{111}_1$ and $M^{112}_2$
increase monotonically with increasing energy, peaking at around about 3.5 MeV. 

\begin{table}[bth]
  \caption{\label{tab:rme-int}
  RMEs in absolute value (in fm$^{3/2}$) corresponding to channels with $J\le 2$ in the $^3$H$+p$
scattering wave function, obtained with either the N3LO500/N2LO500
or NVIa/3NIa chiral interactions
and accompanying electromagnetic charge and current operators at N4LO.  The incident
proton energy is 0.9 MeV and the three-momentum transfer $q$ is 0.1 fm$^{-1}$.}  
  \begin{center}
    \begin{tabular}{l|c|cc}
            \hline
            \hline
      RMEs$\times 10^3$ & $p+\tri$ wave & N3LO500/N2LO500 &  NVIa/3NIa \\      
      \hline
      $|C_0^{000}|$ & ${}^1S_0$ &  $13.74$ & $15.93$ \\
      \hline
      $|M_1^{011}|$ & ${}^3S_1$ & $\n 0.66$ & $\n 0.14$ \\
      $|M_1^{211}|$ & ${}^3D_1$ & $\n 0.02$ & $\n 0.03$ \\
      \hline
      $|C_1^{101}|$ & ${}^1P_1$ & $26.01$  & $25.36$  \\
      $|C_1^{111}|$ & ${}^3P_1$ & $\n5.14$ & $\n4.53$ \\
      $|E_1^{101}|$ & ${}^1P_1$ & $37.89$  & $37.47$ \\
      $|E_1^{111}|$ & ${}^3P_1$ & $\n3.90$ & $\n3.27$ \\
      \hline
      $|C_2^{202}|$ & ${}^1D_2$ & $\n0.54$ & $\n0.82$ \\
      $|C_2^{212}|$ & ${}^3D_2$ & $\n0.01$ & $\n0.01$ \\
      $|E_2^{202}|$ & ${}^1D_2$ & $\n0.89$ & $\n0.73$\\
      $|E_2^{212}|$ & ${}^3D_2$ & $\n0.04$ & $\n0.04$ \\
      \hline
      $|M_2^{112}|$ & ${}^3P_2$ & $\n3.38$ & $\n3.64$ \\
      $|M_2^{312}|$ & ${}^3F_2$ & $\n0.00$ & $\n0.00$ \\
      \hline
      \hline
    \end{tabular}
    \end{center}
  \end{table}
Finally, in Table~\ref{tab:rme-int} we report the RMEs obtained with
the chiral interactions N3LO500/N2LO500 and NVIa/3NIa, and
calculated in all cases using the largest number of HH states for full convergence.
The model dependence is weak for the largest RMEs.  In particular, in the $C_0^{000}$ RME we
do not observe any critical dependence on the input Hamiltonian.  This
is in contrast to what happens in the case of the
corresponding phase-shift~\cite{Viviani:2020gkm}, which is in fact
very sensitive to the Hamiltonian model.   
\subsection{Results for the electromagnetic cross sections}
\label{sec:res_em_rme}

Here we report cross-section results obtained for the
internal pair conversion processes.  The calculations use fully converged
bound- and scattering-state wave functions (with the largest allowed
number of HH states) and the complete N4LO set of electromagnetic
charge and current operators.

\begin{figure}[bth]
\centering
\includegraphics[scale=0.35,clip]{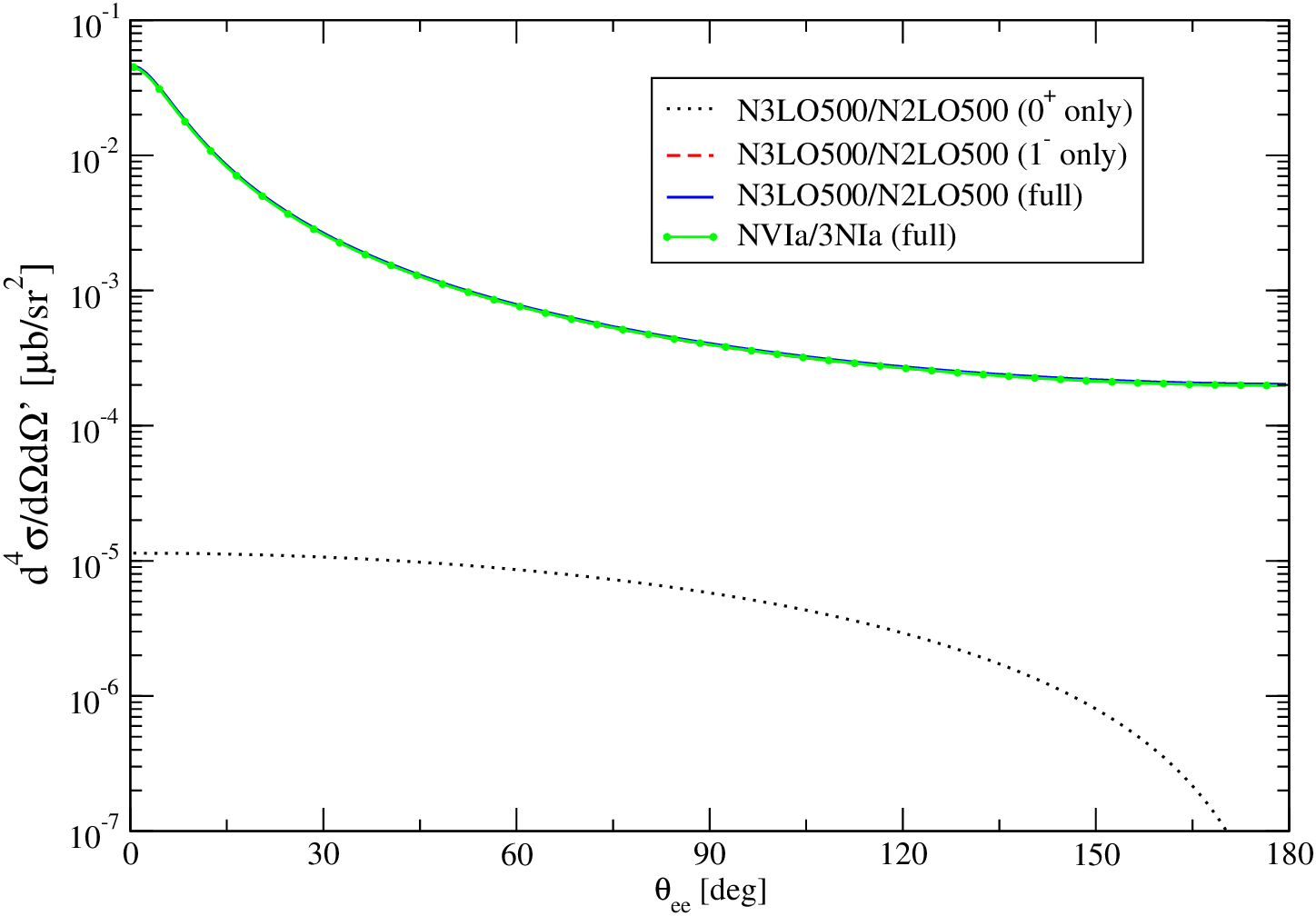}
\caption{The four-fold differential cross
section for the process $\tri(p,e^-e^+)\heq$ calculated at an incident
proton energy of 0.9 MeV; the kinematical configuration corresponds
to the lepton pair being emitted in the plane perpendicular
to the proton incident momentum, and $\theta_{ee}$ is the angle
between the electron and positron momenta. 
Results obtained
with the N3LO500/N2LO500 and NVIa/3NIa Hamiltonians and
under different approximations are shown. The dashed red, dot-dash blue, and
solid green curves are superimposed and cannot be distinguished. See text for further explanations.}
\label{fig:sige_RLRT}
\end{figure}

In Fig.~\ref{fig:sige_RLRT} we show the $\tri(p,e^-e^+)\heq$
four-fold differential cross sections
corresponding to the kinematical configuration in which the lepton pair is
emitted in the plane perpendicular to the incident proton momentum
($\theta\,$=$\,\theta^\prime\,$=$\,90^\circ$) and as function of the relative
angle $\theta_{ee}$, that is, the angle between the electron and positron
momentum.  The model dependence is weak, and the curves obtained
with the N3LO500/N2LO500 and NVIa/3NIa chiral interactions (and
corresponding set of electromagnetic transition operators) essentially
overlap, a result we could have anticipated on the basis of the
RMEs listed in Table~\ref{tab:rme-int}.  

In Fig.~\ref{fig:sige_RLRT}  we also show the differential cross
sections corresponding to the individual $^1S_0$ and $^1P_1$
transitions (again obtained with the N3LO500/N2LO500 model
Hamiltonian).  In terms of the response functions defined in Eq.~(\ref{eq:e31}),
we observe that for the $^1S_0$ transition only $R_1$ is non-vanishing,
\begin{equation}
  R_1(^1S_0;q) = 16\pi^2 \, |C_0^{000}(q)|^2\ . \label{eq:RL}
\end{equation}
In this respect, it is interesting to note that in the cross section
the response function $R_1$ is multiplied by $v_1$, see Eq.~(\ref{eq:e30}). 
In the limit $q\rightarrow 0$ (corresponding to the configuration in which
the electrons are emitted back-to-back with the same energy), it is easily
seen that this kinematical factor is proportional to $1/q^2$; however,
this singularity poses no problem, since $|C_0^{000}(q)|^2 \propto q^4$.
For the $^1P_1$ transitions, the response functions are
\begin{eqnarray}
  R_1(^1P_1;q) &=& 48\pi^2  \cos^2 \theta_q \, |C_1^{101}(q)|^2\ , \nonumber \\
  R_2 (^1P_1;q) &=& 24\pi^2 \sin(2\theta_q) {\rm Re}\, \bigl[ C_1^{101*}(q)\,
   E_1^{101}(q)\bigr]\ ,\nonumber \\
  R_3(^1P_1;q) &=& 0\ , \nonumber \\
  R_4(^1P_1;q)  &=& 24\pi^2 \sin^2\theta_q \, |E_1^{101}(q)|^2 \ , \\
  R_5 (^1P_1;q) &=& -12\pi^2 \sin^2\theta_q |E_1^{101}(q)|^2\ ,\nonumber \\
  R_6 (^1P_1;q) &=& 0\ ,\nonumber \label{eq:RT}
\end{eqnarray}
where $\theta_q$ is the polar angle of the three-momentum transfer
$\bmq=\bmk+\bmk'$ in the lab frame.
For the kinematical configuration in Fig.~\ref{fig:sige_RLRT},
we have $\theta_q\,$=$\,90^\circ$ and hence only $R_4$ and  $R_5$ contribute.

\begin{figure}[bth]
\centering
\includegraphics[scale=0.35,clip]{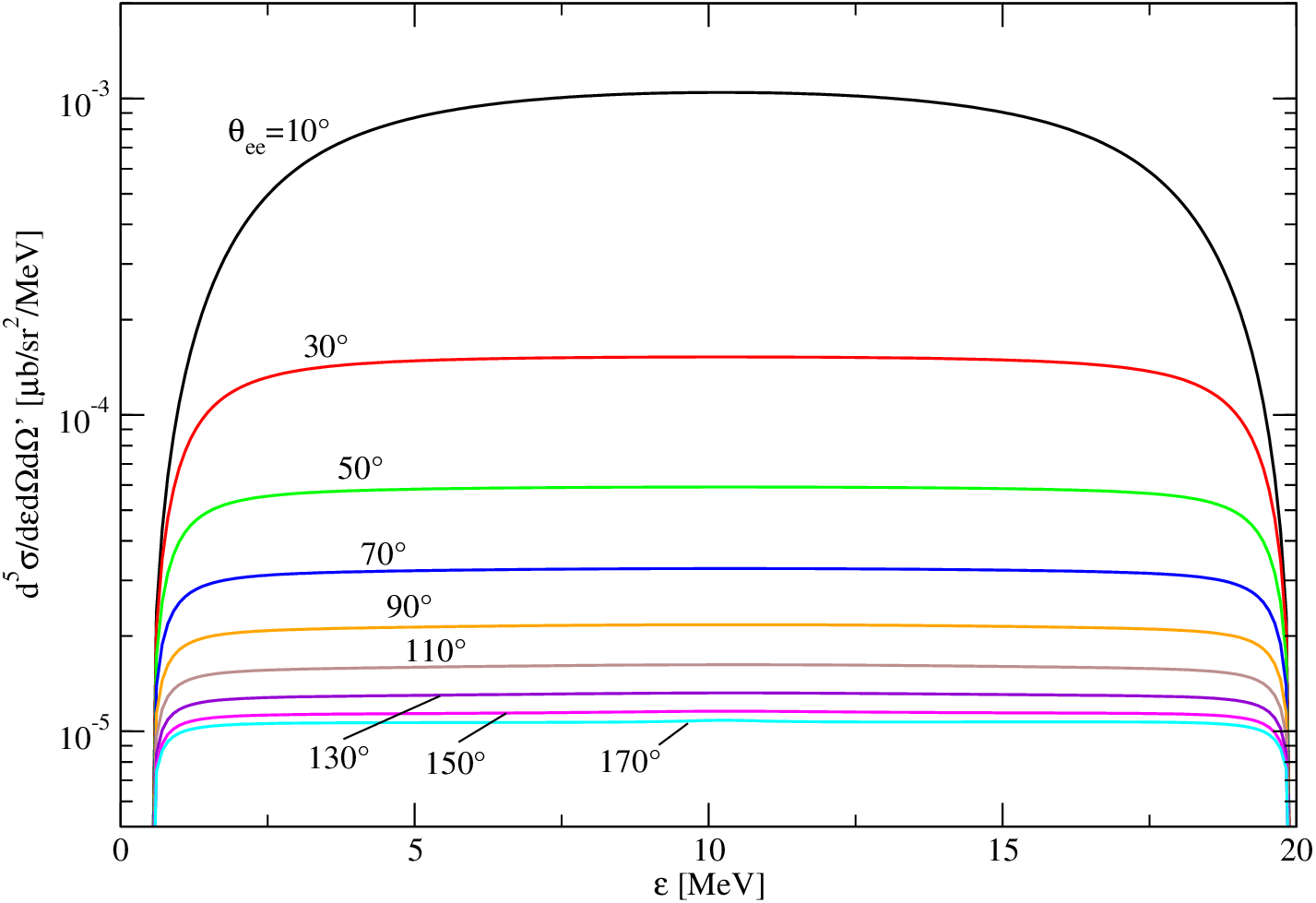}
\caption{The five-fold differential cross
section for the process $\tri(p,e^-e^+)\heq$ calculated with
the N3LO500/N2LO500 Hamiltonian at an incident
proton energy of 0.9 MeV as function of the electron energy
and for selected angles $\theta_{ee}$; the kinematical configuration is
otherwise the same as in Fig.~\ref{fig:sige_RLRT}.}
\label{fig:sig5}
\end{figure}
In Fig.~\ref{fig:sig5} we show the five-fold differential cross section
as function of the electron energy $\epsilon$ with
$m_e \le \epsilon\le \epsilon_{max}$ (approximately 20 MeV) for selected angles
$\theta_{ee}$.  The lepton-pair kinematics is the same as in Fig.~\ref{fig:sige_RLRT}.
As $\theta_{ee}$ increases, the cross section decreases.  Away from the endpoints
at $\epsilon\,$=$\, m_e$ or $\epsilon_{max}$, it is fairly flat.
In the limit $\theta_{ee}\longrightarrow180^\circ$ (leptons emitted back-to-back)
the cross section remains essentially constant as a consequence of the $q$-independence
of the $E_1^{1S1}$ RMEs previously discussed.

\begin{table*}[bth]
  \caption{\label{tab:totxs}Total cross sections (in $\mu$b) for the processes $\het(n,e^-e^+)\heq$ and
    $\het(n,\gamma)\heq$ calculated at a number of incident neutron energies $E_n$ (in MeV)
    with the N3LO500/N2LO500 and NVIa/3NIa Hamiltonians. Note that in the energy conservation relations
    corresponding to pair production and radiative capture, we have used the experimental rather than calculated
    binding energies of $^3$He and $^4$He, so that thresholds are at the observed locations.
}
  \begin{center}
    \begin{tabular}{c|ccccc}
     \hline
     \hline
  & \multicolumn{2}{c}{N3LO500/N2LO500} &  \multicolumn{2}{c}{NVIa/3NIa}\\
    $E_n$ & $\het(n,e^-e^+)\heq$ & $\het(n,\gamma)\heq$  & $\het(n,e^-e^+)\heq$ & $\het(n,\gamma)\heq$ \\
     \hline
      $0.17$  & $0.0431$ & $20.2$ &  $0.0396$  & $18.9$\\
      $0.35$  & $0.0616$ & $29.0$ &  $0.0588$  & $28.3$\\
      $0.70$  & $0.0893$ & $42.0$ &  $0.0868$  & $41.8$\\
      $1.00$  & $0.108\n$ & $50.8$ & $0.106\n$ & $50.8$ \\
      $2.00$  & $0.146\n$ & $67.7$ &  $0.142\n$ & $67.1$ \\
      $3.50$  & $0.160\n$ & $73.1$ &  $0.155\n$ & $72.5$ \\
      \hline
    \end{tabular}
    \end{center}
  \end{table*}
The cross sections at incident proton energies other than 0.9 MeV
display the same qualitative features discussed so far.  At lower
energies, the contribution of the $C_0^{000}$ RME becomes more
important, while at higher energies (above the $n+\het$ threshold)
the cross section is completely dominated by the $C^{101}_1$ and
$E^{101}_1$  RMEs. 

Total cross sections for the $\het(n,e^-e^+)\heq$ process, calculated
with the N3LO500/N2LO500 and NVIa/3NIa Hamiltonian models, are provided
in Table~\ref{tab:totxs} at a number of incident neutron energies.   For comparison,
total cross sections are also reported for the radiative capture process
$\het(n,\gamma)\heq$.  We note that pair production cross sections are
suppressed by a factor of approximately $500$ relative to radiative capture cross sections.

%%%%%%%%%%%%%%%%%%%%%%%%%%%%%%%%%%%%%%%%%%%%%%%%%%

\section{X17-nucleon interactions}
\label{sec:x17lagr}
In this section we obtain, within $\chi$EFT,
the Lagrangians describing the interactions of
the X17 with nucleons.  We consider, in turn,
the cases in which the hypothetical X17 is either a scalar, pseudoscalar, vector or
an axial boson. Conventions and notations are as in Appendix~\ref{sec:chieft},
where we summarize the $\chi$EFT formulation
in the $SU(2)$ framework (with up and down quarks), albeit we only include
couplings to scalar, pseudoscalar, and vector source terms.
The extension of this framework to $SU(3)$ (with up, down, and strange
quarks) as well as the inclusion of an axial source term are briefly
outlined below. 
\subsection{Scalar or pseudoscalar or vector X17}
\label{sec:x17s}
Assuming conservation of parity, the Lagrangian describing the
interactions of a scalar X17 boson with
up and down quarks is taken as
\begin{equation}
  {\cal L}^{S}_{q,X}(x)=\sum_{f=u,d}
  e  \,  {\varepsilon_f\,m_f \over\Lambda} \, \overline{f}(x) f(x)\,X(x) \ ,
\end{equation}
where $f(x)$ is the field of the quark of flavour $f$, $X(x)$ is
the X17 field, and $\Lambda$ is an unknown high-energy mass scale.
Note that we have chosen to rescale the coupling constants by
the unit electric charge $e$, and have introduced explicitly the
quark masses in order to have  renormalization-scale invariant
amplitudes.  This Lagrangian is more conveniently written in terms
of the isodoublet quark fields, defined in Appendix~\ref{sec:chieft}, as
\begin{equation}
 {\cal L}^{S}_{q,X}(x)= e\,{m_q\over\Lambda_S}\, \overline{q}(x)
    (\varepsilon_0 + \varepsilon_z \tau_3)\, q(x)\, X(x)\ ,  \label{eq:Lqs2}
\end{equation}
where we have introduced the coupling constants
\begin{eqnarray}
  \varepsilon_0 &=& {\Lambda_S\over\Lambda}\, {m_u \,\epsilon_u +
   m_d \,\epsilon_d\over 2\,m_q}\ , \label{eq:eps_p_s}\\
  \varepsilon_z &=& {\Lambda_S\over\Lambda}\, {m_u \, \epsilon_u - m_d\, \epsilon_d\over 2\,m_q}\ , \label{eq:eps_m_s}
\end{eqnarray}
and a new scale $\Lambda_S$ which we set
(arbitrarily) at 1 GeV.  We have also defined
\begin{equation}
m_q={m_u+m_d\over 2} \ .
\end{equation}
From Eq.~(\ref{eq:Lqs2}) we identify
the isoscalar and isovector scalar sources $s_0(x)$ and $s_3(x)$,
including quark mass contributions, as
\begin{eqnarray}
s_0(x)&=&m_q-e \, {m_q\over\Lambda_S}  \varepsilon_0\, X(x)\ , \label{eq:s0}\\
s_3(x)&=&{m_u-m_d\over 2} -e\,  {m_q\over\Lambda_S}  \varepsilon_z\, X(x)\ ,\label{eq:s3}
\end{eqnarray}
and $s_1(x)\,$=$\, s_2(x)\,$=$\,0$.  

In the chiral Lagrangians (see Appendix~\ref{sec:chieft}) these scalar
sources enter via $\chi(x)\,$=$\,2 B_c \,s(x)$, where the LEC $B_c$ is
the quark condensate and $m_q \,B_c\,$=$\, m^2_\pi/2$, with $m_\pi$
denoting the pion mass.  Retaining only up to quadratic terms in the
expansion in powers of the pion field yields the Lagrangian
\begin{eqnarray}
{\cal L}^S_{h,X}(x)&=& \overline{N}(x) \left[ 8\,B_c\, c_1\, s_0(x) +4\, B_c \,c_5\, s_3(x) \,
\tau_3 \right] N(x) \nonumber\\
&&-\,B_c \, s_0(x) \, {\bm \pi}^2(x) \ .\label{eq:lagXs}
\end{eqnarray}
We find it convenient to introduce the coupling constants
\begin{equation}
\label{eq:e58a}
\eta^S_0=-\frac{4\, c_1\, m_\pi^2}{\Lambda_S}\, \varepsilon_0 \ ,\qquad
\eta^S_z=-\frac{2\, c_5\, m_\pi^2}{\Lambda_S}\, \varepsilon_z \ ,
\end{equation}
in terms of which the nucleon-X17 interaction Lagrangian reads simply as in Eq.~(\ref{eq:e4s}).

The case in which the X17 is either a pseudoscalar or vector
boson can be treated similarly.  In the former case we consider
the quark-level Lagrangian
\begin{equation}
  {\cal L}^{P}_{q,X}(x)=\sum_{f=u,d}
  e  \, {\varepsilon_f \,m_f \over\Lambda} \, \overline{f}(x) (i\gamma^5) f(x) \, X(x)\ ,
\end{equation}
from which we read off the isoscalar and isovector pseudoscalar
sources $p_0(x)$ and $p_3(x)$ as
\begin{eqnarray}
p_0(x)&=&e \, {m_q\over\Lambda_S}  \varepsilon_0\, X(x)\ , \label{eq:p0}\\
p_3(x)&=&e\,  {m_q\over\Lambda_S}  \varepsilon_z\, X(x)\ ,\label{eq:p3}
\end{eqnarray}
where $\varepsilon_0$ and $\varepsilon_z$ are the combinations of
Eqs.~(\ref{eq:eps_p_s})--(\ref{eq:eps_m_s}).
These pseudoscalar sources enter the chiral Lagrangians
via the term $2\,i B_c\, p(x)$.  Interaction terms come
from ${\cal L}^{(2)}_{\pi \pi}(x)$ and ${\cal L}^{(3)}_{\pi N}(x)$,
giving rise, respectively, to X17-pion and X17-nucleon couplings.
While the latter are nominally suppressed by two orders in the
power counting, we retain them nevertheless,
since it has been speculated~\cite{Alves:2020xhf}
that the X17 may be ``piophobic''.   After integrating by parts and using the
equation of motion to remove derivatives of the nucleon field,
the resulting Lagrangian reads
\begin{eqnarray}
  {\cal L}_{\pi, X}^P(x)&=& 2 f_\pi B_c \,p_3(x) \,\pi_3(x)\nonumber\\
  &+& 4 B_c m_N\, (d_{18}+2\,d_{19})\,\overline{N}(x)i\,\gamma^5\,N(x)\, p_0(x)\nonumber\\
  &+& 4 B_c m_N d_{18}\,\overline{N}(x)i\,\gamma^5\tau_z\, N(x) \,p_3(x) \ . \label{eq:lagXp}
\end{eqnarray}
We define the coupling constants
\begin{eqnarray}
\eta^P_0&=& {2 m_\pi^2 m_N (d_{18}+2\,d_{19})\over \Lambda_S} \, \varepsilon_0\ , \label{eq:e63a}\\
\eta^P_z&=&\frac{f_\pi\, m_\pi^2}{\Lambda_S}\, \varepsilon_z \ ,\label{eq:e63b}
\end{eqnarray}
from which the pseudoscalar interaction Lagrangian follows as in Eq.~(\ref{eq:e4p}).
Note that we have dropped the (direct) isovector X17-nucleon coupling appearing in the third line of Eq.~(\ref{eq:lagXp}).

In the vector boson case, we have
\begin{equation}
  {\cal L}^{V}_{q,X}(x)=\sum_{f=u,d}
  e  \, \varepsilon_f \, \overline{f}(x) \gamma^\mu f(x) \, X_\mu(x)\ ,
\end{equation}
where $X_\mu$ is the X17 vector field, and which can be rewritten as usual as
\begin{equation}
 {\cal L}_{q,X}^V= e\, \overline{q}(x)
    \big(\varepsilon_0 + \varepsilon_z \,\tau_3\big) \gamma^\mu q(x) \, X_\mu(x)\ .  \label{eq:Lqv2}
\end{equation}
The parameters $\varepsilon_0$ and $\varepsilon_z$ are
related to the coupling constants of the X17 to the up and down quarks via
\begin{equation}
\label{eq:e62}
  \varepsilon_0 = {\epsilon_u + \epsilon_d\over 2}\ , \qquad
  \varepsilon_z= {\epsilon_u - \epsilon_d\over 2}\ . 
\end{equation}
The (non-vanishing) vector sources are then given by
\begin{eqnarray}
  v_\mu^{s}(x)&=&  3  \,e  \,\varepsilon_0 \,X_\mu(x) \label{eq:v0} \ ,\\
  v_{\mu,3}(x)  &=& e  \,\varepsilon_z\, X_\mu(x)\ , \label{eq:v3}
\end{eqnarray}
and the ensuing nucleon-X17 interaction Lagrangian (at leading order) follows as
\begin{eqnarray}
{\cal L}_{N,X}^V(x)&=& \overline{N}(x)\gamma^\mu\, [ v_\mu^{s}(x)+v_{\mu,3}(x)\tau_3] N(x) \\
   &+&\!\! {\kappa_0\over 4m_N} \overline{N}(x) \bigl[\partial_\mu v^{s}_{\nu}(x)\!-\!\partial_\nu v^{s}_{\mu}(x)\bigr]\sigma^{\mu\nu} N(x) \nonumber\\
&+&\!\! {\kappa_z\over 4 m_N}\overline{N}(x) \bigl[\partial_\mu v_{\nu,3}(x)\!-\!\partial_\nu v_{\mu,3}(x)\bigr] \sigma^{\mu\nu}\tau_3
   N(x) \ ,\nonumber
 \label{eq:lagXV}  
\end{eqnarray}
which can be expressed as in Eq.~(\ref{eq:e4v}) by defining the
coupling constants
\begin{equation}
\label{eq:e69a}
\eta_0^V=3\,\varepsilon_0\ ,\,\,\,\,\,\,\eta_z^V=\varepsilon_z\ .
\end{equation}
We note that the case of a ``proto-phobic'' coupling of the X17 corresponds
to $\eta_0^V\approx -\eta_z^V$.
\subsection{Axial X17}
\label{sec:sax17}
Because of the isospin singlet axial current anomaly~\cite{Peskin:1995ev}, isoscalar axial sources are absent in the
flavor $SU(2)$ Lagrangian of Appendix~\ref{sec:chieft}.
In order to circumvent this difficulty, we extend
the theory to flavor $SU(3)$~\cite{Scherer:2005ri} by also including
strange quarks,
\begin{equation}
  {\cal L}^{A}_{q,X}(x)=\sum_{f=u,d,s}
  e  \, \varepsilon_f\, \overline{f}(x) \gamma^\mu \gamma^5 \, f(x) \, X_\mu(x)\ ,
\end{equation}
and define the field $q(x)$ as
\begin{equation}
q(x)=\left[\!\begin{array}{c} 
         u(x)\\
         d(x)\\
         s(x)
         \end{array}\!\right].
\end{equation}
If we ignore strange-quark components in the nucleon,\footnote{The
contribution of the strange quark to the axial form factor of the nucleon has been
recently calculated in LQCD, see Ref.~\cite{Green:2017keo}.  However,
experimental knowledge of this contribution from parity-violating
electron scattering at backward angles and from neutrino scattering is
rather uncertain at this point in time.}
\begin{equation}
  \langle N | \overline{s}(x)\gamma_\mu \gamma^5 s(x) | N \rangle=0 \ ,
\end{equation}
we can then identify the isoscalar axial-source term with one of the
$SU(3)$ axial currents, conserved in the chiral limit where the
masses of up, down, and strange quarks vanish, that is,
\begin{equation}
\label{axialA}
\langle N| \overline{u}(x)\gamma_\mu\gamma^5 u(x)\!+\!\overline{d}(x)\gamma_\mu \gamma^5 d (x)| N \rangle\!\longrightarrow\! \langle N|a_{\mu,8}(x) | N \rangle \ ,
\end{equation}
where the current $a_{\mu,8}(x)$ is
\begin{eqnarray}
  a_{\mu,8}(x)\!&=&\!\overline{u}(x)\gamma_\mu\gamma^5 u(x)\!+\overline{d}(x)\gamma_\mu \gamma^5 d(x)
  \!-2\,\overline{s}(x)\gamma_\mu\gamma^5 s(x) \nonumber\\
  &=&\sqrt{3} \,\overline{q}(x) \gamma_\mu \gamma^5 \lambda_8\, q(x) \ ,
\end{eqnarray}
and $\lambda_8$ is the Gell-Mann matrix
\begin{equation}
\lambda_8=\frac{1}{\sqrt{3}}\begin{pmatrix}
1&0&0\\
0&1&0\\
0&0&-2
\end{pmatrix}\ .\label{eq:lambda8}
\end{equation}
The relevant piece of the flavor $SU(3)$ quark-level Lagrangian reads~\cite{Scherer:2005ri}
\begin{equation}
{\cal L}_{q,X}^A=  \overline{q}(x) \,a_\mu(x) \gamma^\mu \gamma^5 \,q(x) \, 
\ ,  \label{eq:Lqa3}
\end{equation}
where
\begin{equation}
\label{eq:e8}
a_\mu(x)=a_{\mu,3}(x) \,\lambda_3 + a_{\mu,8}(x) \,\lambda_8 \ ,
\end{equation}
and
\begin{equation}
   a_{\mu,3}(x) = e\, \varepsilon_z \,X_\mu(x) \ , \qquad
   a_{\mu,8}(x) = \sqrt{3}\,e \, \varepsilon_0 X_\mu(x) \ ,    \label{eq:e8a}
\end{equation}
with $\varepsilon_0$ and $\varepsilon_z$ defined as in Eq.~(\ref{eq:e62}).
The Gell-Mann matrix $\lambda_3$ is the $SU(3)$ extension of the Pauli
matrix $\tau_3$, namely
\begin{equation}
\lambda_3=\begin{pmatrix}
1&0&0\\
0&-1&0\\
0&0&0
\end{pmatrix}.
\end{equation}

At the hadronic level, the flavor $SU(3)$ chiral Lagrangian is written in
terms of the baryon-field matrix
\begin{equation}
B=\begin{pmatrix}
\Sigma^0/\sqrt{2} +\Lambda/\sqrt{6} & \Sigma^+& p\\
\Sigma^- &- \Sigma^0/\sqrt{2} +\Lambda/\sqrt{6} &n\\
\Xi^-&\Xi^0& -2 \,\Lambda/\sqrt{6}
\end{pmatrix}\ ,  \label{eq:bfield}
\end{equation}
and meson-field matrix 
\begin{equation}
\Phi=\begin{pmatrix}
\pi_3+ \eta/\sqrt{3}  & \pi_1+i\, \pi_2 & \sqrt{2} \,K^+\\
\pi_1-i\, \pi_2 & -\pi_3 + \eta/\sqrt{3}& \sqrt{2} \,K^0 \\
\sqrt{2}\, K^- & \sqrt{2} \,\overline{K}^{\,0} &  -2 \,\eta/\sqrt{3}
\end{pmatrix}\ ,
\end{equation}
where $\Sigma^\pm$, $K^\pm$, etc., are the fields associated with the various
strange baryons and mesons.
The building blocks are $3\times3$ matrices, defined as 
\begin{eqnarray}
  U&=&1+{i\over f_\pi} \Phi -{1\over 2\,f_\pi^2}\, \Phi^2 + \cdots
    \ , \label{eq:uumatrix3}\\
    D_\mu B &=& \partial_\mu B + [\Gamma_\mu\, ,\, B]\ , \label{eq:nablab}
\end{eqnarray}
with the remaining auxiliary fields $u$, $u_\mu$, $\Gamma_\mu$, and $F_{\mu\nu}^\pm$ as
given in Eq.~(\ref{eq:ea8}).  Here, we specialize to the case of an external axial current only,
and therefore set $r_\mu(x)\,$=$\,-\ell_\mu(x)\,$=$\,a_\mu(x)$, with $a_\mu(x)$ as in
Eq.~(\ref{eq:e8}).  In this extended framework, the meson-baryon Lagrangian at leading order reads~\cite{Scherer:2005ri}
\begin{eqnarray}
  \mathcal{L}_{mB}^{(1)}&=& \langle \, \overline{B} (i\,\gamma_\mu D^\mu -M_B) B
    +{D\over 2} \overline{B}\gamma^\mu\gamma^5\{u_\mu\, ,\, B \}\nonumber \\
    &&+{F\over 2} \bar{B}\gamma^\mu\gamma^5[ u_\mu\, ,\, B]
    \,\rangle\ ,\label{eq:lagXA}
\end{eqnarray}
where $\langle \cdots\rangle$ indicates a trace in flavor
space, $M_B$ is the mass matrix of the baryon octet,
and $D$ and $F$ are LECs.  Expanding in powers of the meson fields $\Phi$ and considering
only pion-nucleon-X17 interaction terms, we obtain 
\begin{eqnarray}
  \mathcal{L}^{A}_{N,X}(x)&=& (D+F)\, \overline{N}(x)\gamma^\mu\gamma^5 \tau_3\, N(x)\,
  a_{\mu,3}(x) \nonumber \\
 && + \frac{3\,F-D}{\sqrt{3}} \overline{N}(x)\,\gamma^\mu\gamma^5 N(x)\, a_{\mu,8}(x) \ ,
  \label{eq:lagXA2}  
\end{eqnarray}
which can be cast in the form of Eq.~(\ref{eq:e4a}) by defining the
coupling constants
\begin{equation}
\label{eq:e86a}
\eta_0^A=(3\,F-D)\,\varepsilon_0\ ,\qquad \eta_z^A= (F+D)\,\varepsilon_z \ .
\end{equation}
We note in closing that the term $\langle \nabla_\mu U^\dag \nabla^\mu U\rangle$
in the meson-meson Lagrangian at leading order also generates
an interaction term involving the direct coupling of the pion to the axial
field of the form $\partial^\mu \pi_3(x)\, X_\mu(x)$.
As mentioned in Sec.~\ref{sec:s1c}, we ignore it here for
simplicity.
\section{X17-induced nuclear currents}
\label{sec:x17curr}
The pair production amplitude on a single nucleon $T_{fi}^{cX}(N)$ induced by each of the (leading order)
Lagrangians in Eqs.~(\ref{eq:e4s})--(\ref{eq:e4a}) can be easily calculated, for example,
in time-ordered perturbation theory. The general structure is as given in Eq.~(\ref{eq:e8t})
with $j_{fi}^{cX}$ replaced by the single-nucleon current $j^{cX}_{fi}(N)$, that is, 
\begin{equation}
T_{fi}^{cX}(N)=4\pi\alpha \, \frac{\varepsilon_e\,\overline{u}({{\bm k}s}) \,\Gamma_{c} \,
v({{\bm k}^\prime s^\prime})\, j_{fi}^{c X}(N)}{q^\mu\, q_\mu-M_X^2}\ ,
\end{equation} 
where\footnote{We should note here that the
vector and axial amplitudes obtained in time-ordered perturbation theory
also include a contact term of the form $\sim (\overline{u}\,\gamma^0\, v) j^X_{0,fi}/M_X^2$,
involving the time components of the electron and nucleon currents.  This term
is, however, exactly cancelled by a corresponding term present in the interaction Hamiltonians of the
X17 vector and axial boson with electrons and
nucleons~\cite{Weinberg1995}, not shown in Eqs.~(\ref{eq:e4v})--(\ref{eq:e4a}).
} 
\begin{eqnarray}
j_{fi}^{S X}(N)&=&\overline{u}({{\bm p}^\prime s_N^\prime}) \, u({{\bm p} s_N}) \,
\chi^\dagger_{t_N^\prime} P^{SX} \chi_{t_N} \ ,\\
j_{fi}^{P X}(N)&=&\frac{\overline{u}({{\bm p}^\prime s_N^\prime})
\gamma^\mu\gamma_5  \, i\,q_\mu\,u({{\bm p} s_N})}{q^\mu q_\mu -m_\pi^2}\,
\chi^\dagger_{t_N^\prime} P^{PX} \chi_{t_N}\nonumber\\
     &+& \overline{u}({{\bm p}^\prime s_N^\prime}) \, i\gamma^5\, u({{\bm p} s_N}) \,
\chi^\dagger_{t_N^\prime} \overline P^{PX} \chi_{t_N}\ ,\\
j_{fi}^{V X}(N)&=&-\overline{u}({{\bm p}^\prime s_N^\prime}) \, 
\gamma^\mu\, u({{\bm p} s_N}) \,\chi^\dagger_{t_N^\prime} P^{VX} \chi_{t_N}  \\
&&+\frac{i}{2\,m_N}\,\overline{u}({{\bm p}^\prime s_N^\prime}) \, 
\sigma^{\mu\nu}q_\nu\, u({{\bm p} s_N}) \,\chi^\dagger_{t_N^\prime} \overline{P}^{VX} \chi_{t_N}
\ , \nonumber\\
j_{fi}^{A X}(N)&=&-\overline{u}({{\bm p}^\prime s_N^\prime}) \, \gamma^\mu\,\gamma_5\, u({{\bm p} s_N})\,
\chi^\dagger_{t_N^\prime} P^{AX} \chi_{t_N} \ ,
\end{eqnarray}
and $\chi_{t_N}$ and $\chi_{t^\prime_N}$ denote the initial and final nucleon
isospin states, respectively.  We have defined the isospin operators
\begin{equation}
P^{SX}=\eta_0^S+\eta_z^S\, \tau_3 \ ,
\end{equation}
and similarly for $P^{VX}$  and
$P^{AX}$ with the $\eta^S_0$ and $\eta_z^S$ replaced
by the corresponding set of coupling constants,
and 
\begin{eqnarray}
  P^{PX} &=&\frac{g_A}{2\,f_\pi}\, \eta^P_z\,\tau_3 \ , \\
  \overline P^{PX} &=& \eta^P_0 \ , \\
  \overline{P}^{VX}&=& \kappa_0\,\eta_0^V+\kappa_z\,
  \eta_z^V\, \tau_3 \ .
\end{eqnarray}

The nuclear currents follow by retaining the leading-order term
in the non-relativistic expansion of $j^{cX}(N)$ and
by summing over the individual nucleons (impulse approximation).
We define matrix elements of these currents
between the initial 3+1 scattering state and final $^4$He ground
state as in Eq.~(\ref{eq:ampli1})
\begin{equation}
 j^{cX}_{fi}= \langle \Psi({\bm p}-\bmq) | 
 j^{cX \,\dagger}(\bmq) | \Psi^{(\gamma)}_{m_3,m_1}({\bm p})\rangle \ ,
    \label{eq:amplix}
\end{equation}
where the $j^{cX}({\bm q})$ are single-nucleon operators
which in configuration-space can be written in terms of the following operator structures, stripped
of the coupling constants $\eta^c_{0,z}$,
\begin{eqnarray}
\rho^{S\lambda}(\bmq)&=&\sum_{i=1}^A e^{i\bmq\cdot\bmr_i}\, P^{\lambda}_i \ ,\label{eq:e94}\\ 
\rho^{P\lambda}(\bmq)&=&\sum_{i=1}^A e^{i\bmq\cdot\bmr_i}\,i\,\hat{\bm q}\cdot \bmsi_i \,P^{\lambda}_i \ ,\label{eq:e95} \\
 \rho^{A\lambda}(\bmq)&=&\sum_{i=1}^A  \frac{1}{2\, m_N} \, \left[ e^{i\bmq
 \cdot\bmr_i}\, , \,\bmp_i\cdot\bmsi_i \right ]_+ \,P^{\lambda}_i  \ ,\label{eq:e96} \\
 {\bm j}^{V\lambda}(\bmq)&=&\sum_{i=1}^A\frac{1}{2\,m_N} \left[ e^{i\bmq\cdot\bmr_i}\, , \,\bmp_i \right ]_+ P^{\lambda}_i\ ,\label{eq:97}\\
  \overline{{\bm j}}^{V\lambda}(\bmq) &=&  \sum_{i=1}^A\frac{i}{2\,m_N} \,e^{i\bmq\cdot\bmr_i}\, \bmq\times \bmsi_i \, P^{\lambda}_i  \ ,\label{eq:e98}\\
{\bm j}^{A\lambda}(\bmq)&=&\sum_{i=1}^A e^{i\bmq\cdot\bmr_i}\, {\bm \sigma}_i\,P^{\lambda}_i \ ,
\label{eq:e99}
\end{eqnarray}
where $\lambda=\pm$ with $P^+_i\,$=$\,1$ and $P^-_i\,$=$\,\tau_{i,3}$,
${\bm p}_i$ is the momentum operator,
and $[\cdots  ]_+$ denotes the anticommutator.
Note that the operators are defined to be adimensional.\footnote{The scalar-exchange
current follows as $j^{SX}=\eta^S_0\, \rho^{S+} +\eta_z^S\, \rho^{S-}$,
and similarly for the time component of the vector-exchange, and for
the time and space components of the axial-exchange currents.  The pseudoscalar-exchange and space
component of the vector-exchange currents read, respectively, 
\[
{j}^{PX}={g_A\over 2 f_\pi}\eta_z^P\,\frac{q}{q^2+m_\pi^2}\, {\rho}^{P-}  + \eta_0^P\, \frac{q}{2 m_N} \, {\rho}^{P+}\ ,
\]
and
\[
{\bm j}^{VX}= \eta_0^V\left( {\bm j}^{V+}+\kappa_0\, \overline{\bm j}^{V+}\right)
+\eta_z^V\left( {\bm j}^{V-}+\kappa_z\, \overline{\bm j}^{V-}\right) \ .
\]}

In the calculation we have retained the space and time components of,
respectively, the vector and axial currents, even though they are suppressed in the
power counting relative to the corresponding time and space components.  As a matter
of fact, these terms give important contributions by connecting the 3+1 $^1P_1$ and
$^3P_0$ continuum states to the $^4$He ground state via $E_1(V)$ and $C_0(A)$
multipole transitions.
\section{X17-induced cross sections}
\label{sec:x17cross}
In this section we compute the cross section including the contribution
of the X17 boson.  This cross section consists of a purely electromagnetic term, which
we have already analyzed in Sec.~\ref{sec:s-em}, an interference term
between the virtual photon and X17 amplitudes,
\begin{equation}
 d\sigma_X={1\over 4}\sum_{m_3 m_1}\sum_{s s'} 
 \frac{T^*_{fi} T^{cX}_{fi}+{\rm c.c.}}{v_r}\,d\phi \ ,
\end{equation}
and a term associated with the X17 exchange,
\begin{equation}
 d\sigma_{XX}={1\over 4}\sum_{m_3 m_1}\sum_{s s'} 
 \frac{|T^{cX}_{fi}|^2}{v_r}\,d\phi \ ,
\end{equation}
where $T_{fi}$ and $T^{cX}_{fi}$ are the amplitudes given in Eqs.~(\ref{eq:e1})
and~(\ref{eq:e8t}), the phase-space factor $d\phi$ has been defined in Eq.~(\ref{eq:e28a}),
and the superscript $\gamma$ identifying the initial 3+1 nuclear state ($p$+$^3$H or $n$+$^3$He)
is understood.  Standard trace theorems are used to evaluate the lepton tensors entering these
cross sections.  Noting that $\Gamma^{c\,\dagger}=\gamma^0\, \Gamma^c\, \gamma^0$, we find
\begin{itemize}
\item for the scalar case:
\begin{eqnarray}
\sum_{s s'}\ell^\mu_{\bmk s,\bmk' s'}\,
\ell^{S\,*}_{\bmk s,\bmk' s'} &=& {m_e\over \epsilon\epsilon'} \, ( k^{\prime\mu} -k^\mu )\ , \label{eq:traceVS}\\
\sum_{s s'} \ell^{S}_{\bmk s,\bmk' s'} \ell^{S\,*}_{\bmk s,\bmk' s'} 
 &=&  {1\over \epsilon\epsilon'} \, (k\cdot k'-m_e^2)\ ;
 \label{eq:traceSS}
\end{eqnarray}
\item for the pseudoscalar case, the mixed trace and hence the interference term in the
cross section vanish, while
\begin{equation}
\sum_{s s'} \ell^{P}_{\bmk s,\bmk' s'} \ell^{P\,*}_{\bmk s,\bmk' s'} 
 = {1\over \epsilon\epsilon'} \, (k\cdot k'+m_e^2)\ ;
 \label{eq:tracePP}
\end{equation}
\item for the vector case, the mixed and direct traces are of course the same, and
\begin{eqnarray}
\sum_{s s'}\ell^\mu_{\bmk s,\bmk' s'}\,
\ell^{\nu\,*}_{\bmk s,\bmk' s'} &=& {1\over \epsilon\epsilon'}[k^\mu k^{\prime\nu}+k^{\prime \mu} k^{\nu}
\nonumber\\
&&\,-\, g^{\mu\nu} (k\cdot k'+m_e^2)] \ ; \label{eq:traceVV}
\end{eqnarray}
\item for the axial case:
\begin{eqnarray}
\sum_{s s'}\ell^\mu_{\bmk s,\bmk' s'}\,
\ell^{\nu A\,*}_{\bmk s,\bmk' s'} &=&{i\over \epsilon\epsilon'} \, \epsilon^{\mu\nu\alpha\beta} k_\alpha k'_\beta \label{eq:traceVA}  \ , \\
\sum_{s s'} \ell^{\mu A}_{\bmk s,\bmk' s'} \ell^{\nu A\,*}_{\bmk s,\bmk' s'} 
 &=& {1\over \epsilon\epsilon'}[k^\mu k^{\prime\nu}+k^{\prime \mu} k^{\nu}
\nonumber\\
&&\,-\, g^{\mu\nu} (k\cdot k'-m_e^2)] \ . \label{eq:traceAA}
 \end{eqnarray}
\end{itemize}

\begin{table*}[bth]
\renewcommand{\arraystretch}{2.0}
 \label{tab:vxrx}
  \caption{The reduced response functions $R^{X}_{n,\lambda}$ with $\lambda\,$=$\pm$ entering
  the interference cross section $\sigma_X$ and the kinematical factors $v^X_n$.  The matrix elements
  $\rho_{fi}$ and $j^{\pm}_{fi}$ of the time and space components of the electromagnetic current are
  defined as in Eq.~(\ref{eq:e34a}).  The matrix elements $\rho^X_{fi}$ and $j^{X,\pm}_{fi,\lambda}$ are
  defined similarly with the X17-induced current replacing the electromagnetic one; of course, since the
  X17 current is not assumed to be conserved, a longitudinal matrix element
$j^{X,z}_{fi,\lambda}$ is also present (see text).  Note that there is no contribution to the
inference cross section in the case of a pseudoscalar X17.  We have defined $u\,$=$\,\omega/q$
and the spherical components $P_\pm \,$=$\,\mp (P_x \pm i\,P_y)/\sqrt{2}$ with ${\bm P}\,$=${\bm k}-{\bm k}^\prime$.
Moreover, $P_z=\hat\bme_z\cdot\bmP$. 
Lastly, the second column specifies the character---purely longitudinal (L), purely transverse (T), or
longitudinal-transverse (LT)---of the given $R^{X}_{n,\lambda}$.}
  \begin{center}
    \begin{tabular}{l|c|c|c|c|c}
      \hline
      $n$ & Character & $R_{n,\lambda}^{X}$ & \multicolumn{3}{c}{$v_n^X$}  \\
          & & & $S$ & $V$ & $A$ \\
      \hline
      $1$ & L & $\sum_{m_3 m_1}\rho_{fi}^* \,\, \rho^{X}_{fi,\lambda}$  & $m_e \,(P_0 -u P_z)$ & $(P_0 -u P_z)P_0/2+(m_e^2+k\cdot k')$ & \\
      $2$ & L & $\sum_{m_3 m_1}\rho_{fi}^*\,\, j^{X,z}_{fi,\lambda}$  & &$-(P_0 -u P_z)P_z/2-u (m_e^2+k\cdot k')$
      & \\
      \hline
      $3$ & LT & $\sum_{m_3 m_1}\rho^*_{fi} \,\, j^{X,+}_{fi,\lambda}$ & & $ (P_0 -u P_z) P_- /2$
      & $-Q^2 P_-/(2\,q)$  \\
      $4$ & LT & $\sum_{m_3 m_1}\rho^{*}_{fi} \,\,j^{X,-}_{fi,\lambda}$ & & $(P_0 -u P_z) P_+/2 $ & $Q^2 P_+/(2\,q)$  \\
      $5$ & LT & $\sum_{m_3 m_1}j^{+\,*}_{fi}\,\, \rho^{X}_{fi,\lambda}$ & $-m_e\, P_+$ & $-P_0 P_+/2$ & $-qP_+/2 $\\
      $6$ & LT & $\sum_{m_3 m_1} j^{-\,*}_{fi}\,\, \rho^{X}_{fi,\lambda}$ & $-m_e\, P_-$ & $-P_0 P_-/2$  & $q P_- /2$\\
      $7$ & LT & $\sum_{m_3 m_1} j^{+\,*}_{fi} \,\, j^{X,z}_{fi,\lambda}$ & & $P_z P_+/2$ & $ \omega P_+/2$\\
      $8$ & LT & $\sum_{m_3 m_1} j^{-\,*}_{fi}\,\, j^{X,z}_{fi,\lambda}$ & & $P_z P_-/2$ & $-\omega  P_- /2 $\\
      \hline
      $9$ & T & $\sum_{m_3 m_1}\bigl(j^{+\,*}_{fi} \,j^{X,+}_{fi,\lambda} +j^{-\,*}_{fi}\, j^{X,-}_{fi,\lambda}\bigr)$  & &$(P_x^2 +P_y^2)/4 - k\cdot k'-m_e^2$  &  \\
     $10$ & T & $\sum_{m_3 m_1}\bigl(j^{+\,*}_{fi} \,j^{X,+}_{fi,\lambda} -j^{-\,*}_{fi}\, j^{X,-}_{fi,\lambda}\bigr)$ & & &  $(P_0 \,q - P_z\, \omega )/2$ \\
     $11$ & T & $\sum_{m_3 m_1}j^{+\,*}_{fi} \,\, j^{X,-}_{fi,\lambda}$    & & $-P_+^2/2 $ & \\
     $12$ & T & $\sum_{m_3 m_1}j^{-\,*}_{fi}\,\, j^{X,+}_{fi,\lambda}$    & & $-P_-^2/2$ & \\
      \hline
    \end{tabular}
    \end{center}
  \end{table*}
In analogy to the purely electromagnetic case in Eq.~(\ref{eq:e26a}), we express the
interference and direct five-fold differential cross sections as
\begin{eqnarray}
\label{eq:e110}
\hspace{-0.55cm}&& \sigma_X(\epsilon,\hat{\bmk},\hat{\bmk}^\prime)\! =\!\frac{\alpha^2}{ (2\pi)^3}
   {k\,k'\over Q^2} \frac{f_{\rm rec}}{v_r} \,\varepsilon_e\bigg[\frac{R^X_{fi}}{Q^2-M_X^2}+{\rm c.c.}\bigg] \ , \\
\hspace{-0.55cm}&&\sigma_{XX}(\epsilon,\hat{\bmk},\hat{\bmk}^\prime)\! =\!\frac{\alpha^2}{ (2\pi)^3}
   {k\,k'\over |Q^2-M_X^2|^2}  \frac{f_{\rm rec}}{v_r} \,\varepsilon_e^2\,R^{XX}_{fi} \ ,
 \end{eqnarray}
where $\varepsilon_e$ is the coupling constant of the electron to the X17 particle,
and $R^X_{fi}$ and $R^{XX}_{fi}$ denote the nuclear responses associated
with, respectively, the interference and direct terms (the former one is generally complex).
Note that, when accounting for the X17 width, we make the replacement of Eq.~(\ref{eq:e9}) in the X17
propagator, yielding a complex function of the mass $M_X$ and width $\Gamma_X$.
Of course, the full cross section results from Eq.~(\ref{eq:e13}).

The interference and direct nuclear responses can be conveniently cast into
the forms
\begin{eqnarray}
R_{fi}^X\!&=&\!\sum_{n=1}^{12}v_n^X \left( \eta_0\, R^X_{n,+} +\eta_z R_{n,-}^{X}\right) \ , \\
R_{fi}^{XX}\!\!&=&\!\!\sum_{n=1}^{10}v_n^{XX} \left( \eta^2_0\, R^{XX}_{n,++} +
2\, \eta_0\,\eta_z R_{n,+-}^{XX}+ \eta^2_z\, R^{XX}_{n,--}\right) \ ,\nonumber
\end{eqnarray}
where the $v_n^X$ and $v_n^{XX}$ involve the lepton kinematic variables and
the reduced response functions $R^{X}_{n,\lambda}$ and $R^{XX}_{n,\lambda\lambda'}$
denote appropriate combinations of the matrix elements of the electromagnetic
and X17-induced currents, as indicated in Tables~\ref{tab:vxrx}
and~\ref{tab:vxxrxx} for the various possibilities
($S$, $P$, $V$, and $A$).  The dependence of
the reduced response functions on the coupling constants $\varepsilon_0$ and
$\varepsilon_z$ of the X17 to the quarks can be 
obtained using Eqs.~(\ref{eq:e58a}), (\ref{eq:e63a}), (\ref{eq:e63b}), (\ref{eq:e69a}),
and~(\ref{eq:e86a}). The index $\lambda=\pm$ specifies the matrix elements calculated
using either the isoscalar ($+$) and isovector ($-$)
component of the various operators. 
\begin{table*}[bth]
 \renewcommand{\arraystretch}{2.0}
   \caption{The reduced response functions $R^{XX}_{n,\lambda\lambda^\prime}$
  with $\lambda,\lambda^\prime\,$=$\pm$ entering the direct cross section $\sigma_{XX}$ and
  the kinematical factors $v^{XX}_n$; remaining notation as in Table~\ref{tab:vxrx}.  The upper and lower signs
  in the $v_n^{XX}$ correspond to the $S/V$ or $P/A$ cases, respectively. \label{tab:vxxrxx}
}
  \begin{center}
    \begin{tabular}{l|c|c|c|c}
      \hline
      $n$ & Character & $R^{XX}_{n,\lambda\lambda'}$ & \multicolumn{2}{c}{$v_n^{XX}$}  \\
          & & & $S,P$ & $V,A$  \\
      \hline
      $1$ & L & $\sum_{m_3 m_1}{\rm Re}\,\big(\rho^{X *}_{fi,\lambda} \,\,\rho^{X}_{fi,\lambda'}\big)$  & $k\cdot k'\mp m_e^2$ & $2\epsilon\epsilon'-k\cdot k'\mp m_e^2$ \\
      $2$ & L & $\sum_{m_3 m_1}{\rm Re}\, \bigl(\rho^{X*}_{fi,\lambda} \,\,j^{X,z}_{fi,\lambda'}+\rho^{X*}_{fi,\lambda'} \,\,j^{X,z}_{fi,\lambda} \bigr)/2$  && $ -\omega q + P_0 P_z$ \\
      $3$ & L & $\sum_{m_3 m_1}{\rm Re}\,\bigl( j^{X,z\,*}_{fi,\lambda} \,\, j^{X,z}_{fi,\lambda'} \bigr)$  && $(q^2-P_z^2)/2 + k\cdot k' \pm m_e^2$ \\
      \hline
      $4$ & LT & $\sum_{m_3 m_1}{\rm Re}\,\bigl[\rho^{X\,*}_{fi,\lambda} \, \bigl( j^{X,+}_{fi,\lambda'} - j^{X,-}_{fi,\lambda'}\bigr) + \rho^{X\,*}_{fi,\lambda'}\bigl (j^{X,+}_{fi,\lambda} - j^{X,-}_{fi,\lambda}\bigr)\bigr]/2$ && $-P_0 P_x/\sqrt{2} $ \\
      $5$ & LT &$\sum_{m_3 m_1}{\rm Im}\,\bigl[\rho^{X\,*}_{fi,\lambda} \, \bigl( j^{X,+}_{fi,\lambda'} + j^{X,-}_{fi,\lambda'}\bigr) + \rho^{X\,*}_{fi,\lambda'}\bigl (j^{X,+}_{fi,\lambda} + j^{X,-}_{fi,\lambda}\bigr)\bigr]/2$ && $ -P_0 P_y/\sqrt{2} $ \\
      $6$ & LT &  $\sum_{m_3 m_1}{\rm Re}\,\bigl[j^{X,z\,*}_{fi,\lambda} \, \bigl( j^{X,+}_{fi,\lambda'} - j^{X,-}_{fi,\lambda'}\bigr) + j^{X,z\,*}_{fi,\lambda'}\bigl (j^{X,+}_{fi,\lambda} - j^{X,-}_{fi,\lambda}\bigr)\bigr]/2$  && $ P_z P_x/\sqrt{2}$ \\
      $7$ & LT & $\sum_{m_3 m_1}{\rm Im}\,\bigl[j^{X,z\,*}_{fi,\lambda} \, \bigl( j^{X,+}_{fi,\lambda'} + j^{X,-}_{fi,\lambda'}\bigr) + j^{X,z\,*}_{fi,\lambda'}\bigl (j^{X,+}_{fi,\lambda} + j^{X,-}_{fi,\lambda}\bigr)\bigr]/2$ && $ P_z P_y/\sqrt{2}$ \\
      \hline
      $8$ & T & $\sum_{m_3 m_1}{\rm Re}\,\bigl(j^{X,+\,*}_{fi,\lambda}\,\, j^{X,+}_{fi,\lambda'} + j^{X,-\,*}_{fi,\lambda}\,\, j^{X,-}_{fi,\lambda'} \bigr)$  && $ -(P_x^2+P_y^2)/4 + k\cdot k' \pm m_e^2$ \\
      $9$ & T & $\sum_{m_3 m_1}{\rm Re}\, \bigl(j^{X,+\,*}_{fi,\lambda}\,\, j^{X,-}_{fi,\lambda'} + j^{X,+\,*}_{fi,\lambda'} \,\,j^{X,-}_{fi,\lambda}\bigr)/2$  && $ (P_x^2-P_y^2)/2 $\\
      $10$ & T &  $\sum_{m_3 m_1}{\rm Im}\, \bigl(j^{X,+\,*}_{fi,\lambda}\,\, j^{X,-}_{fi,\lambda'} + j^{X,+\,*}_{fi,\lambda'} \,\,j^{X,-}_{fi,\lambda}\bigr)/2$  && $ -P_x P_y$\\
      \hline
    \end{tabular}
    \end{center}
  \end{table*}

Since the vector and axial X17-induced currents are not
assumed to be conserved, in the reduced response functions
listed in Tables~\ref{tab:vxrx} and~\ref{tab:vxxrxx} there is a contribution
involving the longitudinal component of the current,
\begin{equation}
j^{X,z}_{fi,\lambda}=\langle \Psi |\hat{\bm e}_z\cdot {\bm j}^{cX\,\dagger}_\lambda(\bmq)|\Psi^{(\gamma)}\rangle \ ,
\end{equation}
which is of course absent in the purely electromagnetic case.  The matrix elements of the charge and
those of the transverse components of the current are expanded in RMEs of multipole
operators as in Eqs.~(\ref{eq:c})--(\ref{eq:me}).
The longitudinal current matrix element has a similar expansion
\begin{equation}
j^{X,z}_{fi,\lambda}= \sqrt{4\pi} (-{i})^{J} 
(-)^{J-J_z} D_{-J_{z},0}^{J}(-\phi,-\theta, 0) \,
L_{J}^{LSJ}(q) \ . \label{eq:l}
\end{equation}
Of course, the multipole parities for the $S$ and $V$ currents
are opposite to those of the $P$ and $A$ currents.

Lastly, the contribution resulting from the contraction of the $q^\mu\, q^\nu/M_X^2$
term in the propagator of the X17 boson with the leptonic tensor vanishes in the vector
case and is neglected in the axial case, since it is proportional to $(m_e/M_X)^2 \approx 0.0009$.

\section{Results including the X17 boson}
\label{sec:resX}
In this section we provide further numerical details in support of the cross
section results reported in Sec.~\ref{sec:s1d}.  We begin by analyzing the relevant
RMEs contributing to the transition mediated by the X17 boson.  The RMEs
contributing to the purely electromagnetic transition have already been analyzed in
Sec.~\ref{sec:res_em_rmeaa}. 
\subsection{Numerical results for the RMEs}
\label{sec:rmex}
The relevant operators are given in
Eqs.~(\ref{eq:e94})--(\ref{eq:e99}).  The RMEs associated with a
$S$ or $V$ exchange are those in Table~\ref{tab:em-sr} (except that the longitudinal
RMEs are not listed; they connect to the same states as the charge RMEs), while
the RMEs contributing to a transition mediated by a $P$ or $A$ exchange
are listed in Table~\ref{tab:axial_rme}. 
\begin{table}[bth]
  \caption{\label{tab:axial_rme}
    The RMEs contributing to a transition mediated by a pseudoscalar or axial
    X17 exchange. Remaining notation as in Table~\ref{tab:rme-em}.}
    \begin{center}
      \begin{tabular}{l|ccc}
        \hline\hline
        state & ${}^{2S+1}L_J$ & charge multipoles & current multipoles  \\
        \hline
        $0^+$  &  ${}^1S_0$ & $-$ & $-$ \\
        $0^-$  &  ${}^3P_0$ & $C_0^{110}$  &  $L_0^{110}$ \\
        \hline
        $1^+$  &  ${}^3S_1, {}^3D_1$  & $C_1^{LS1}$  & $E_1^{LS1},L_1^{LS1}$ \\
        $1^-$  &  ${}^1P_1, {}^3P_1$  &  $-$ & $M_1^{1S1}$ \\
        \hline
        $2^+$  &  ${}^1D_2, {}^3D_2$  &  $-$ & $M_2^{2S2}$ \\
        $2^-$  &  ${}^3P_2, {}^3F_2$  &  $C_2^{LS2}$  & $E_2^{LS2},L_2^{LS2}$ \\
        \hline
      \end{tabular}    
    \end{center}
  \end{table}
In the following figures we consider separately the isoscalar ($+$) and isovector ($-$) components
of the operators as given
explicitly in Eqs.~(\ref{eq:e94})--(\ref{eq:e99}). The corresponding RMEs carry
a superscript $+$ or $-$. 
Of course, the isoscalar and isovector terms
in the $S$ operator are identical to the corresponding ones in the $V$ charge operator.
However, the isovector $P$ operator and isovector term in the $A$ charge operator differ,
the former involving $\hat\bmq\cdot {\bm\sigma}_i$ and the latter involving
${\bm p}_i\cdot {\bm\sigma}_i$.
  
\begin{figure}[bth]
\centering
 \includegraphics[scale=0.35,clip]{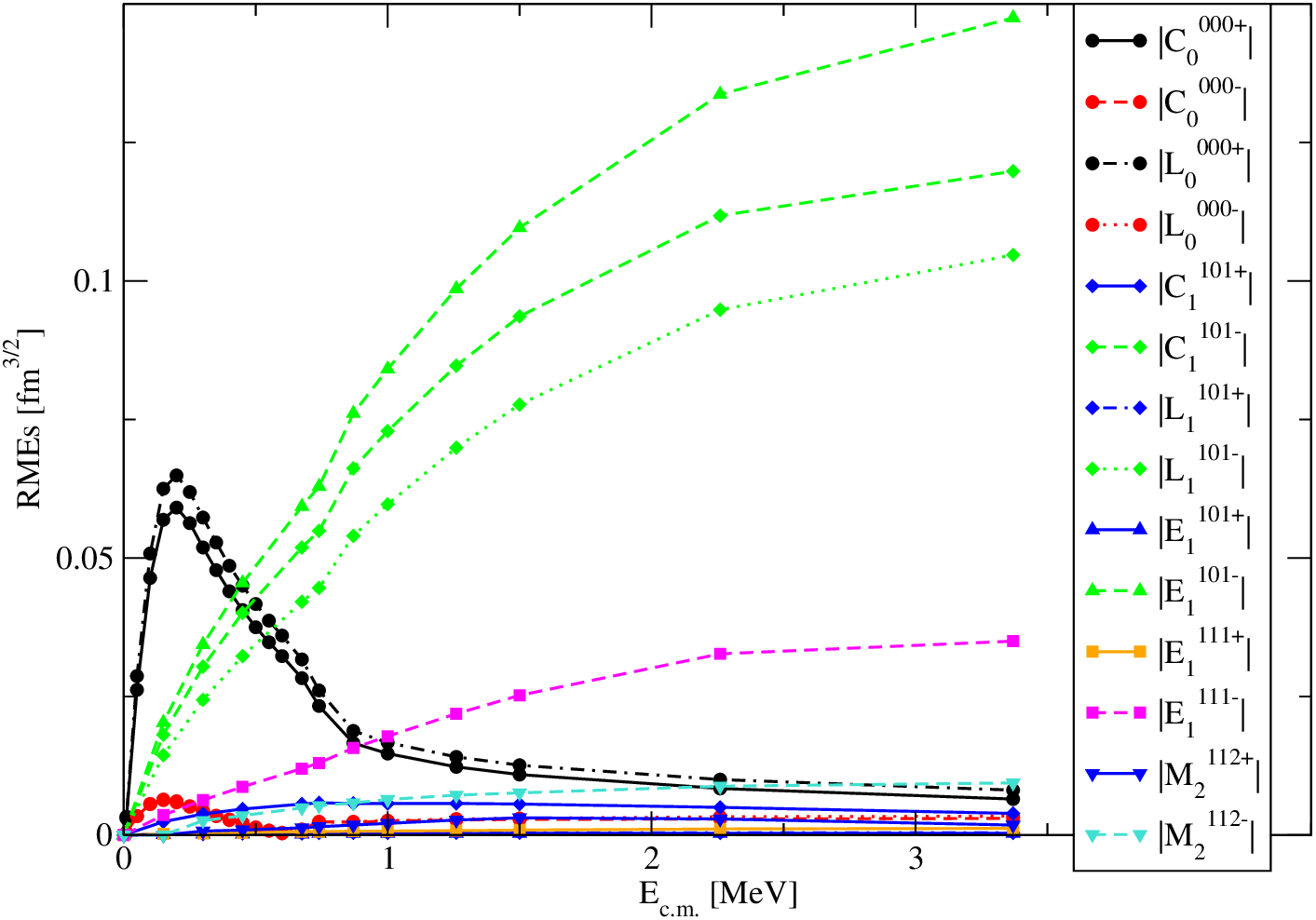}
   \caption{RMEs (in absolute value) resulting from $V$-induced transitions, see text
for additional information.  The $C_J^{LSJ\pm}$ ($L_J^{LSJ\pm}$,
$E_J^{LSJ\pm}$, and $M_J^{LSJ\pm}$) RMEs come from matrix elements
of the isoscalar ($+$) and isovector ($-$) charge (current) operator.
The calculations are for the $^3$H($p,e^+ e^-)^4$He process and are
based on the N3LO500/N2LO500 interactions.}
\label{fig:X_SV}
\end{figure}  
In Fig.~\ref{fig:X_SV}, we show the RMEs, as functions of the relative energy,
calculated with the $V$ operators.  The results are similar to those
shown in Fig.~\ref{fig:Edep}.  However, since we separate the isoscalar and
isovector components of the transition operators, we can observe the effect
of the different isospin content of the various initial states.  For example, in the
low-energy regime only the RME associated with the isoscalar component of the
$V$ charge operator displays a resonant behavior, confirming the predominant $T\,$=$\,$0
character of the $0^+$ resonance.  At higher energies, by contrast, the dominant contribution
is from the RMEs induced by the isovector $V$ charge and current operators (specifically,
the RMEs $C_1^{101-}$, $L_1^{101-}$, and $E_1^{101 -}$), attesting to the dominant
$T\,$=$\,$1 nature of the $1^-$ resonance; indeed, the corresponding isoscalar RMEs are
roughly an order of magnitude smaller. This is also related to the fact that,
for small $q$, the electric dipole operator is dominated by the
isovector component~\cite{Walecka1995}.

\begin{figure}[bth]
\centering
 \includegraphics[scale=0.35,clip]{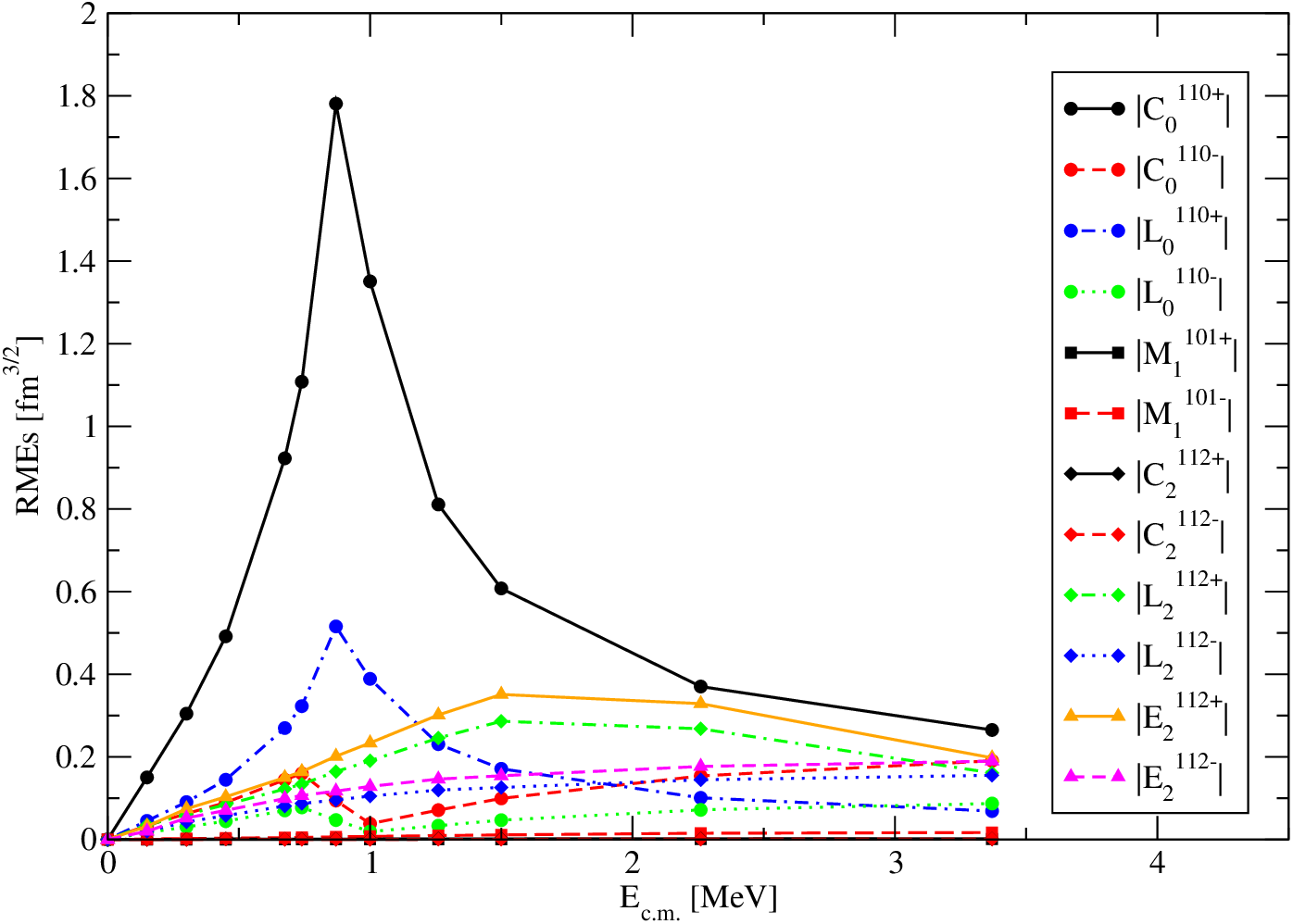}
   \caption{Same as in Fig.~\ref{fig:X_SV} bur for $A$-induced transitions.}
\label{fig:X_A}
\end{figure}
\begin{figure}[bth]
\centering
 \includegraphics[scale=0.35,clip]{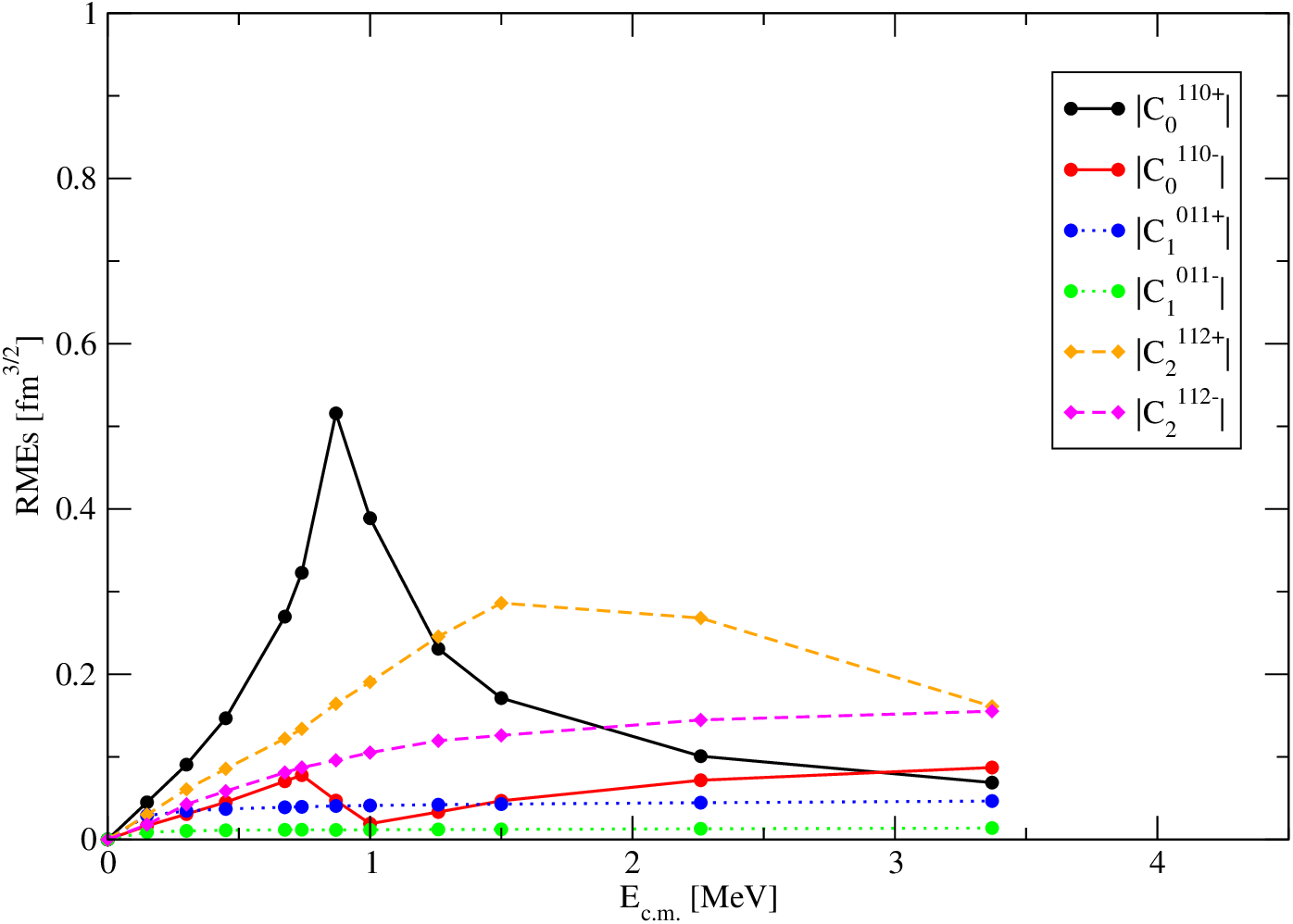}
   \caption{Same as in Fig.~\ref{fig:X_SV} bur for $P$-induced transitions.}
\label{fig:X_P}
\end{figure}

  
In Figs.~\ref{fig:X_A} and~\ref{fig:X_P}, we show the RMEs, as functions of the
relative energy, calculated only with the $A$ and $P$ operators. Note that the
charge RMEs are calculated using the operator of Eq.~(\ref{eq:e95}) for $P$-exchange
and the operator of Eq.~(\ref{eq:e96}) for $A$-exchange. 
These RMEs are qualitatively similar.  The resonant behavior of the 
$C_0^{110+}$ RME in both figures, due to the transition from the initial $^3P_0$
state to the final ground state, is evident.  This behavior is related to the 
presence in the $\heq$ spectrum of a $0^-$ resonance just above the
opening of the $n+^3$He channel.  The very broad peaks in the $C_2^{LS2\pm}$ 
and $E_2^{LS2\pm}$ RMEs, connected to the presence of $2^-$ resonant states 
in the $\heq$ spectrum, should also be noted.

\begin{figure}[bth]
\centering
 \includegraphics[scale=0.40]{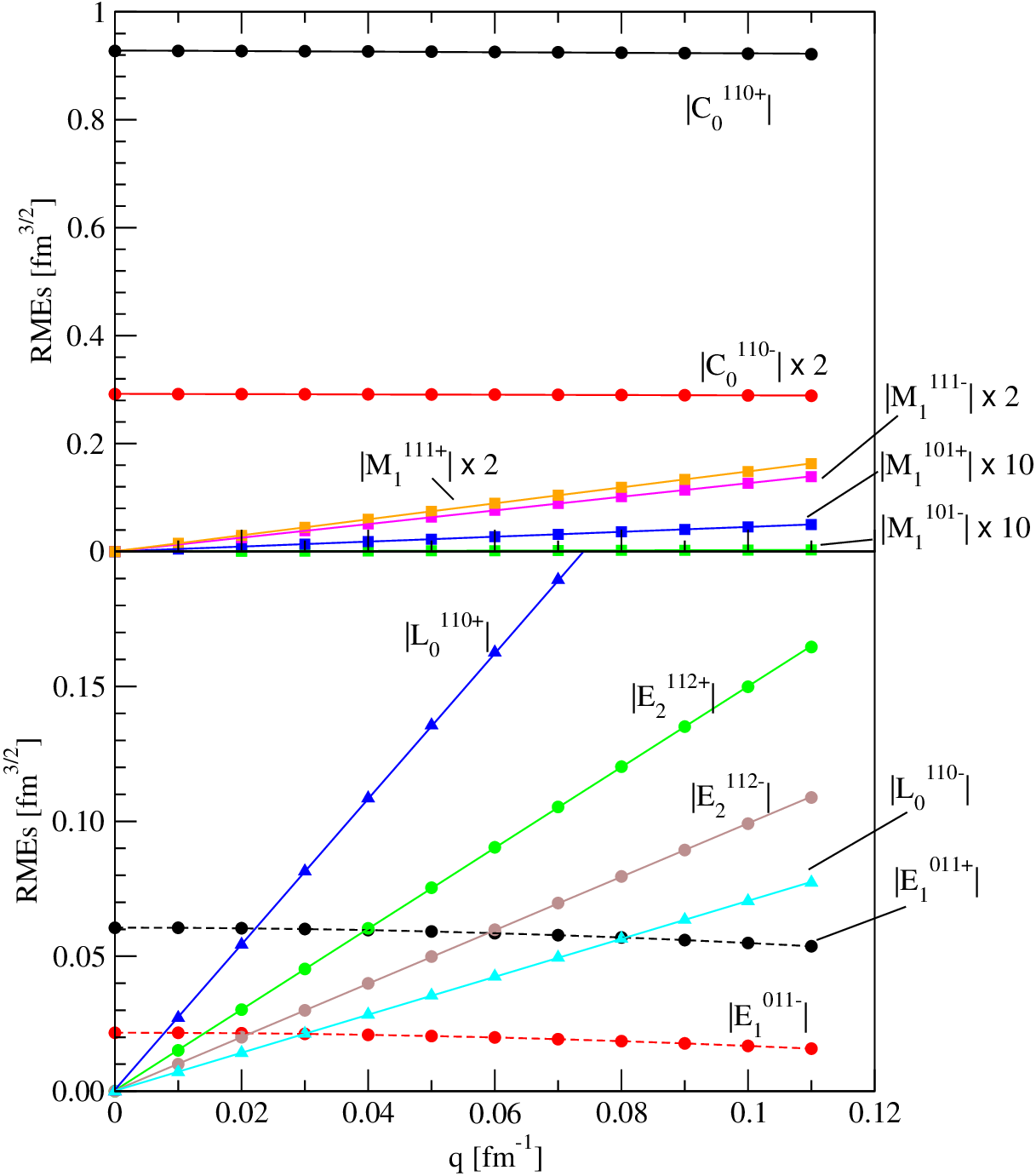}
   \caption{The dependence on the three-momentum transfer $q$ of some RMEs for the axial operators; the calculations are at incident proton
energy of 0.9 MeV and use the N3LO500/N2LO500 chiral
interactions.  The solid (dashed) lines show fits of the calculated values using linear (quadratic) parametrizations.
The RMEs not shown in this plot are negligible.}
\label{fig:cqq3}
\end{figure}
\begin{figure}[bth]
\centering
 \includegraphics[scale=0.40,clip]{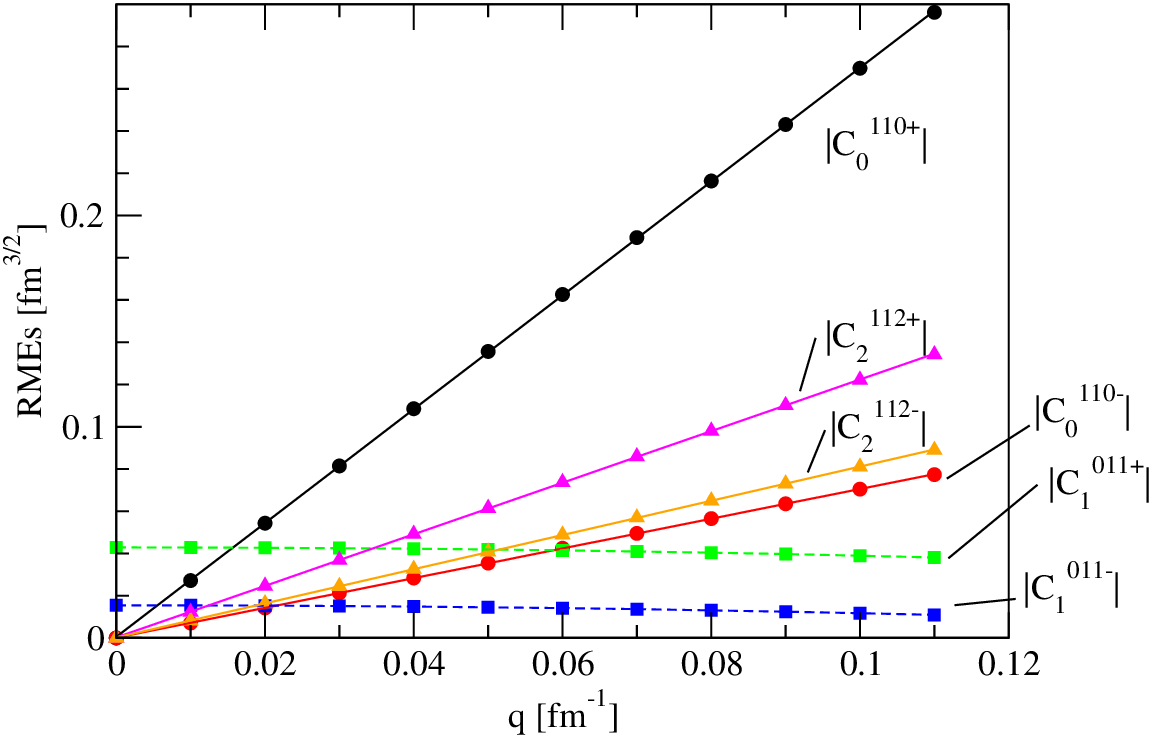}
   \caption{The same as in Fig.~\protect\ref{fig:cqq3}
   but for the pseudoscalar operator}
\label{fig:cqq4}
\end{figure}

  
Lastly, in Figs.~\ref{fig:cqq3} and~\ref{fig:cqq4} we report the dependence on the three-momentum transfer $q$ of
some of the RMEs associated with the axial- and pseudoscalar-exchange. Note that the RMEs associated with the pseudoscalar and  time component of the
axial operators behave differently,
in particular $C_0^{110\pm}$ are $q$-independent and linear in $q$ in the axial and pseudoscalar case,
respectively.

\subsection{Further comments on cross section results}
\label{sec:s7b}
Cross section results for the processes $^3$H$(p,e^+ e^-)^4$He and $^3$He$(n,e^+ e^-)^4$He
have been presented and discussed in Sec.~\ref{sec:s1d}.  Here, 
we only report the values we have adopted for the isoscalar and
isovector combinations, respectively $\eta^{c}_0$ and $\eta^{c}_z$,
of coupling constants modulo the unknown $\varepsilon_0$ and $\varepsilon_z$,
\begin{eqnarray}
\eta_0^{S}/\varepsilon_0 &=& - 4 m_\pi^2 c_1/ \Lambda_S\approx 0.085\ , \label{eq:csp}\nonumber \\
\eta_z^{S}/\varepsilon_z&=& -2 m_\pi^2 c_5/\Lambda_S\approx 0.00387\ , \label{eq:csm}\nonumber \\
\eta_0^{P}/\varepsilon_0&=& 2 m_\pi^2 \,m_N (d_{18}+2\,d_{19})/\Lambda_S \approx 0.038 \ , \label{eq:cpp}\nonumber\\
(g_A\,\eta_z^{P}/m_\pi f_\pi)/\varepsilon_z&=& g_A m_\pi / \Lambda_S \approx 0.175 \ , \label{eq:cpm}\nonumber\\
\eta_0^{V}/\varepsilon_0  &=& 3\ , \label{eq:cvp}\\
\eta_z^{V}/\varepsilon_z  &=& 1\ ,\qquad  \label{eq:cvm}\nonumber\\|
\eta_0^{A}/\varepsilon_0 &=& 3\,F-D\approx 0.55\ , \label{eq:cap}\nonumber\\
\eta_z^{A}/\varepsilon_z &=& F+D\approx 1.25\ , \nonumber\label{eq:cam}
\end{eqnarray}
see Eqs.~(\ref{eq:e58a}), (\ref{eq:e63a}), (\ref{eq:e63b}), (\ref{eq:e69a}), and (\ref{eq:e86a}).
The value of the LEC $c_1\,$=$\,-1.10$ GeV$^{-1}$ is taken
from Ref.~\cite{Hoferichter:2015hva}, where it has been
extracted from an analysis of $\pi N$ scattering data.  The value of the LEC $c_5$ is related to
the $n$-$p$ mass difference $\delta m^{\rm str}$ induced by the strong interactions~\cite{Bsaisou:2014oka}, that is,
$c_5\,$=$\,\delta m^{\rm str}/(4\,B_c (m_u-m_d)\,)\approx -9.9\times 10^{-2}$ GeV$^{-1}$. The values of $\delta m^{\rm str}$, and $m_u$ and $m_d$ are
taken from Lattice QCD calculations~\cite{Borsanyi:2014jba,Walker-Loud:2014iea,Aoki:2019cca}.
Note that $\eta^S_z/\varepsilon_z \ll \eta^S_0/\varepsilon_0$.  
The combination $\eta^S_0/\varepsilon_0$ is actually
related to the so-called $\sigma_{\pi N}$ term, that is, $\eta^S_0/\varepsilon_0\,$=$\,\sigma_{\pi N}/\Lambda_S$.
The adopted value is equivalent to approximating $\sigma_{\pi N}\,$=$\,-4 m_\pi^2 c_1$
which represents the LO contribution as determined in a $\chi$PT analysis~\cite{Crivellin:2013ipa}.  Since
this expansion is poorly convergent for $\sigma_{\pi N}$, perhaps a better approximation would
be to use the empirical value, as derived from the analysis of
$\pi N$ scattering data or directly from LQCD calculations (see, for example, Ref.~~\cite{RuizdeElvira:2017stg}).
This would give a value for $\eta^S_0/\varepsilon_0\,$ that is about 30\% smaller from that
reported in Eq.~(\ref{eq:csp}). The estimate for the pseudoscalar coupling constant
$\eta_0^{P}$ follows from taking $m_N(d_{18}+2\,d_{19})\,$=$\,1$ GeV$^{-1}$, with
$d_{18}\approx -1$ GeV$^{-2}$ from the Goldberger-Treiman
discrepancy~\cite{Fettes:1998ud} and the poorly known $d_{19} \approx 1$ GeV$^{-2}$.
Given that we have no knowledge
of the coupling constants of the X17 to the electron and nucleon,
the above uncertainties are unimportant at the present time. 

In computing the four-fold differential cross section, special
care must be exercised in carrying out the integration over the electron energy,
particularly when the width $\Gamma_X$ of the X17 is small. The
origin of this difficulty is easily understood.  The interference and
direct pieces of the cross section are proportional to powers of
\begin{equation}
\frac{1}{Q^2-M_X^2} \longrightarrow \frac{1}{Q^2-M_X^2+i\, M_X\, \Gamma_X} \equiv \frac{1}{D_X}\ ,
\end{equation}
where it has been assumed $\Gamma_X^2\ll M_X^2$.  When the lepton-pair kinematics
is such that $Q^2\approx M_X^2$, the X17
propagator reduces to $-i/(M_X\Gamma_X)$, and this behavior
is at the origin of the peaks observed in the five-fold differential
cross section for large $\theta_{ee}$.  We elaborate on this aspect
of the calculations in the following.
\subsubsection{Energy and angular dependence of the cross section}
\label{sec:crossX}
In general the five-fold differential cross section can be
schematically written as (we only indicate explicitly the energy
dependence)
\begin{eqnarray}
\label{eq:sigma5aa}
\hspace{-0.35cm} {d^5\sigma\over d\epsilon\, d\hat\bmk\, d\hat\bmk'}&=&
  \sigma(\epsilon)+ \varepsilon_e \left[ {R_X(\epsilon)\over D_X}+{\rm  c.c.}\right] 
   + \varepsilon_e^2 \, {R_{XX}(\epsilon)\over |D_X|^2} \\
&=&\sigma(\epsilon)+\frac{\varepsilon_e \left[ R_X(\epsilon)\, D_X^*+{\rm c.c.}\right]+\varepsilon_e^2\,
R_{XX}(\epsilon)}{|D_X|^2} \nonumber\ ,
\end{eqnarray}
where $\sigma(\epsilon)$, $R_X(\epsilon)$, and $R_{XX}(\epsilon)$
denote, respectively, the purely electromagnetic term, the interference term between electromagnetic
and X17-induced amplitudes, and the purely X17 term.  Note that
we have made explicit the dependence on the coupling constant $\varepsilon_e$.

\begin{figure}[bth]
\centering
 \includegraphics[scale=0.35,clip]{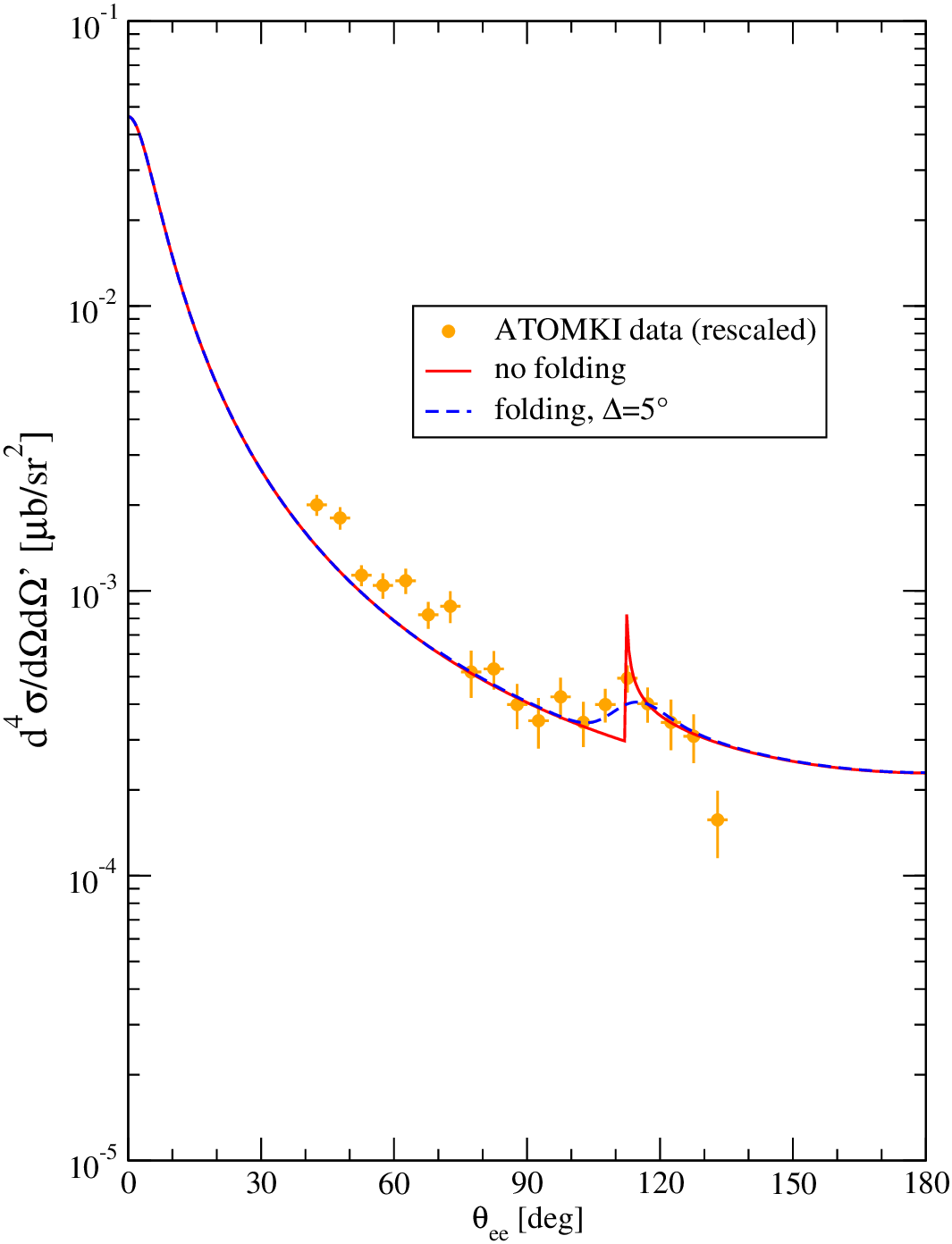}
   \caption{Example of the folding of the four-fold differential cross section for the process $\tri(p,e^-e^+)\heq$ calculated with
   the N3LO500/N2LO500 Hamiltonian at an incident
    proton energy of 0.9 MeV. }
\label{fig:sige_X17_dephi}
\end{figure}  

Using the Lagrangians given in Eq.~(\ref{eq:LeX}), the width of the X17 is obtained (in
first-order perturbation theory) as
\begin{eqnarray}
  \Gamma^S_X&=& \alpha\, \varepsilon_e^2 \, {M_X^2-4 m_e^2\over 2M_X}\sqrt{1-{4\,m_e^2\over M_X^2}}\ , 
  \nonumber\\
    \Gamma^P_X&=& \alpha\, \varepsilon_e^2 \, {M_X\over 2}\sqrt{1-{4\, m_e^2\over M_X^2}}\ ,
    \nonumber \\
      \Gamma^V_X&=& \alpha\, \varepsilon_e^2 \, {M_X^2+2\, m_e^2\over 3\, M_X}\sqrt{1-{4\, m_e^2\over M_X^2}}\ , \\
        \Gamma^A_X&=& \alpha\, \varepsilon_e^2 \, { M_X^2-4\, m_e^2\over 3\, M_X}\sqrt{1-{4\,m_e^2\over M_X^2}}\ . \nonumber
\end{eqnarray}
In the limit $M_X^2\gg m_e^2$, the width $\Gamma^c_X$  reduces to
\begin{equation}
\Gamma^c_X=x_c \, \alpha\, \varepsilon_e^2 \, M_X\equiv \varepsilon^2_e\, \gamma_X^c \ ,
\end{equation}
where $x_c$ is a numerical factor of order unity.  Available
current bounds on $\varepsilon_e$ suggest $\Gamma_X^c\ll M_X$~\cite{Feng:2016jff,DelleRose:2018pgm}.

As noted in Sec.~\ref{sec:s1d}, the condition $Q^2\,$=$\,M_X^2$ is
satisfied by two different values, $\epsilon_1$ and $\epsilon_2$, of the electron energy,
and according to Eq.~(\ref{eq:e16aa}) for $\epsilon$ close to $\epsilon_i$ we can
approximate $1/|D_X|^2$ as
\begin{equation}
\frac{1}{|D_X|^2} \longrightarrow \frac{\pi}{|\alpha_i|}\,\frac{1}{\varepsilon_e^2\, \gamma^c_X\, M_X}\, \delta(\epsilon-\epsilon_i) \ .
\end{equation}
In such a limit, in the cross section of Eq.~(\ref{eq:sigma5aa}),
the interference and direct terms are proportional to $\delta(\epsilon-\epsilon_i)$
with the electron energy argument evaluated at $\epsilon_i$.
It is worthwhile pointing out here that in the direct term proportional to $R_{XX}$
the dependence on $\varepsilon_e^2$ is removed (of course, in the present
tree-level treatment of the X17 width).  It turns out that for $\varepsilon_e\approx 10^{-3}$
the interference contribution (proportional to $R_X$) is always negligible
relative to the direct one. 

In order to obtain the four-fold differential cross section, the integration
over $\epsilon$ is carried out numerically for the electromagnetic
term, and analytically for the interference and direct terms.
We account roughly for the finite angular resolution of the detector
employed in the experiment by folding the $\theta_{ee}$ dependence
of this (four-fold) cross section with a normalized Gaussian of width $\Delta$.
The cross sections plotted in Figs.~\ref{fig:sige_X17_6ene_v6} and \ref{fig:sige_X17_theta}
have been obtained in this way, using typical values of $\Delta$ around $3$--$5^\circ$. 
In Fig.~\ref{fig:sige_X17_dephi}, we show an
example of a folded and non-folded cross section.
In principle, we should have accounted for the finite resolution
in the determination of the individual electron and positron angles
$\theta$ and $\theta'$ (rather than $\theta_{ee}$).  However,
in this first study we are not concerned with refinements such as
this. 
\subsubsection{Determining the X17 coupling constants from the 2019 ATOMKI data}
We now turn our attention to the comparison with the ATOMKI data for the $\tri(p,e^+ e^- )\heq$ process
at $E_p\,$=$\,0.90$ MeV~\cite{Krasznahorkay:2019lyl}.  As noted earlier, the cross section is
essentially independent of $\varepsilon_e$, but does depend
on the mass $M_X$ and the coupling constants $\eta^c_{\alpha}$.
We have taken $M_X\,$=$\,17$ MeV, while we have fixed the $\eta^c_{\alpha}$
by fitting the ATOMKI data in the peak region (the X17 contribution is negligible
for $\theta_{ee}< 90^\circ$) as follows: for $S$-exchange, we have determined
only $\eta_0^S$, and have set $\eta^S_z\,$=$\,0$, since (i) this coupling constant,
being proportional to the LEC $c_5$, is expected to be much smaller than $\eta_0^S$, and (ii) matrix
elements of the isovector $S$ current are much smaller than those of the isoscalar one;
for $P$ exchange, we have considered two
possibilities: in the first, we have set $\eta^P_0\,$=$\,0$, and have determined the isovector
coupling constant $\eta^P_z$ (this would be the leading-order contribution in $\chi$EFT), whereas
in the second we have determined $\eta^P_0$ by setting $\eta^P_z\,$ to zero, in accordance
with the piophobic hypothesis of Ref.~\cite{Alves:2020xhf};
for $V$ exchange we have taken $\eta^V_z\,$=$\,-\eta_0^V$ (proto-phobic
assumption) and have determined  $\eta^V_0$;  for $A$ exchange, we have
determined $\eta_0^A$, while setting $\eta^A_z\,$=$\,0$, since isoscalar matrix elements
of the $A$ current are much larger than isovector ones, see Figs.~\ref{fig:X_SV} and~\ref{fig:X_A}.

\begin{table}[bth]
\caption{\label{tab:eta_values}
Values of the coupling constants $\varepsilon_0$ and $\varepsilon_z$ obtained from the
fit of the 2019 ATOMKI angular distribution~\protect\cite{Krasznahorkay:2019lyl} for the two Hamiltonians 
employed in the present paper. The electromagnetic amplitudes are calculated using the 
corresponding accompanying electromagnetic currents.
The values in boldface have been fixed as discussed in the main text;
furthermore, $M_X\,$=$\,17$ MeV and $\varepsilon_e\,$=$\,10^{-3}$.
The relations between these parameters and the coupling constants $\eta^c_{\alpha}$
are given in Eqs.~(\ref{eq:e58a}), (\ref{eq:e63a}), (\ref{eq:e63b}),
(\ref{eq:e69a}), and (\ref{eq:e86a}). Note that the proto-phobic condition
for the vector case is equivalent to requiring $3\,\varepsilon_0+\varepsilon_z\,$=$\,0$.}
\begin{center}
\begin{tabular}{l|cc | cc}
\hline 
\hline
 & \multicolumn{2}{c|}{N3LO500/N2LO500} & \multicolumn{2}{c}{NVIa/3NIa}\\
\hline 
Case & $\varepsilon_0$  & $\varepsilon_z$ &   $\varepsilon_0$  & $\varepsilon_z$ \\
\hline
$S$  & $0.86\times 10^0$ & ${\bm 0}$ & $0.75\times 10^0$ & ${\bm 0}$ \\
$P$  & ${\bm 0}$ & $5.06\times 10^0$ & ${\bm 0}$  & $4.82\times 10^0$  \\
$P$  & $2.55\times 10^1$ & ${\bm 0}$ & $2.72\times 10^1$ & ${\bm 0}$  \\
$V$  & $2.56\times 10^{-3}$ & $-{\bm 3} \,{\bm \varepsilon}_0$ &  $2.66\times 10^{-3}$ & $-{\bm 3}\,{\bm  \varepsilon}_0$ \\
$A$  & $2.58\times 10^{-3}$ & ${\bm 0}$ & $2.89\times10^{-3}$ & ${\bm 0}$ \\
\hline 
\hline 
\end{tabular}
\end{center}
\end{table}
The values of the coupling constants resulting from
the fits\footnote{There are 19 data points; however, one of these points is excluded from the fit, since it is
used for rescaling the experimental cross sections
in the region of small $\theta_{ee}$.} are reported in Table~\ref{tab:eta_values} for the
two different Hamiltonian models (and accompanying electromagnetic currents), considered in the present work.  Of course, these
values depend on the choice of the parameter $\Delta$ of the
Gaussian function used in the folding procedure (see above); those in the table correspond to a $\Delta$ of
$5^\circ$. Nevertheless,
Table~\ref{tab:eta_values} indicates that the model dependence
is weak; indeed, the coupling constants vary at most of of about 10\%.
We have made no attempt to estimate the uncertainty coming
from the fits to the experimental data.

%In future, we plan to perform
%a complete study of the dependence on the nuclear interaction, 
%investigate the convergence with respect to the chiral order, etc. 
%However, we can safely affirm that the dependence on the nuclear
%dynamics is weak. In this exploratory study we also refrain ourselves from 
%estimating the uncertainty coming from the fit of the experimental data. 
%
\subsubsection{Comparing to previous determinations of the X17 coupling constants}
Of course, the coupling constants reported in Table~\ref{tab:eta_values} are rather uncertain
(they depend on the specific values assumed for $M_X$ and $\varepsilon_e$,
on the uncertainties in the experimental data, on the unsubtracted
background of EPC events, etc.).  Nevertheless, here we
compare them to previous determinations, where available. 
In Ref.~\cite{Feng:2016jff} the coupling constants $\epsilon_{u}$ and $\epsilon_{d}$,
entering $V$ exchange, were obtained by fitting the branching ratio for the $\beo$ anomaly,
with typical values of the order $\sim\,10^{-3}$.
The values for $\epsilon_{u,d}\,$=$\,(\varepsilon_0\pm \varepsilon_z)/2$
inferred from Table~\ref{tab:eta_values} are consistent (order of magnitude) with those
reported in that work.

For the case of $A$ exchange, we can compare the results of our fit with the $\epsilon_{u,d}$
extracted in Ref.~\cite{Kozaczuk:2016nma} (again, from the $\beo$ anomaly); there, they
were reported to be of order $10^{-5}$--$10^{-4}$.  We find them to be at least one order of
magnitude larger than in that work.  However, it is worthwhile pointing out that matrix elements
of the axial current, in particular its longitudinal component, between the $0^-$ and $0^+$ states
in helium are suppressed by $q$ relative to those between the $1^+$ and $0^+$ states in beryllium.
As a matter of fact, in helium these matrix elements are even smaller than those induced
by the axial charge, see curves labeled $L_0^{110+}$ and $C_0^{110+}$ in Fig.~\ref{fig:X_A},
even though, at least nominally, the latter operator
is subleading in the power counting of $\chi$EFT.
This suppression might explain why
the coupling constants extracted from the
helium and beryllium data differ by such a large
factor.

In reference to $P$ exchange, on the other hand, in the literature only the
  case of an isoscalar coupling is considered~\cite{DelleRose:2018pgm,Alves:2020xhf}.
 From our fit to the 2019 ATOMKI data, the nucleon coupling constant, Eq.~(\ref{eq:e63b}),
 is estimated to be $g^P_0\equiv e\,\eta_0^P\approx0.3$, two orders of magnitude
larger than obtained in Refs.~\cite{DelleRose:2018pgm,Alves:2020xhf}.  The origin
of this difference in the estimates of  $g^P_0$ is unclear at this point in time.

The exchange of a scalar X17 was already excluded by the analysis of the $\beo$ anomaly
in Refs.~\cite{Feng:2016jff,DelleRose:2018pgm}; we have been unable to find any estimates in
literature.

We conclude that both the $\beo$ and $\heq$ anomalies
can be explained simultaneously by the exchange of a proto-phobic vector X17.
For an axial X17 exchange, the coupling constants appear to be inconsistent with each other,
proviso the cautionary note above.
Clearly, these conclusions are somewhat uncertain, due to various assumptions
we made in fitting the data.  A better strategy would be to perform simulations of the experimental
data by including all relevant contributions for each of the four possibilities,
that the X17 be a $S$, $P$, $V$, and $A$ boson.
Work along these lines is in progress~\cite{Krasznahorkay:2015iga:2021abcd}.

\subsubsection{Comparing to the 2021 ATOMKI data}

\begin{figure}[bth]
  \centering
  \includegraphics[scale=0.50,clip]{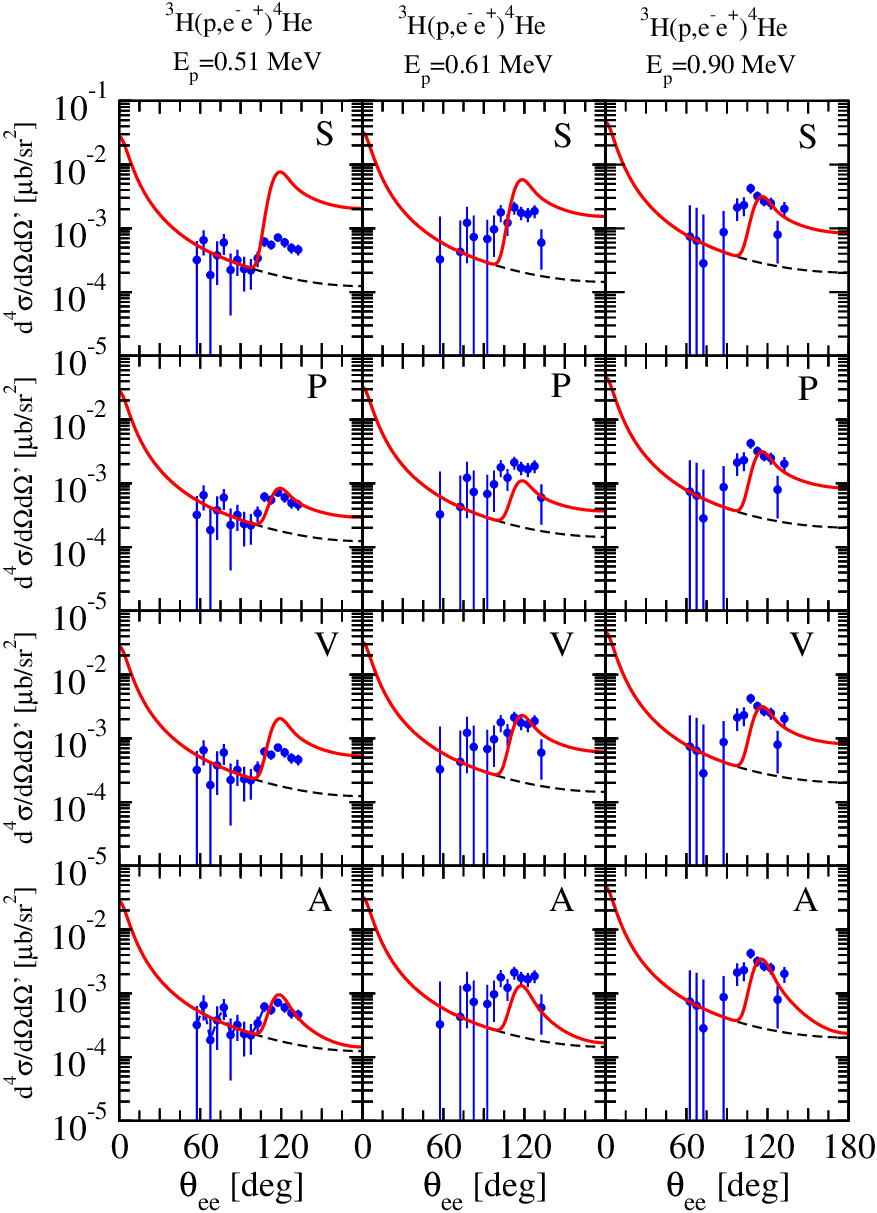}
  \caption{The four-fold differential cross section for the $\tri(p,e^-e^+)\heq$
   process at three different incident nucleon energies for the
   configuration in which the $e^+$ and $e^-$ momenta are
   in the plane orthogonal to the incident nucleon momentum and as function
   of the angle $\theta_{ee}$ between them, compared with the 2021 data of the ATOMKI
   group~\protect\cite{Krasznahorkay:2021joi}. Notation as in Fig.~\ref{fig:sige_X17_6ene_v6}.
   The coupling constants for the $S$, piophobic $P$, protophobic $V$, and $A$ X17 boson
   have been fixed by reproducing the
   data at $E_p\,$=$\,0.90$ MeV.}  
    \label{fig:sige_X17_3ene_v6_new}
\end{figure}

In this brief section we provide a preliminary analysis of the 2021 $\tri(p,e^+ e^- )\heq$ data
by the ATOMKI group~\cite{Krasznahorkay:2021joi}.  As noted earlier, in this
new publication the authors also report background-free data (i.e., data from
which EPC spurious events have been subtracted out).  Unfortunately, these
data have rather large errors, especially at smaller angles ($\theta_{ee}\lesssim 90^\circ$).
On a more positive note, however, measurements are carried out at three different beam
energies, thus permitting a first test of how the assumed nature for the X17 boson would
impact the energy dependence of the peak structure, see Fig.~\ref{fig:sige_X17_3ene_v6_new}.
  
In this figure, the coupling constants for the $S$ (with $\eta_z^S\,$=$\,0$),
piophobic $P$, proto-phobic $V$, and $A$ (with $\eta_z^A\,$=$\,0$) exchange
have been fixed by reproducing the data at $E_p\,$=$\,0.90$ MeV.  The
results at $E_p\,$=$\,0.51$ and $0.61$ MeV are therefore predictions, and seem to
better reproduce the corresponding data when the X17 is
either a pseudoscalar or an axial boson.  As the energy decreases, the prominence
of the peak relative to the baseline of $e^+$-$e^-$ pairs produced by purely
electromagnetic transitions is reduced, suggesting that the relevant
matrix element may be that connecting the $0^-$ resonance to
the $\heq$ ground state.  However, we note that (i)
the large errors in the experimental data at $E_p\,$=$\,0.90$ MeV render problematic
the matching of these data to the photon-only cross sections, and (ii) the coupling
constants extracted from the present fit are much larger (in absolute value) than
those reported in Table~\ref{tab:eta_values}.  In view of these considerations,
we do not list these values here.
  
In conclusion, in order to clarify the present situation, we believe it would be very helpful
to have more accurate measurements of these pair-production cross sections, by performing experimental 
studies in a wider range of energy and $e^+$-$e^-$ angles, as outlined in Sec.~\ref{sec:s1e}.
\acknowledgments
We wish to thank A.J. Krasznahorkay and the $n\_ {\rm TOF}$ collaboration for many useful discussions. 
+We also wish to thank D.~Alves for alerting us to the possibility that
the X17 may be piophobic and for prompting us to include an isoscalar coupling in the
Lagrangian ${\cal L}_X^P(x)$. 
The work of R.S. is supported by the U.S.~Department of Energy, Office
of Nuclear Science, under contract DE-AC05-06OR23177.
The calculations were made possible by grants of computing time
from the National Energy Research Supercomputer Center (NERSC) and
from the Italian National Supercomputing Center CINECA. We also gratefully acknowledge
the support of the INFN-Pisa computing center.

%\newpage
\appendix
\section{Chiral Lagrangians }
\label{sec:chieft}
The QCD Lagrangian ${\cal L}_{q}(x)$ in the $SU(2)$ formulation
with up and down quark flavors
and including couplings to external scalar, pseudoscalar, and vector
sources is written as
%\footnote{For the treatment of axial sources in the
%extended flavor $SU(3)$ framework, we defer to %Sec.~\ref{sec:sax17}.}
\begin{eqnarray}\label{eq:Lq}
  \mathcal{L}_{q}(x)&=&\mathcal{L}_{q}^{0}(x)
  +\overline{q}(x)\,\gamma^\mu \left[v_\mu(x)+\frac{1}{3}\,v_\mu^{s}(x)\right] q(x)\nonumber\\
  &-&\overline{q}(x)[ s(x)-i\gamma^5 \,p(x) ]q(x)\ , 
\end{eqnarray}
where $\mathcal{L}_q^{0}(x)$ is the Lagrangian for massless quarks,
$q(x)$ is the two-component flavor vector
\begin{equation}
q(x)=\left[\!\begin{array}{c} 
         u(x)\\
         d(x)
         \end{array}\!\right]\ ,
\end{equation}
and $u(x)$ and $d(x)$ are the up and down  quark fields,
respectively.  The external scalar, pseudoscalar,
and vector sources $s(x)$, $p(x)$, and $v_\mu(x)$
have the flavor (or isospin) structure
\begin{equation}
s(x)=\sum_{i=0}^3\tau_i\, s_i(x) \ ,\qquad p(x)=\sum_{i=0}^3\tau_i\, p_i(x) \ ,
\end{equation}
and
\begin{equation}
v_\mu(x)=\sum_{i=1}^3\tau_i\, v_{\mu,i}(x) \ ,
\end{equation}
where $\tau_0$ is the
identity matrix (in isospin space) and $\tau_i$ are standard Pauli matrices.
The external (isoscalar) vector source $v_\mu^{s}(x)$ is multiplied
by the identity matrix $\tau_0$.
Quark masses are reintroduced as part of the scalar source
$s(x)=M_q+\cdots$ via
\begin{equation}
M_q=\left(\begin{array}{cc} 
         m_u & 0\\
         0 & m_d
         \end{array}\right)=m_q\,\tau_0+\delta m_q\, \tau_3\ ,
\end{equation}
where $m_q=(m_u+m_d)/2$ and $\delta m_q=(m_u-m_d)/2$,
and $m_u$ and $m_d$ are the up- and down-quark mass, respectively.

The chiral Lagrangian ${\cal L}_\chi(x)$ consists of terms 
${\cal L}_{\pi \pi}(x)$ and ${\cal L}_{\pi N}(x)$ describing, respectively,
the interactions of pions and those of pions with nucleons.  These
terms have the expansions
\begin{eqnarray}
{\cal L}_{\pi\pi}(x)&=& {\cal L}_{\pi\pi}^{(2)}(x)
%+{\cal L}_{\pi\pi}^{(4)}(x)
+\cdots \ , \\
{\cal L}_{\pi N}(x)&=&{\cal L}_{\pi N}^{(1)}(x)+{\cal L}_{\pi N}^{(2)}(x)+
{\cal L}_{\pi N}^{(3)}(x)
+\cdots\ ,\label{eq:Lpc_pin}
\end{eqnarray}
where
\begin{eqnarray}
  {\cal L}_{\pi\pi}^{(2)}&=& {f_\pi^2\over 4} \langle\nabla_\mu U^\dag
  \nabla^\mu U + \chi^\dag U+\chi U^\dag\rangle\ , \label{eq:Lpc_pipi2} \\
  {\cal L}_{\pi N}^{(1)}&=& \overline{N}\Bigl(i\gamma^\mu D_\mu -m_N \,\tau_0
  +{g_A\over 2} \gamma^\mu\gamma^5 u_\mu \Bigr)N\ , \label{eq:Lpc_pin1} \\
  {\cal L}_{\pi N}^{(2)}  &=& \overline{N}\Bigl(c_1\langle \chi_+\rangle
    +\cdots + c_5\hat\chi +{\kappa_z\over 8\,m_N}\sigma^{\mu\nu}F_{\mu\nu}^+\nonumber\\
    &&\quad  +{\kappa_0\over 4\,m_N}\sigma^{\mu\nu}F_{\mu\nu}^s+\cdots\Bigr) N
     \ ,\label{eq:Lpc_pin2}\\
  {\cal L}_{\pi N}^{(3)}  &=& \cdots+ i\, {d_{18} \over 2}\,\overline{N}\,
  \gamma^\mu\gamma^5 [D_\mu\, , \,\chi_-] \,N \nonumber\\
      &&\quad  +i\, {d_{19} \over 2}\,\overline{N}\,
  \gamma^\mu\gamma^5 [D_\mu\, , \,\langle\chi_-\rangle] \,N+
  \cdots \ ,\label{eq:Lpc_pin3}
\end{eqnarray}
where $f_\pi\approx 92.4$ MeV is the pion decay constant,
$g_A=1.26$ is the nucleon axial coupling constant, $m_N$ is the nucleon mass,
$\langle\cdots\rangle$ denotes a trace over isospin, and $\hat\chi=\chi-\langle\chi\rangle/2$.
The isotriplet of pion fields is denoted below with ${\bm \pi}(x)$, while the isodoublet
nucleon field $N(x)$ is given by
\begin{equation}
N(x) =\left[\begin{array}{c} 
         p(x)\\
         n(x)
         \end{array}\right]\ .
\end{equation}
In the previous expressions we have omitted terms not relevant in the present work; the complete
${\cal L}_{\pi\pi}^{(4)}$ can be found in Ref.~\cite{Gasser:1983yg}, and the complete
${\cal L}_{\pi N}^{(2)}$ and successive term in Ref.~\cite{Fettes:2000gb}.  Here,
we adopt the notation and conventions of this latter work for the
various fields and covariant derivatives, which we summarize below:
\begin{eqnarray}
  U&=&1+{i\over f_\pi} {\bm \tau} \cdot {\bm \pi}-{1\over 2f_\pi^2}\, {\bm \pi}^{2}+ \cdots
  \ , \nonumber \\
  \nabla_\mu U  &=& \partial_\mu
  U -i\, r_\mu\, U + i\, U \,\ell_\mu \ ,\nonumber \\
  u&=& \sqrt{U}\ , \nonumber \\
  D_\mu N &=& (\partial_\mu +\Gamma_\mu - i \,v_\mu^{s})\,N\ , \nonumber\\
  u_\mu &=& i(u^\dag\partial_\mu u - u\, \partial_\mu u^\dag)
    +u^\dag r_\mu u - u\, \ell_\mu u^\dag\ , \\
  \Gamma_\mu &=& {1\over 2} (u^\dag\partial_\mu u + u\, \partial_\mu
  u^\dag) - \frac{i}{2}(     u^\dag r_\mu u + u\,  \ell_\mu u^\dag) \ , \nonumber\\
  \chi_\pm &=& u^\dag \chi \,u^\dag \pm u\, \chi^\dag u\ ,\nonumber\\
  F_{\mu\nu}^\pm&=& u^\dag \, F_{\mu\nu}^R \, u \pm u\, F_{\mu\nu}^L \, u^\dag
  \ .\nonumber
  \label{eq:ea8}
\end{eqnarray}
For the cases of interest in the present work, we have
\begin{eqnarray}
   r_\mu &=& \ell_\mu = v_\mu \ , \quad \chi = 2\,B_c\,(s + i p)\ ,\nonumber\\
   F_{\mu\nu}^R &=& F_{\mu\nu}^L = \partial_\mu v_\nu - \partial_\nu v_\mu - i\,
   [v_\mu\, ,\,v_\nu] \ , \label{eq:fmunuV}\\
   F_{\mu\nu}^s &=& \partial_\mu v_\nu^{s} - \partial_\nu v_\mu^{s} \ ,\nonumber
\end{eqnarray}
The parameters $B_c$, $c_1$, etc., entering the Lagrangians are the so-called low energy constants (LECs),
to be determined from experimental data and/or (possibly) lattice QCD calculations. 

The nucleon and pion fields (in interaction picture) read 
\begin{eqnarray}
  N_t(x) & = & \sum_{\bmp s} {1\over \sqrt{\Omega} }
   \Bigl[ b_{\bmp s t}\, u_{\bmp s}\, e^{-i p\cdot x}
     + d_{\bmp s t}^\dag \, v_{\bmp s} \,e^{i p\cdot x}\Bigr]\ ,\label{eq:nfield} \nonumber \\
  \pi_a(x) & = & \sum_{\bmk} {1\over \sqrt{2\omega_k\, \Omega} }
   \Bigl[ a_{\bmk a}\,  e^{-ik\cdot x}
     + a_{\bmk a}^\dag \, e^{i k\cdot x}\Bigr]\ ,\label{eq:pifield}
\end{eqnarray}
where the various momenta are discretized by assuming periodic boundary
conditions in a box of volume $\Omega$, $b_{\bmp s t}$ and $d_{\bmp s t}$
are the annihilation operators of, respectively, a nucleon and antinucleon having spin
and isospin projections $s$ and $t$ ($t\,$=$\,1/2$ for a proton and $t\,$=$\,-1/2$ for a neutron),
$u$ and $v$ are the corresponding Dirac spinors with the (non-standard)
normalization $u^\dagger u\,$=$\,v^\dagger v\,$=$\,1$,
and $a_{\bmk a}$ is the annihilation operator of a pion
of isospin projection $a$.

\bibliographystyle{apsrev4-1}
\bibliography{x17-paper}{}

\begin{thebibliography}{93}%
\makeatletter
\providecommand \@ifxundefined [1]{%
 \@ifx{#1\undefined}
}%
\providecommand \@ifnum [1]{%
 \ifnum #1\expandafter \@firstoftwo
 \else \expandafter \@secondoftwo
 \fi
}%
\providecommand \@ifx [1]{%
 \ifx #1\expandafter \@firstoftwo
 \else \expandafter \@secondoftwo
 \fi
}%
\providecommand \natexlab [1]{#1}%
\providecommand \enquote  [1]{``#1''}%
\providecommand \bibnamefont  [1]{#1}%
\providecommand \bibfnamefont [1]{#1}%
\providecommand \citenamefont [1]{#1}%
\providecommand \href@noop [0]{\@secondoftwo}%
\providecommand \href [0]{\begingroup \@sanitize@url \@href}%
\providecommand \@href[1]{\@@startlink{#1}\@@href}%
\providecommand \@@href[1]{\endgroup#1\@@endlink}%
\providecommand \@sanitize@url [0]{\catcode `\\12\catcode `\$12\catcode
  `\&12\catcode `\#12\catcode `\^12\catcode `\_12\catcode `\%12\relax}%
\providecommand \@@startlink[1]{}%
\providecommand \@@endlink[0]{}%
\providecommand \url  [0]{\begingroup\@sanitize@url \@url }%
\providecommand \@url [1]{\endgroup\@href {#1}{\urlprefix }}%
\providecommand \urlprefix  [0]{URL }%
\providecommand \Eprint [0]{\href }%
\providecommand \doibase [0]{http://dx.doi.org/}%
\providecommand \selectlanguage [0]{\@gobble}%
\providecommand \bibinfo  [0]{\@secondoftwo}%
\providecommand \bibfield  [0]{\@secondoftwo}%
\providecommand \translation [1]{[#1]}%
\providecommand \BibitemOpen [0]{}%
\providecommand \bibitemStop [0]{}%
\providecommand \bibitemNoStop [0]{.\EOS\space}%
\providecommand \EOS [0]{\spacefactor3000\relax}%
\providecommand \BibitemShut  [1]{\csname bibitem#1\endcsname}%
\let\auto@bib@innerbib\@empty
%</preamble>
\bibitem [{\citenamefont {Battaglieri}\ \emph {et~al.}(2017)\citenamefont
  {Battaglieri} \emph {et~al.}}]{Battaglieri:2017aum}%
  \BibitemOpen
  \bibfield  {author} {\bibinfo {author} {\bibfnamefont {M.}~\bibnamefont
  {Battaglieri}} \emph {et~al.},\ }in\ \href@noop {} {\emph {\bibinfo
  {booktitle} {{U.S . Cosmic Visions: New Ideas in Dark Matter}}}}\ (\bibinfo
  {year} {2017})\ \Eprint {http://arxiv.org/abs/1707.04591} {arXiv:1707.04591
  [hep-ph]} \BibitemShut {NoStop}%
\bibitem [{\citenamefont {Bertone}\ and\ \citenamefont
  {Hooper}(2018)}]{Bertone:2018aaa}%
  \BibitemOpen
  \bibfield  {author} {\bibinfo {author} {\bibfnamefont {G.}~\bibnamefont
  {Bertone}}\ and\ \bibinfo {author} {\bibfnamefont {D.}~\bibnamefont
  {Hooper}},\ }\href {\doibase 10.1103/RevModPhys.90.045002} {\bibfield
  {journal} {\bibinfo  {journal} {Rev. Mod. Phys.}\ }\textbf {\bibinfo {volume}
  {90}},\ \bibinfo {pages} {045002} (\bibinfo {year} {2018})}\BibitemShut
  {NoStop}%
\bibitem [{\citenamefont {Krasznahorkay}\ \emph {et~al.}(2016)\citenamefont
  {Krasznahorkay} \emph {et~al.}}]{Krasznahorkay:2015iga}%
  \BibitemOpen
  \bibfield  {author} {\bibinfo {author} {\bibfnamefont {A.~J.}\ \bibnamefont
  {Krasznahorkay}} \emph {et~al.},\ }\href {\doibase
  10.1103/PhysRevLett.116.042501} {\bibfield  {journal} {\bibinfo  {journal}
  {Phys. Rev. Lett.}\ }\textbf {\bibinfo {volume} {116}},\ \bibinfo {pages}
  {042501} (\bibinfo {year} {2016})},\ \Eprint
  {http://arxiv.org/abs/1504.01527} {arXiv:1504.01527 [nucl-ex]} \BibitemShut
  {NoStop}%
\bibitem [{\citenamefont {Pospelov}\ \emph {et~al.}(2008)\citenamefont
  {Pospelov}, \citenamefont {Ritz},\ and\ \citenamefont
  {Voloshin}}]{Pospelov:2008aaa}%
  \BibitemOpen
  \bibfield  {author} {\bibinfo {author} {\bibfnamefont {M.}~\bibnamefont
  {Pospelov}}, \bibinfo {author} {\bibfnamefont {A.}~\bibnamefont {Ritz}}, \
  and\ \bibinfo {author} {\bibfnamefont {M.}~\bibnamefont {Voloshin}},\ }\href
  {\doibase https://doi.org/10.1016/j.physletb.2008.02.052} {\bibfield
  {journal} {\bibinfo  {journal} {Physics Letters B}\ }\textbf {\bibinfo
  {volume} {662}},\ \bibinfo {pages} {53 } (\bibinfo {year}
  {2008})}\BibitemShut {NoStop}%
\bibitem [{\citenamefont {Banerjee}\ \emph {et~al.}(2018)\citenamefont
  {Banerjee} \emph {et~al.}}]{Banerjee:2018vgk}%
  \BibitemOpen
  \bibfield  {author} {\bibinfo {author} {\bibfnamefont {D.}~\bibnamefont
  {Banerjee}} \emph {et~al.} (\bibinfo {collaboration} {NA64}),\ }\href
  {\doibase 10.1103/PhysRevLett.120.231802} {\bibfield  {journal} {\bibinfo
  {journal} {Phys. Rev. Lett.}\ }\textbf {\bibinfo {volume} {120}},\ \bibinfo
  {pages} {231802} (\bibinfo {year} {2018})},\ \Eprint
  {http://arxiv.org/abs/1803.07748} {arXiv:1803.07748 [hep-ex]} \BibitemShut
  {NoStop}%
\bibitem [{\citenamefont {Batley}\ \emph {et~al.}(2015)\citenamefont {Batley}
  \emph {et~al.}}]{Batley:2015lha}%
  \BibitemOpen
  \bibfield  {author} {\bibinfo {author} {\bibfnamefont {J.}~\bibnamefont
  {Batley}} \emph {et~al.} (\bibinfo {collaboration} {NA48/2}),\ }\href
  {\doibase 10.1016/j.physletb.2015.04.068} {\bibfield  {journal} {\bibinfo
  {journal} {Phys. Lett. B}\ }\textbf {\bibinfo {volume} {746}},\ \bibinfo
  {pages} {178} (\bibinfo {year} {2015})},\ \Eprint
  {http://arxiv.org/abs/1504.00607} {arXiv:1504.00607 [hep-ex]} \BibitemShut
  {NoStop}%
\bibitem [{\citenamefont {Andreas}\ \emph {et~al.}(2010)\citenamefont
  {Andreas}, \citenamefont {Lebedev}, \citenamefont {Ramos-Sanchez},\ and\
  \citenamefont {Ringwald}}]{Andreas:2010ms}%
  \BibitemOpen
  \bibfield  {author} {\bibinfo {author} {\bibfnamefont {S.}~\bibnamefont
  {Andreas}}, \bibinfo {author} {\bibfnamefont {O.}~\bibnamefont {Lebedev}},
  \bibinfo {author} {\bibfnamefont {S.}~\bibnamefont {Ramos-Sanchez}}, \ and\
  \bibinfo {author} {\bibfnamefont {A.}~\bibnamefont {Ringwald}},\ }\href
  {\doibase 10.1007/JHEP08(2010)003} {\bibfield  {journal} {\bibinfo  {journal}
  {J.High Energ. Phys.}\ }\textbf {\bibinfo {volume} {2010}},\ \bibinfo {pages}
  {3} (\bibinfo {year} {2010})},\ \Eprint {http://arxiv.org/abs/1005.3978}
  {arXiv:1005.3978 [hep-ph]} \BibitemShut {NoStop}%
\bibitem [{\citenamefont {Feng}\ \emph {et~al.}(2016)\citenamefont {Feng},
  \citenamefont {Fornal}, \citenamefont {Galon}, \citenamefont {Gardner},
  \citenamefont {Smolinsky}, \citenamefont {Tait},\ and\ \citenamefont
  {Tanedo}}]{Feng:2016jff}%
  \BibitemOpen
  \bibfield  {author} {\bibinfo {author} {\bibfnamefont {J.~L.}\ \bibnamefont
  {Feng}}, \bibinfo {author} {\bibfnamefont {B.}~\bibnamefont {Fornal}},
  \bibinfo {author} {\bibfnamefont {I.}~\bibnamefont {Galon}}, \bibinfo
  {author} {\bibfnamefont {S.}~\bibnamefont {Gardner}}, \bibinfo {author}
  {\bibfnamefont {J.}~\bibnamefont {Smolinsky}}, \bibinfo {author}
  {\bibfnamefont {T.~M.~P.}\ \bibnamefont {Tait}}, \ and\ \bibinfo {author}
  {\bibfnamefont {P.}~\bibnamefont {Tanedo}},\ }\href {\doibase
  10.1103/PhysRevLett.117.071803} {\bibfield  {journal} {\bibinfo  {journal}
  {Phys. Rev. Lett.}\ }\textbf {\bibinfo {volume} {117}},\ \bibinfo {pages}
  {071803} (\bibinfo {year} {2016})},\ \Eprint
  {http://arxiv.org/abs/1604.07411} {arXiv:1604.07411 [hep-ph]} \BibitemShut
  {NoStop}%
\bibitem [{\citenamefont {Kozaczuk}\ \emph {et~al.}(2017)\citenamefont
  {Kozaczuk}, \citenamefont {Morrissey},\ and\ \citenamefont
  {Stroberg}}]{Kozaczuk:2016nma}%
  \BibitemOpen
  \bibfield  {author} {\bibinfo {author} {\bibfnamefont {J.}~\bibnamefont
  {Kozaczuk}}, \bibinfo {author} {\bibfnamefont {D.~E.}\ \bibnamefont
  {Morrissey}}, \ and\ \bibinfo {author} {\bibfnamefont {S.}~\bibnamefont
  {Stroberg}},\ }\href {\doibase 10.1103/PhysRevD.95.115024} {\bibfield
  {journal} {\bibinfo  {journal} {Phys. Rev. D}\ }\textbf {\bibinfo {volume}
  {95}},\ \bibinfo {pages} {115024} (\bibinfo {year} {2017})},\ \Eprint
  {http://arxiv.org/abs/1612.01525} {arXiv:1612.01525 [hep-ph]} \BibitemShut
  {NoStop}%
\bibitem [{\citenamefont {Krasznahorkay}\ \emph {et~al.}(2019)\citenamefont
  {Krasznahorkay} \emph {et~al.}}]{Krasznahorkay:2019lyl}%
  \BibitemOpen
  \bibfield  {author} {\bibinfo {author} {\bibfnamefont {A.}~\bibnamefont
  {Krasznahorkay}} \emph {et~al.},\ }\href@noop {} {\  (\bibinfo {year}
  {2019})},\ \Eprint {http://arxiv.org/abs/1910.10459} {arXiv:1910.10459
  [nucl-ex]} \BibitemShut {NoStop}%
\bibitem [{\citenamefont {Krasznahorkay}\ \emph
  {et~al.}(2021{\natexlab{a}})\citenamefont {Krasznahorkay}, \citenamefont
  {Csatl\'os}, \citenamefont {Csige}, \citenamefont {Guly\'as}, \citenamefont
  {Krasznahorkay}, \citenamefont {Nyak\'o}, \citenamefont {Rajta},
  \citenamefont {Tim\'ar}, \citenamefont {Vajda},\ and\ \citenamefont
  {Sas}}]{Krasznahorkay:2021joi}%
  \BibitemOpen
  \bibfield  {author} {\bibinfo {author} {\bibfnamefont {A.~J.}\ \bibnamefont
  {Krasznahorkay}}, \bibinfo {author} {\bibfnamefont {M.}~\bibnamefont
  {Csatl\'os}}, \bibinfo {author} {\bibfnamefont {L.}~\bibnamefont {Csige}},
  \bibinfo {author} {\bibfnamefont {J.}~\bibnamefont {Guly\'as}}, \bibinfo
  {author} {\bibfnamefont {A.}~\bibnamefont {Krasznahorkay}}, \bibinfo {author}
  {\bibfnamefont {B.~M.}\ \bibnamefont {Nyak\'o}}, \bibinfo {author}
  {\bibfnamefont {I.}~\bibnamefont {Rajta}}, \bibinfo {author} {\bibfnamefont
  {J.}~\bibnamefont {Tim\'ar}}, \bibinfo {author} {\bibfnamefont
  {I.}~\bibnamefont {Vajda}}, \ and\ \bibinfo {author} {\bibfnamefont {N.~J.}\
  \bibnamefont {Sas}},\ }\href@noop {} {\  (\bibinfo {year}
  {2021}{\natexlab{a}})},\ \Eprint {http://arxiv.org/abs/2104.10075}
  {arXiv:2104.10075 [nucl-ex]} \BibitemShut {NoStop}%
\bibitem [{\citenamefont {Feng}\ \emph {et~al.}(2020)\citenamefont {Feng},
  \citenamefont {Tait},\ and\ \citenamefont {Verhaaren}}]{Feng:2020mbt}%
  \BibitemOpen
  \bibfield  {author} {\bibinfo {author} {\bibfnamefont {J.~L.}\ \bibnamefont
  {Feng}}, \bibinfo {author} {\bibfnamefont {T.~M.}\ \bibnamefont {Tait}}, \
  and\ \bibinfo {author} {\bibfnamefont {C.~B.}\ \bibnamefont {Verhaaren}},\
  }\href {\doibase 10.1103/PhysRevD.102.036016} {\bibfield  {journal} {\bibinfo
   {journal} {Phys. Rev. D}\ }\textbf {\bibinfo {volume} {102}},\ \bibinfo
  {pages} {036016} (\bibinfo {year} {2020})},\ \Eprint
  {http://arxiv.org/abs/2006.01151} {arXiv:2006.01151 [hep-ph]} \BibitemShut
  {NoStop}%
\bibitem [{\citenamefont {Zhang}\ and\ \citenamefont
  {Miller}(2021)}]{Zhang:2020ukq}%
  \BibitemOpen
  \bibfield  {author} {\bibinfo {author} {\bibfnamefont {X.}~\bibnamefont
  {Zhang}}\ and\ \bibinfo {author} {\bibfnamefont {G.~A.}\ \bibnamefont
  {Miller}},\ }\href@noop {} {\bibfield  {journal} {\bibinfo  {journal}
  {Physics Letters B}\ }\textbf {\bibinfo {volume} {813}},\ \bibinfo {pages}
  {136061} (\bibinfo {year} {2021})},\ \Eprint
  {http://arxiv.org/abs/2008.11288} {arXiv:2008.11288 [hep-ph]} \BibitemShut
  {NoStop}%
\bibitem [{\citenamefont {Hayes}\ \emph {et~al.}(2021)\citenamefont {Hayes},
  \citenamefont {Friar}, \citenamefont {Hale},\ and\ \citenamefont
  {Garvey}}]{Hayes:2021hin}%
  \BibitemOpen
  \bibfield  {author} {\bibinfo {author} {\bibfnamefont {A.~C.}\ \bibnamefont
  {Hayes}}, \bibinfo {author} {\bibfnamefont {J.~L.}\ \bibnamefont {Friar}},
  \bibinfo {author} {\bibfnamefont {G.}~\bibnamefont {Hale}}, \ and\ \bibinfo
  {author} {\bibfnamefont {G.}~\bibnamefont {Garvey}},\ }\href@noop {} {\
  (\bibinfo {year} {2021})},\ \Eprint {http://arxiv.org/abs/2106.06834}
  {arXiv:2106.06834 [nucl-th]} \BibitemShut {NoStop}%
\bibitem [{\citenamefont {Andreas}\ \emph {et~al.}(2012)\citenamefont
  {Andreas}, \citenamefont {Niebuhr},\ and\ \citenamefont
  {Ringwald}}]{PhysRevD.86.095019}%
  \BibitemOpen
  \bibfield  {author} {\bibinfo {author} {\bibfnamefont {S.}~\bibnamefont
  {Andreas}}, \bibinfo {author} {\bibfnamefont {C.}~\bibnamefont {Niebuhr}}, \
  and\ \bibinfo {author} {\bibfnamefont {A.}~\bibnamefont {Ringwald}},\ }\href
  {\doibase 10.1103/PhysRevD.86.095019} {\bibfield  {journal} {\bibinfo
  {journal} {Phys. Rev. D}\ }\textbf {\bibinfo {volume} {86}},\ \bibinfo
  {pages} {095019} (\bibinfo {year} {2012})}\BibitemShut {NoStop}%
\bibitem [{\citenamefont {Riordan}\ \emph {et~al.}(1987)\citenamefont
  {Riordan}, \citenamefont {Krasny}, \citenamefont {Lang}, \citenamefont
  {de~Barbaro}, \citenamefont {Bodek}, \citenamefont {Dasu}, \citenamefont
  {Varelas}, \citenamefont {Wang}, \citenamefont {Arnold}, \citenamefont
  {Benton}, \citenamefont {Bosted}, \citenamefont {Clogher}, \citenamefont
  {Lung}, \citenamefont {Rock}, \citenamefont {Szalata}, \citenamefont
  {Filippone}, \citenamefont {Walker}, \citenamefont {Bjorken}, \citenamefont
  {Crisler}, \citenamefont {Para}, \citenamefont {Lambert}, \citenamefont
  {Button-Shafer}, \citenamefont {Debebe}, \citenamefont {Frodyma},
  \citenamefont {Hicks}, \citenamefont {Peterson},\ and\ \citenamefont
  {Gearhart}}]{PhysRevLett.59.755}%
  \BibitemOpen
  \bibfield  {author} {\bibinfo {author} {\bibfnamefont {E.~M.}\ \bibnamefont
  {Riordan}}, \bibinfo {author} {\bibfnamefont {M.~W.}\ \bibnamefont {Krasny}},
  \bibinfo {author} {\bibfnamefont {K.}~\bibnamefont {Lang}}, \bibinfo {author}
  {\bibfnamefont {P.}~\bibnamefont {de~Barbaro}}, \bibinfo {author}
  {\bibfnamefont {A.}~\bibnamefont {Bodek}}, \bibinfo {author} {\bibfnamefont
  {S.}~\bibnamefont {Dasu}}, \bibinfo {author} {\bibfnamefont {N.}~\bibnamefont
  {Varelas}}, \bibinfo {author} {\bibfnamefont {X.}~\bibnamefont {Wang}},
  \bibinfo {author} {\bibfnamefont {R.}~\bibnamefont {Arnold}}, \bibinfo
  {author} {\bibfnamefont {D.}~\bibnamefont {Benton}}, \bibinfo {author}
  {\bibfnamefont {P.}~\bibnamefont {Bosted}}, \bibinfo {author} {\bibfnamefont
  {L.}~\bibnamefont {Clogher}}, \bibinfo {author} {\bibfnamefont
  {A.}~\bibnamefont {Lung}}, \bibinfo {author} {\bibfnamefont {S.}~\bibnamefont
  {Rock}}, \bibinfo {author} {\bibfnamefont {Z.}~\bibnamefont {Szalata}},
  \bibinfo {author} {\bibfnamefont {B.~W.}\ \bibnamefont {Filippone}}, \bibinfo
  {author} {\bibfnamefont {R.~C.}\ \bibnamefont {Walker}}, \bibinfo {author}
  {\bibfnamefont {J.~D.}\ \bibnamefont {Bjorken}}, \bibinfo {author}
  {\bibfnamefont {M.}~\bibnamefont {Crisler}}, \bibinfo {author} {\bibfnamefont
  {A.}~\bibnamefont {Para}}, \bibinfo {author} {\bibfnamefont {J.}~\bibnamefont
  {Lambert}}, \bibinfo {author} {\bibfnamefont {J.}~\bibnamefont
  {Button-Shafer}}, \bibinfo {author} {\bibfnamefont {B.}~\bibnamefont
  {Debebe}}, \bibinfo {author} {\bibfnamefont {M.}~\bibnamefont {Frodyma}},
  \bibinfo {author} {\bibfnamefont {R.~S.}\ \bibnamefont {Hicks}}, \bibinfo
  {author} {\bibfnamefont {G.~A.}\ \bibnamefont {Peterson}}, \ and\ \bibinfo
  {author} {\bibfnamefont {R.}~\bibnamefont {Gearhart}},\ }\href {\doibase
  10.1103/PhysRevLett.59.755} {\bibfield  {journal} {\bibinfo  {journal} {Phys.
  Rev. Lett.}\ }\textbf {\bibinfo {volume} {59}},\ \bibinfo {pages} {755}
  (\bibinfo {year} {1987})}\BibitemShut {NoStop}%
\bibitem [{\citenamefont {Anastasi}\ \emph {et~al.}(2015)\citenamefont
  {Anastasi} \emph {et~al.}}]{Anastasi:2015qla}%
  \BibitemOpen
  \bibfield  {author} {\bibinfo {author} {\bibfnamefont {A.}~\bibnamefont
  {Anastasi}} \emph {et~al.},\ }\href {\doibase 10.1016/j.physletb.2015.10.003}
  {\bibfield  {journal} {\bibinfo  {journal} {Phys. Lett. B}\ }\textbf
  {\bibinfo {volume} {750}},\ \bibinfo {pages} {633} (\bibinfo {year}
  {2015})},\ \Eprint {http://arxiv.org/abs/1509.00740} {arXiv:1509.00740
  [hep-ex]} \BibitemShut {NoStop}%
\bibitem [{\citenamefont {Abi}\ \emph {et~al.}(2021)\citenamefont {Abi} \emph
  {et~al.}}]{Abi:2021gix}%
  \BibitemOpen
  \bibfield  {author} {\bibinfo {author} {\bibfnamefont {B.}~\bibnamefont
  {Abi}} \emph {et~al.} (\bibinfo {collaboration} {Muon g-2}),\ }\href
  {\doibase 10.1103/PhysRevLett.126.141801} {\bibfield  {journal} {\bibinfo
  {journal} {Phys. Rev. Lett.}\ }\textbf {\bibinfo {volume} {126}},\ \bibinfo
  {pages} {141801} (\bibinfo {year} {2021})},\ \Eprint
  {http://arxiv.org/abs/2104.03281} {arXiv:2104.03281 [hep-ex]} \BibitemShut
  {NoStop}%
\bibitem [{\citenamefont {Borsanyi}\ \emph {et~al.}(2021)\citenamefont
  {Borsanyi}, \citenamefont {Fodor}, \citenamefont {Guenther}, \citenamefont
  {Hoelbling}, \citenamefont {Katz}, \citenamefont {Lellouch}, \citenamefont
  {Lippert}, \citenamefont {Miura}, \citenamefont {Parato}, \citenamefont
  {Szabo}, \citenamefont {Stokes}, \citenamefont {Toth}, \citenamefont
  {Torok},\ and\ \citenamefont {Varnhorst}}]{Borsanyi:2021}%
  \BibitemOpen
  \bibfield  {author} {\bibinfo {author} {\bibfnamefont {S.}~\bibnamefont
  {Borsanyi}}, \bibinfo {author} {\bibfnamefont {Z.}~\bibnamefont {Fodor}},
  \bibinfo {author} {\bibfnamefont {J.}~\bibnamefont {Guenther}}, \bibinfo
  {author} {\bibfnamefont {C.}~\bibnamefont {Hoelbling}}, \bibinfo {author}
  {\bibfnamefont {S.}~\bibnamefont {Katz}}, \bibinfo {author} {\bibfnamefont
  {L.}~\bibnamefont {Lellouch}}, \bibinfo {author} {\bibfnamefont
  {T.}~\bibnamefont {Lippert}}, \bibinfo {author} {\bibfnamefont
  {K.}~\bibnamefont {Miura}}, \bibinfo {author} {\bibfnamefont
  {L.}~\bibnamefont {Parato}}, \bibinfo {author} {\bibfnamefont
  {K.}~\bibnamefont {Szabo}}, \bibinfo {author} {\bibfnamefont
  {F.}~\bibnamefont {Stokes}}, \bibinfo {author} {\bibfnamefont
  {B.}~\bibnamefont {Toth}}, \bibinfo {author} {\bibfnamefont {C.}~\bibnamefont
  {Torok}}, \ and\ \bibinfo {author} {\bibfnamefont {L.}~\bibnamefont
  {Varnhorst}},\ }\href {\doibase 10.1038/s41586-021-03418-1} {\bibfield
  {journal} {\bibinfo  {journal} {Nature}\ }\textbf {\bibinfo {volume} {593}},\
  \bibinfo {pages} {51—55} (\bibinfo {year} {2021})}\BibitemShut {NoStop}%
\bibitem [{\citenamefont {Fayet}(2007)}]{PhysRevD.75.115017}%
  \BibitemOpen
  \bibfield  {author} {\bibinfo {author} {\bibfnamefont {P.}~\bibnamefont
  {Fayet}},\ }\href {\doibase 10.1103/PhysRevD.75.115017} {\bibfield  {journal}
  {\bibinfo  {journal} {Phys. Rev. D}\ }\textbf {\bibinfo {volume} {75}},\
  \bibinfo {pages} {115017} (\bibinfo {year} {2007})}\BibitemShut {NoStop}%
\bibitem [{\citenamefont {Morel}\ \emph {et~al.}(2020)\citenamefont {Morel},
  \citenamefont {Yao}, \citenamefont {Clad\'e},\ and\ \citenamefont
  {Guellati-Kh\'elifa}}]{Morel:2020dww}%
  \BibitemOpen
  \bibfield  {author} {\bibinfo {author} {\bibfnamefont {L.}~\bibnamefont
  {Morel}}, \bibinfo {author} {\bibfnamefont {Z.}~\bibnamefont {Yao}}, \bibinfo
  {author} {\bibfnamefont {P.}~\bibnamefont {Clad\'e}}, \ and\ \bibinfo
  {author} {\bibfnamefont {S.}~\bibnamefont {Guellati-Kh\'elifa}},\ }\href
  {\doibase 10.1038/s41586-020-2964-7} {\bibfield  {journal} {\bibinfo
  {journal} {Nature}\ }\textbf {\bibinfo {volume} {588}},\ \bibinfo {pages}
  {61} (\bibinfo {year} {2020})}\BibitemShut {NoStop}%
\bibitem [{\citenamefont {Zhang}\ and\ \citenamefont
  {Miller}(2017)}]{Zhang:2017zap}%
  \BibitemOpen
  \bibfield  {author} {\bibinfo {author} {\bibfnamefont {X.}~\bibnamefont
  {Zhang}}\ and\ \bibinfo {author} {\bibfnamefont {G.~A.}\ \bibnamefont
  {Miller}},\ }\href {\doibase 10.1016/j.physletb.2017.08.013} {\bibfield
  {journal} {\bibinfo  {journal} {Phys. Lett. B}\ }\textbf {\bibinfo {volume}
  {773}},\ \bibinfo {pages} {159} (\bibinfo {year} {2017})},\ \Eprint
  {http://arxiv.org/abs/1703.04588} {arXiv:1703.04588 [nucl-th]} \BibitemShut
  {NoStop}%
\bibitem [{\citenamefont {Ellwanger}\ and\ \citenamefont
  {Moretti}(2016)}]{Ellwanger:2016wfe}%
  \BibitemOpen
  \bibfield  {author} {\bibinfo {author} {\bibfnamefont {U.}~\bibnamefont
  {Ellwanger}}\ and\ \bibinfo {author} {\bibfnamefont {S.}~\bibnamefont
  {Moretti}},\ }\href {\doibase 10.1007/JHEP11(2016)039} {\bibfield  {journal}
  {\bibinfo  {journal} {J. High Energ. Phys.}\ }\textbf {\bibinfo {volume}
  {2016}},\ \bibinfo {pages} {39} (\bibinfo {year} {2016})},\ \Eprint
  {http://arxiv.org/abs/1609.01669} {arXiv:1609.01669 [hep-ph]} \BibitemShut
  {NoStop}%
\bibitem [{\citenamefont {Dror}\ \emph {et~al.}(2017)\citenamefont {Dror},
  \citenamefont {Lasenby},\ and\ \citenamefont {Pospelov}}]{Dror:2017ehi}%
  \BibitemOpen
  \bibfield  {author} {\bibinfo {author} {\bibfnamefont {J.~A.}\ \bibnamefont
  {Dror}}, \bibinfo {author} {\bibfnamefont {R.}~\bibnamefont {Lasenby}}, \
  and\ \bibinfo {author} {\bibfnamefont {M.}~\bibnamefont {Pospelov}},\ }\href
  {\doibase 10.1103/PhysRevLett.119.141803} {\bibfield  {journal} {\bibinfo
  {journal} {Phys. Rev. Lett.}\ }\textbf {\bibinfo {volume} {119}},\ \bibinfo
  {pages} {141803} (\bibinfo {year} {2017})},\ \Eprint
  {http://arxiv.org/abs/1705.06726} {arXiv:1705.06726 [hep-ph]} \BibitemShut
  {NoStop}%
\bibitem [{\citenamefont {Delle~Rose}\ \emph {et~al.}(2017)\citenamefont
  {Delle~Rose}, \citenamefont {Khalil},\ and\ \citenamefont
  {Moretti}}]{DelleRose:2017xil}%
  \BibitemOpen
  \bibfield  {author} {\bibinfo {author} {\bibfnamefont {L.}~\bibnamefont
  {Delle~Rose}}, \bibinfo {author} {\bibfnamefont {S.}~\bibnamefont {Khalil}},
  \ and\ \bibinfo {author} {\bibfnamefont {S.}~\bibnamefont {Moretti}},\ }\href
  {\doibase 10.1103/PhysRevD.96.115024} {\bibfield  {journal} {\bibinfo
  {journal} {Phys. Rev. D}\ }\textbf {\bibinfo {volume} {96}},\ \bibinfo
  {pages} {115024} (\bibinfo {year} {2017})},\ \Eprint
  {http://arxiv.org/abs/1704.03436} {arXiv:1704.03436 [hep-ph]} \BibitemShut
  {NoStop}%
\bibitem [{\citenamefont {Delle~Rose}\ \emph
  {et~al.}(2019{\natexlab{a}})\citenamefont {Delle~Rose}, \citenamefont
  {Khalil}, \citenamefont {King}, \citenamefont {Moretti},\ and\ \citenamefont
  {Thabt}}]{DelleRose:2018eic}%
  \BibitemOpen
  \bibfield  {author} {\bibinfo {author} {\bibfnamefont {L.}~\bibnamefont
  {Delle~Rose}}, \bibinfo {author} {\bibfnamefont {S.}~\bibnamefont {Khalil}},
  \bibinfo {author} {\bibfnamefont {S.~J.}\ \bibnamefont {King}}, \bibinfo
  {author} {\bibfnamefont {S.}~\bibnamefont {Moretti}}, \ and\ \bibinfo
  {author} {\bibfnamefont {A.~M.}\ \bibnamefont {Thabt}},\ }\href {\doibase
  10.1103/PhysRevD.99.055022} {\bibfield  {journal} {\bibinfo  {journal} {Phys.
  Rev. D}\ }\textbf {\bibinfo {volume} {99}},\ \bibinfo {pages} {055022}
  (\bibinfo {year} {2019}{\natexlab{a}})},\ \Eprint
  {http://arxiv.org/abs/1811.07953} {arXiv:1811.07953 [hep-ph]} \BibitemShut
  {NoStop}%
\bibitem [{\citenamefont {Delle~Rose}\ \emph
  {et~al.}(2019{\natexlab{b}})\citenamefont {Delle~Rose}, \citenamefont
  {Khalil}, \citenamefont {King},\ and\ \citenamefont
  {Moretti}}]{DelleRose:2018pgm}%
  \BibitemOpen
  \bibfield  {author} {\bibinfo {author} {\bibfnamefont {L.}~\bibnamefont
  {Delle~Rose}}, \bibinfo {author} {\bibfnamefont {S.}~\bibnamefont {Khalil}},
  \bibinfo {author} {\bibfnamefont {S.~J.}\ \bibnamefont {King}}, \ and\
  \bibinfo {author} {\bibfnamefont {S.}~\bibnamefont {Moretti}},\ }\href
  {\doibase 10.3389/fphy.2019.00073} {\bibfield  {journal} {\bibinfo  {journal}
  {Front. in Phys.}\ }\textbf {\bibinfo {volume} {7}},\ \bibinfo {pages} {73}
  (\bibinfo {year} {2019}{\natexlab{b}})},\ \Eprint
  {http://arxiv.org/abs/1812.05497} {arXiv:1812.05497 [hep-ph]} \BibitemShut
  {NoStop}%
\bibitem [{\citenamefont {Alves}\ and\ \citenamefont
  {Weiner}(2018)}]{Alves:2017avw}%
  \BibitemOpen
  \bibfield  {author} {\bibinfo {author} {\bibfnamefont {D.~S.~M.}\
  \bibnamefont {Alves}}\ and\ \bibinfo {author} {\bibfnamefont
  {N.}~\bibnamefont {Weiner}},\ }\href {\doibase 10.1007/JHEP07(2018)092}
  {\bibfield  {journal} {\bibinfo  {journal} {JHEP}\ }\textbf {\bibinfo
  {volume} {07}},\ \bibinfo {pages} {092} (\bibinfo {year} {2018})},\ \Eprint
  {http://arxiv.org/abs/1710.03764} {arXiv:1710.03764 [hep-ph]} \BibitemShut
  {NoStop}%
\bibitem [{\citenamefont {Alves}(2021)}]{Alves:2020xhf}%
  \BibitemOpen
  \bibfield  {author} {\bibinfo {author} {\bibfnamefont {D.~S.~M.}\
  \bibnamefont {Alves}},\ }\href {\doibase 10.1103/PhysRevD.103.055018}
  {\bibfield  {journal} {\bibinfo  {journal} {Phys. Rev. D}\ }\textbf {\bibinfo
  {volume} {103}},\ \bibinfo {pages} {055018} (\bibinfo {year} {2021})},\
  \Eprint {http://arxiv.org/abs/2009.05578} {arXiv:2009.05578 [hep-ph]}
  \BibitemShut {NoStop}%
\bibitem [{\citenamefont {Bordes}\ \emph {et~al.}(2019)\citenamefont {Bordes},
  \citenamefont {Chan},\ and\ \citenamefont {Tsou}}]{Bordes:2019wcp}%
  \BibitemOpen
  \bibfield  {author} {\bibinfo {author} {\bibfnamefont {J.}~\bibnamefont
  {Bordes}}, \bibinfo {author} {\bibfnamefont {H.-M.}\ \bibnamefont {Chan}}, \
  and\ \bibinfo {author} {\bibfnamefont {S.~T.}\ \bibnamefont {Tsou}},\ }\href
  {\doibase 10.1142/S0217751X19501409} {\bibfield  {journal} {\bibinfo
  {journal} {Int. J. Mod. Phys. A}\ }\textbf {\bibinfo {volume} {34}},\
  \bibinfo {pages} {1950140} (\bibinfo {year} {2019})},\ \Eprint
  {http://arxiv.org/abs/1906.09229} {arXiv:1906.09229 [hep-ph]} \BibitemShut
  {NoStop}%
\bibitem [{\citenamefont {Nam}(2020)}]{Nam:2019osu}%
  \BibitemOpen
  \bibfield  {author} {\bibinfo {author} {\bibfnamefont {C.~H.}\ \bibnamefont
  {Nam}},\ }\href {\doibase 10.1140/epjc/s10052-020-7794-0} {\bibfield
  {journal} {\bibinfo  {journal} {Eur. Phys. J. C}\ }\textbf {\bibinfo {volume}
  {80}},\ \bibinfo {pages} {231} (\bibinfo {year} {2020})},\ \Eprint
  {http://arxiv.org/abs/1907.09819} {arXiv:1907.09819 [hep-ph]} \BibitemShut
  {NoStop}%
\bibitem [{\citenamefont {Kirpichnikov}\ \emph {et~al.}(2020)\citenamefont
  {Kirpichnikov}, \citenamefont {Lyubovitskij},\ and\ \citenamefont
  {Zhevlakov}}]{Kirpichnikov:2020tcf}%
  \BibitemOpen
  \bibfield  {author} {\bibinfo {author} {\bibfnamefont {D.}~\bibnamefont
  {Kirpichnikov}}, \bibinfo {author} {\bibfnamefont {V.~E.}\ \bibnamefont
  {Lyubovitskij}}, \ and\ \bibinfo {author} {\bibfnamefont {A.~S.}\
  \bibnamefont {Zhevlakov}},\ }\href {\doibase 10.1103/PhysRevD.102.095024}
  {\bibfield  {journal} {\bibinfo  {journal} {Phys. Rev. D}\ }\textbf {\bibinfo
  {volume} {102}},\ \bibinfo {pages} {095024} (\bibinfo {year} {2020})},\
  \Eprint {http://arxiv.org/abs/2002.07496} {arXiv:2002.07496 [hep-ph]}
  \BibitemShut {NoStop}%
\bibitem [{\citenamefont {Fayet}(2021)}]{Fayet:2020bmb}%
  \BibitemOpen
  \bibfield  {author} {\bibinfo {author} {\bibfnamefont {P.}~\bibnamefont
  {Fayet}},\ }\href@noop {} {\bibfield  {journal} {\bibinfo  {journal}
  {Physical Review D}\ }\textbf {\bibinfo {volume} {103}},\ \bibinfo {pages}
  {035034} (\bibinfo {year} {2021})},\ \Eprint
  {http://arxiv.org/abs/2010.04673} {arXiv:2010.04673 [hep-ph]} \BibitemShut
  {NoStop}%
\bibitem [{\citenamefont {Fornal}(2017)}]{Fornal:2017msy}%
  \BibitemOpen
  \bibfield  {author} {\bibinfo {author} {\bibfnamefont {B.}~\bibnamefont
  {Fornal}},\ }\href {\doibase 10.1142/S0217751X17300204} {\bibfield  {journal}
  {\bibinfo  {journal} {Int. J. Mod. Phys. A}\ }\textbf {\bibinfo {volume}
  {32}},\ \bibinfo {pages} {1730020} (\bibinfo {year} {2017})},\ \Eprint
  {http://arxiv.org/abs/1707.09749} {arXiv:1707.09749 [hep-ph]} \BibitemShut
  {NoStop}%
\bibitem [{\citenamefont {Baldini}\ \emph {et~al.}(2018)\citenamefont {Baldini}
  \emph {et~al.}}]{Baldini:2018nnn}%
  \BibitemOpen
  \bibfield  {author} {\bibinfo {author} {\bibfnamefont {A.~M.}\ \bibnamefont
  {Baldini}} \emph {et~al.} (\bibinfo {collaboration} {MEG II}),\ }\href
  {\doibase 10.1140/epjc/s10052-018-5845-6} {\bibfield  {journal} {\bibinfo
  {journal} {Eur. Phys. J. C}\ }\textbf {\bibinfo {volume} {78}},\ \bibinfo
  {pages} {380} (\bibinfo {year} {2018})},\ \Eprint
  {http://arxiv.org/abs/1801.04688} {arXiv:1801.04688 [physics.ins-det]}
  \BibitemShut {NoStop}%
\bibitem [{\citenamefont {Balewski}\ \emph {et~al.}(2014)\citenamefont
  {Balewski} \emph {et~al.}}]{Balewski:2014pxa}%
  \BibitemOpen
  \bibfield  {author} {\bibinfo {author} {\bibfnamefont {J.}~\bibnamefont
  {Balewski}} \emph {et~al.}\ }(\bibinfo {year} {2014})\ \Eprint
  {http://arxiv.org/abs/1412.4717} {arXiv:1412.4717 [physics.ins-det]}
  \BibitemShut {NoStop}%
\bibitem [{\citenamefont {Ahdida}\ \emph {et~al.}(2020)\citenamefont {Ahdida}
  \emph {et~al.}}]{SHiP:2020noy}%
  \BibitemOpen
  \bibfield  {author} {\bibinfo {author} {\bibfnamefont {C.}~\bibnamefont
  {Ahdida}} \emph {et~al.} (\bibinfo {collaboration} {SHiP}),\ }\href@noop {}
  {\  (\bibinfo {year} {2020})},\ \Eprint {http://arxiv.org/abs/2010.11057}
  {arXiv:2010.11057 [hep-ex]} \BibitemShut {NoStop}%
\bibitem [{\citenamefont {Kou}\ \emph {et~al.}(2019)\citenamefont {Kou},
  \citenamefont {Urquijo}, \citenamefont {Altmannshofer} \emph
  {et~al.}}]{Kou:2018nap}%
  \BibitemOpen
  \bibfield  {author} {\bibinfo {author} {\bibfnamefont {E.}~\bibnamefont
  {Kou}}, \bibinfo {author} {\bibfnamefont {P.}~\bibnamefont {Urquijo}},
  \bibinfo {author} {\bibfnamefont {W.}~\bibnamefont {Altmannshofer}},  \emph
  {et~al.} (\bibinfo {collaboration} {Belle-II}),\ }\href {\doibase
  10.1093/ptep/ptz106} {\bibfield  {journal} {\bibinfo  {journal} {Prog. Theor.
  Exp. Phys.}\ }\textbf {\bibinfo {volume} {2019}},\ \bibinfo {pages} {123C01}
  (\bibinfo {year} {2019})},\ \bibinfo {note} {[Erratum: PTEP 2020, 029201
  (2020)]},\ \Eprint {http://arxiv.org/abs/1808.10567} {arXiv:1808.10567
  [hep-ex]} \BibitemShut {NoStop}%
\bibitem [{\citenamefont {Banerjee}\ \emph {et~al.}(2020)\citenamefont
  {Banerjee}, \citenamefont {Bernhard}, \citenamefont {Burtsev}, \citenamefont
  {Chumakov}, \citenamefont {Cooke}, \citenamefont {Crivelli}, \citenamefont
  {Depero}, \citenamefont {Dermenev}, \citenamefont {Donskov}, \citenamefont
  {Dusaev}, \citenamefont {Enik}, \citenamefont {Charitonidis}, \citenamefont
  {Feshchenko}, \citenamefont {Frolov}, \citenamefont {Gardikiotis},
  \citenamefont {Gerassimov}, \citenamefont {Gninenko}, \citenamefont
  {H\"osgen}, \citenamefont {Jeckel}, \citenamefont {Kachanov}, \citenamefont
  {Karneyeu}, \citenamefont {Kekelidze}, \citenamefont {Ketzer}, \citenamefont
  {Kirpichnikov}, \citenamefont {Kirsanov}, \citenamefont {Kolosov},
  \citenamefont {Konorov}, \citenamefont {Kovalenko}, \citenamefont
  {Kramarenko}, \citenamefont {Kravchuk}, \citenamefont {Krasnikov},
  \citenamefont {Kuleshov}, \citenamefont {Lyubovitskij}, \citenamefont
  {Lysan}, \citenamefont {Matveev}, \citenamefont {Mikhailov}, \citenamefont
  {Molina~Bueno}, \citenamefont {Peshekhonov}, \citenamefont {Polyakov},
  \citenamefont {Radics}, \citenamefont {Rojas}, \citenamefont {Rubbia},
  \citenamefont {Samoylenko}, \citenamefont {Shchukin}, \citenamefont
  {Tikhomirov}, \citenamefont {Tlisova}, \citenamefont {Tlisov}, \citenamefont
  {Toropin}, \citenamefont {Trifonov}, \citenamefont {Vasilishin},
  \citenamefont {Vasquez~Arenas}, \citenamefont {Volkov}, \citenamefont
  {Volkov},\ and\ \citenamefont {Ulloa}}]{Banerjee:2020bbb}%
  \BibitemOpen
  \bibfield  {author} {\bibinfo {author} {\bibfnamefont {D.}~\bibnamefont
  {Banerjee}}, \bibinfo {author} {\bibfnamefont {J.}~\bibnamefont {Bernhard}},
  \bibinfo {author} {\bibfnamefont {V.~E.}\ \bibnamefont {Burtsev}}, \bibinfo
  {author} {\bibfnamefont {A.~G.}\ \bibnamefont {Chumakov}}, \bibinfo {author}
  {\bibfnamefont {D.}~\bibnamefont {Cooke}}, \bibinfo {author} {\bibfnamefont
  {P.}~\bibnamefont {Crivelli}}, \bibinfo {author} {\bibfnamefont
  {E.}~\bibnamefont {Depero}}, \bibinfo {author} {\bibfnamefont {A.~V.}\
  \bibnamefont {Dermenev}}, \bibinfo {author} {\bibfnamefont {S.~V.}\
  \bibnamefont {Donskov}}, \bibinfo {author} {\bibfnamefont {R.~R.}\
  \bibnamefont {Dusaev}}, \bibinfo {author} {\bibfnamefont {T.}~\bibnamefont
  {Enik}}, \bibinfo {author} {\bibfnamefont {N.}~\bibnamefont {Charitonidis}},
  \bibinfo {author} {\bibfnamefont {A.}~\bibnamefont {Feshchenko}}, \bibinfo
  {author} {\bibfnamefont {V.~N.}\ \bibnamefont {Frolov}}, \bibinfo {author}
  {\bibfnamefont {A.}~\bibnamefont {Gardikiotis}}, \bibinfo {author}
  {\bibfnamefont {S.~G.}\ \bibnamefont {Gerassimov}}, \bibinfo {author}
  {\bibfnamefont {S.~N.}\ \bibnamefont {Gninenko}}, \bibinfo {author}
  {\bibfnamefont {M.}~\bibnamefont {H\"osgen}}, \bibinfo {author}
  {\bibfnamefont {M.}~\bibnamefont {Jeckel}}, \bibinfo {author} {\bibfnamefont
  {V.~A.}\ \bibnamefont {Kachanov}}, \bibinfo {author} {\bibfnamefont {A.~E.}\
  \bibnamefont {Karneyeu}}, \bibinfo {author} {\bibfnamefont {G.}~\bibnamefont
  {Kekelidze}}, \bibinfo {author} {\bibfnamefont {B.}~\bibnamefont {Ketzer}},
  \bibinfo {author} {\bibfnamefont {D.~V.}\ \bibnamefont {Kirpichnikov}},
  \bibinfo {author} {\bibfnamefont {M.~M.}\ \bibnamefont {Kirsanov}}, \bibinfo
  {author} {\bibfnamefont {V.~N.}\ \bibnamefont {Kolosov}}, \bibinfo {author}
  {\bibfnamefont {I.~V.}\ \bibnamefont {Konorov}}, \bibinfo {author}
  {\bibfnamefont {S.~G.}\ \bibnamefont {Kovalenko}}, \bibinfo {author}
  {\bibfnamefont {V.~A.}\ \bibnamefont {Kramarenko}}, \bibinfo {author}
  {\bibfnamefont {L.~V.}\ \bibnamefont {Kravchuk}}, \bibinfo {author}
  {\bibfnamefont {N.~V.}\ \bibnamefont {Krasnikov}}, \bibinfo {author}
  {\bibfnamefont {S.~V.}\ \bibnamefont {Kuleshov}}, \bibinfo {author}
  {\bibfnamefont {V.~E.}\ \bibnamefont {Lyubovitskij}}, \bibinfo {author}
  {\bibfnamefont {V.}~\bibnamefont {Lysan}}, \bibinfo {author} {\bibfnamefont
  {V.~A.}\ \bibnamefont {Matveev}}, \bibinfo {author} {\bibfnamefont {Y.~V.}\
  \bibnamefont {Mikhailov}}, \bibinfo {author} {\bibfnamefont {L.}~\bibnamefont
  {Molina~Bueno}}, \bibinfo {author} {\bibfnamefont {D.~V.}\ \bibnamefont
  {Peshekhonov}}, \bibinfo {author} {\bibfnamefont {V.~A.}\ \bibnamefont
  {Polyakov}}, \bibinfo {author} {\bibfnamefont {B.}~\bibnamefont {Radics}},
  \bibinfo {author} {\bibfnamefont {R.}~\bibnamefont {Rojas}}, \bibinfo
  {author} {\bibfnamefont {A.}~\bibnamefont {Rubbia}}, \bibinfo {author}
  {\bibfnamefont {V.~D.}\ \bibnamefont {Samoylenko}}, \bibinfo {author}
  {\bibfnamefont {D.}~\bibnamefont {Shchukin}}, \bibinfo {author}
  {\bibfnamefont {V.~O.}\ \bibnamefont {Tikhomirov}}, \bibinfo {author}
  {\bibfnamefont {I.}~\bibnamefont {Tlisova}}, \bibinfo {author} {\bibfnamefont
  {D.~A.}\ \bibnamefont {Tlisov}}, \bibinfo {author} {\bibfnamefont {A.~N.}\
  \bibnamefont {Toropin}}, \bibinfo {author} {\bibfnamefont {A.~Y.}\
  \bibnamefont {Trifonov}}, \bibinfo {author} {\bibfnamefont {B.~I.}\
  \bibnamefont {Vasilishin}}, \bibinfo {author} {\bibfnamefont
  {G.}~\bibnamefont {Vasquez~Arenas}}, \bibinfo {author} {\bibfnamefont
  {P.~V.}\ \bibnamefont {Volkov}}, \bibinfo {author} {\bibfnamefont {V.~Y.}\
  \bibnamefont {Volkov}}, \ and\ \bibinfo {author} {\bibfnamefont
  {P.}~\bibnamefont {Ulloa}} (\bibinfo {collaboration} {The NA64
  Collaboration}),\ }\href {\doibase 10.1103/PhysRevD.101.071101} {\bibfield
  {journal} {\bibinfo  {journal} {Phys. Rev. D}\ }\textbf {\bibinfo {volume}
  {101}},\ \bibinfo {pages} {071101} (\bibinfo {year} {2020})}\BibitemShut
  {NoStop}%
\bibitem [{\citenamefont {Tilley}\ \emph {et~al.}(1992)\citenamefont {Tilley},
  \citenamefont {Weller},\ and\ \citenamefont {Hale}}]{Tilley:1992zz}%
  \BibitemOpen
  \bibfield  {author} {\bibinfo {author} {\bibfnamefont {D.~R.}\ \bibnamefont
  {Tilley}}, \bibinfo {author} {\bibfnamefont {H.~R.}\ \bibnamefont {Weller}},
  \ and\ \bibinfo {author} {\bibfnamefont {G.~M.}\ \bibnamefont {Hale}},\
  }\href {\doibase 10.1016/0375-9474(92)90635-W} {\bibfield  {journal}
  {\bibinfo  {journal} {Nucl. Phys. A}\ }\textbf {\bibinfo {volume} {541}},\
  \bibinfo {pages} {1} (\bibinfo {year} {1992})}\BibitemShut {NoStop}%
\bibitem [{\citenamefont {K\'alm\'an}\ and\ \citenamefont
  {Keszthelyi}(2020)}]{Kalman:2020meg}%
  \BibitemOpen
  \bibfield  {author} {\bibinfo {author} {\bibfnamefont {P.}~\bibnamefont
  {K\'alm\'an}}\ and\ \bibinfo {author} {\bibfnamefont {T.}~\bibnamefont
  {Keszthelyi}},\ }\href {\doibase 10.1140/epja/s10050-020-00202-z} {\bibfield
  {journal} {\bibinfo  {journal} {Eur. Phys. J. A}\ }\textbf {\bibinfo {volume}
  {56}},\ \bibinfo {pages} {205} (\bibinfo {year} {2020})},\ \Eprint
  {http://arxiv.org/abs/2005.10643} {arXiv:2005.10643 [nucl-th]} \BibitemShut
  {NoStop}%
\bibitem [{\citenamefont {Aleksejevs}\ \emph {et~al.}(2021)\citenamefont
  {Aleksejevs}, \citenamefont {Barkanova}, \citenamefont {Kolomensky},\ and\
  \citenamefont {Sheff}}]{Aleksejevs:2021zjw}%
  \BibitemOpen
  \bibfield  {author} {\bibinfo {author} {\bibfnamefont {A.}~\bibnamefont
  {Aleksejevs}}, \bibinfo {author} {\bibfnamefont {S.}~\bibnamefont
  {Barkanova}}, \bibinfo {author} {\bibfnamefont {Y.~G.}\ \bibnamefont
  {Kolomensky}}, \ and\ \bibinfo {author} {\bibfnamefont {B.}~\bibnamefont
  {Sheff}},\ }\href@noop {} {\  (\bibinfo {year} {2021})},\ \Eprint
  {http://arxiv.org/abs/2102.01127} {arXiv:2102.01127 [hep-ph]} \BibitemShut
  {NoStop}%
\bibitem [{\citenamefont {Koch}(2021)}]{Koch:2020ouk}%
  \BibitemOpen
  \bibfield  {author} {\bibinfo {author} {\bibfnamefont {B.}~\bibnamefont
  {Koch}},\ }\href {\doibase 10.1016/j.nuclphysa.2021.122143} {\bibfield
  {journal} {\bibinfo  {journal} {Nucl. Phys. A}\ }\textbf {\bibinfo {volume}
  {1008}},\ \bibinfo {pages} {122143} (\bibinfo {year} {2021})},\ \Eprint
  {http://arxiv.org/abs/2003.05722} {arXiv:2003.05722 [hep-ph]} \BibitemShut
  {NoStop}%
\bibitem [{\citenamefont {Viviani}\ \emph {et~al.}(2020)\citenamefont
  {Viviani}, \citenamefont {Girlanda}, \citenamefont {Kievsky},\ and\
  \citenamefont {Marcucci}}]{Viviani:2020gkm}%
  \BibitemOpen
  \bibfield  {author} {\bibinfo {author} {\bibfnamefont {M.}~\bibnamefont
  {Viviani}}, \bibinfo {author} {\bibfnamefont {L.}~\bibnamefont {Girlanda}},
  \bibinfo {author} {\bibfnamefont {A.}~\bibnamefont {Kievsky}}, \ and\
  \bibinfo {author} {\bibfnamefont {L.~E.}\ \bibnamefont {Marcucci}},\ }\href
  {\doibase 10.1103/PhysRevC.102.034007} {\bibfield  {journal} {\bibinfo
  {journal} {Phys. Rev. C}\ }\textbf {\bibinfo {volume} {102}},\ \bibinfo
  {pages} {034007} (\bibinfo {year} {2020})},\ \Eprint
  {http://arxiv.org/abs/2003.14059} {arXiv:2003.14059 [nucl-th]} \BibitemShut
  {NoStop}%
\bibitem [{\citenamefont {Entem}\ and\ \citenamefont
  {Machleidt}(2003)}]{Entem:2003ft}%
  \BibitemOpen
  \bibfield  {author} {\bibinfo {author} {\bibfnamefont {D.}~\bibnamefont
  {Entem}}\ and\ \bibinfo {author} {\bibfnamefont {R.}~\bibnamefont
  {Machleidt}},\ }\href {\doibase 10.1103/PhysRevC.68.041001} {\bibfield
  {journal} {\bibinfo  {journal} {Phys. Rev. C}\ }\textbf {\bibinfo {volume}
  {68}},\ \bibinfo {pages} {041001} (\bibinfo {year} {2003})},\ \Eprint
  {http://arxiv.org/abs/nucl-th/0304018} {arXiv:nucl-th/0304018} \BibitemShut
  {NoStop}%
\bibitem [{\citenamefont {Machleidt}\ and\ \citenamefont
  {Entem}(2011)}]{Machleidt:2011zz}%
  \BibitemOpen
  \bibfield  {author} {\bibinfo {author} {\bibfnamefont {R.}~\bibnamefont
  {Machleidt}}\ and\ \bibinfo {author} {\bibfnamefont {D.~R.}\ \bibnamefont
  {Entem}},\ }\href {\doibase 10.1016/j.physrep.2011.02.001} {\bibfield
  {journal} {\bibinfo  {journal} {Phys. Rept.}\ }\textbf {\bibinfo {volume}
  {503}},\ \bibinfo {pages} {1} (\bibinfo {year} {2011})},\ \Eprint
  {http://arxiv.org/abs/1105.2919} {arXiv:1105.2919 [nucl-th]} \BibitemShut
  {NoStop}%
%%CITATION = ARXIV:1105.2919;%%
\bibitem [{\citenamefont {Epelbaum}\ \emph {et~al.}(2002)\citenamefont
  {Epelbaum}, \citenamefont {Nogga}, \citenamefont {Gloeckle}, \citenamefont
  {Kamada}, \citenamefont {Meissner},\ and\ \citenamefont
  {Witala}}]{Epelbaum:2002vt}%
  \BibitemOpen
  \bibfield  {author} {\bibinfo {author} {\bibfnamefont {E.}~\bibnamefont
  {Epelbaum}}, \bibinfo {author} {\bibfnamefont {A.}~\bibnamefont {Nogga}},
  \bibinfo {author} {\bibfnamefont {W.}~\bibnamefont {Gloeckle}}, \bibinfo
  {author} {\bibfnamefont {H.}~\bibnamefont {Kamada}}, \bibinfo {author}
  {\bibfnamefont {U.-G.}\ \bibnamefont {Meissner}}, \ and\ \bibinfo {author}
  {\bibfnamefont {H.}~\bibnamefont {Witala}},\ }\href {\doibase
  10.1103/PhysRevC.66.064001} {\bibfield  {journal} {\bibinfo  {journal} {Phys.
  Rev. C}\ }\textbf {\bibinfo {volume} {66}},\ \bibinfo {pages} {064001}
  (\bibinfo {year} {2002})},\ \Eprint {http://arxiv.org/abs/0208023}
  {arXiv:0208023 [nucl-th]} \BibitemShut {NoStop}%
\bibitem [{\citenamefont {Piarulli}\ \emph {et~al.}(2016)\citenamefont
  {Piarulli}, \citenamefont {Girlanda}, \citenamefont {Schiavilla},
  \citenamefont {Kievsky}, \citenamefont {Lovato}, \citenamefont {Marcucci},
  \citenamefont {Pieper}, \citenamefont {Viviani},\ and\ \citenamefont
  {Wiringa}}]{Piarulli:2016vel}%
  \BibitemOpen
  \bibfield  {author} {\bibinfo {author} {\bibfnamefont {M.}~\bibnamefont
  {Piarulli}}, \bibinfo {author} {\bibfnamefont {L.}~\bibnamefont {Girlanda}},
  \bibinfo {author} {\bibfnamefont {R.}~\bibnamefont {Schiavilla}}, \bibinfo
  {author} {\bibfnamefont {A.}~\bibnamefont {Kievsky}}, \bibinfo {author}
  {\bibfnamefont {A.}~\bibnamefont {Lovato}}, \bibinfo {author} {\bibfnamefont
  {L.~E.}\ \bibnamefont {Marcucci}}, \bibinfo {author} {\bibfnamefont {S.~C.}\
  \bibnamefont {Pieper}}, \bibinfo {author} {\bibfnamefont {M.}~\bibnamefont
  {Viviani}}, \ and\ \bibinfo {author} {\bibfnamefont {R.~B.}\ \bibnamefont
  {Wiringa}},\ }\href {\doibase 10.1103/PhysRevC.94.054007} {\bibfield
  {journal} {\bibinfo  {journal} {Phys. Rev. C}\ }\textbf {\bibinfo {volume}
  {94}},\ \bibinfo {pages} {054007} (\bibinfo {year} {2016})},\ \Eprint
  {http://arxiv.org/abs/1606.06335} {arXiv:1606.06335 [nucl-th]} \BibitemShut
  {NoStop}%
\bibitem [{\citenamefont {Piarulli}\ \emph {et~al.}(2018)\citenamefont
  {Piarulli} \emph {et~al.}}]{Piarulli:2017dwd}%
  \BibitemOpen
  \bibfield  {author} {\bibinfo {author} {\bibfnamefont {M.}~\bibnamefont
  {Piarulli}} \emph {et~al.},\ }\href {\doibase 10.1103/PhysRevLett.120.052503}
  {\bibfield  {journal} {\bibinfo  {journal} {Phys. Rev. Lett.}\ }\textbf
  {\bibinfo {volume} {120}},\ \bibinfo {pages} {052503} (\bibinfo {year}
  {2018})},\ \Eprint {http://arxiv.org/abs/1707.02883} {arXiv:1707.02883
  [nucl-th]} \BibitemShut {NoStop}%
\bibitem [{\citenamefont {Pastore}\ \emph {et~al.}(2008)\citenamefont
  {Pastore}, \citenamefont {Schiavilla},\ and\ \citenamefont
  {Goity}}]{Pastore:2008ui}%
  \BibitemOpen
  \bibfield  {author} {\bibinfo {author} {\bibfnamefont {S.}~\bibnamefont
  {Pastore}}, \bibinfo {author} {\bibfnamefont {R.}~\bibnamefont {Schiavilla}},
  \ and\ \bibinfo {author} {\bibfnamefont {J.~L.}\ \bibnamefont {Goity}},\
  }\href {\doibase 10.1103/PhysRevC.78.064002} {\bibfield  {journal} {\bibinfo
  {journal} {Phys. Rev. C}\ }\textbf {\bibinfo {volume} {78}},\ \bibinfo
  {pages} {064002} (\bibinfo {year} {2008})},\ \Eprint
  {http://arxiv.org/abs/0810.1941} {arXiv:0810.1941 [nucl-th]} \BibitemShut
  {NoStop}%
%%CITATION = ARXIV:0810.1941;%%
\bibitem [{\citenamefont {Pastore}\ \emph {et~al.}(2009)\citenamefont
  {Pastore}, \citenamefont {Girlanda}, \citenamefont {Schiavilla},
  \citenamefont {Viviani},\ and\ \citenamefont {Wiringa}}]{Pastore:2009is}%
  \BibitemOpen
  \bibfield  {author} {\bibinfo {author} {\bibfnamefont {S.}~\bibnamefont
  {Pastore}}, \bibinfo {author} {\bibfnamefont {L.}~\bibnamefont {Girlanda}},
  \bibinfo {author} {\bibfnamefont {R.}~\bibnamefont {Schiavilla}}, \bibinfo
  {author} {\bibfnamefont {M.}~\bibnamefont {Viviani}}, \ and\ \bibinfo
  {author} {\bibfnamefont {R.}~\bibnamefont {Wiringa}},\ }\href {\doibase
  10.1103/PhysRevC.80.034004} {\bibfield  {journal} {\bibinfo  {journal} {Phys.
  Rev. C}\ }\textbf {\bibinfo {volume} {80}},\ \bibinfo {pages} {034004}
  (\bibinfo {year} {2009})},\ \Eprint {http://arxiv.org/abs/0906.1800}
  {arXiv:0906.1800 [nucl-th]} \BibitemShut {NoStop}%
\bibitem [{\citenamefont {Pastore}\ \emph {et~al.}(2011)\citenamefont
  {Pastore}, \citenamefont {Girlanda}, \citenamefont {Schiavilla},\ and\
  \citenamefont {Viviani}}]{Pastore:2011ip}%
  \BibitemOpen
  \bibfield  {author} {\bibinfo {author} {\bibfnamefont {S.}~\bibnamefont
  {Pastore}}, \bibinfo {author} {\bibfnamefont {L.}~\bibnamefont {Girlanda}},
  \bibinfo {author} {\bibfnamefont {R.}~\bibnamefont {Schiavilla}}, \ and\
  \bibinfo {author} {\bibfnamefont {M.}~\bibnamefont {Viviani}},\ }\href
  {\doibase 10.1103/PhysRevC.84.024001} {\bibfield  {journal} {\bibinfo
  {journal} {Phys. Rev. C}\ }\textbf {\bibinfo {volume} {84}},\ \bibinfo
  {pages} {024001} (\bibinfo {year} {2011})},\ \Eprint
  {http://arxiv.org/abs/1106.4539} {arXiv:1106.4539 [nucl-th]} \BibitemShut
  {NoStop}%
%%CITATION = ARXIV:1106.4539;%%
\bibitem [{\citenamefont {Koelling}\ \emph {et~al.}(2009)\citenamefont
  {Koelling}, \citenamefont {Epelbaum}, \citenamefont {Krebs},\ and\
  \citenamefont {Meissner}}]{Kolling:2009iq}%
  \BibitemOpen
  \bibfield  {author} {\bibinfo {author} {\bibfnamefont {S.}~\bibnamefont
  {Koelling}}, \bibinfo {author} {\bibfnamefont {E.}~\bibnamefont {Epelbaum}},
  \bibinfo {author} {\bibfnamefont {H.}~\bibnamefont {Krebs}}, \ and\ \bibinfo
  {author} {\bibfnamefont {U.-G.}\ \bibnamefont {Meissner}},\ }\href {\doibase
  10.1103/PhysRevC.80.045502} {\bibfield  {journal} {\bibinfo  {journal} {Phys.
  Rev. C}\ }\textbf {\bibinfo {volume} {80}},\ \bibinfo {pages} {045502}
  (\bibinfo {year} {2009})},\ \Eprint {http://arxiv.org/abs/0907.3437}
  {arXiv:0907.3437 [nucl-th]} \BibitemShut {NoStop}%
\bibitem [{\citenamefont {Koelling}\ \emph {et~al.}(2011)\citenamefont
  {Koelling}, \citenamefont {Epelbaum}, \citenamefont {Krebs},\ and\
  \citenamefont {Meissner}}]{Kolling:2011mt}%
  \BibitemOpen
  \bibfield  {author} {\bibinfo {author} {\bibfnamefont {S.}~\bibnamefont
  {Koelling}}, \bibinfo {author} {\bibfnamefont {E.}~\bibnamefont {Epelbaum}},
  \bibinfo {author} {\bibfnamefont {H.}~\bibnamefont {Krebs}}, \ and\ \bibinfo
  {author} {\bibfnamefont {U.-G.}\ \bibnamefont {Meissner}},\ }\href {\doibase
  10.1103/PhysRevC.84.054008} {\bibfield  {journal} {\bibinfo  {journal} {Phys.
  Rev. C}\ }\textbf {\bibinfo {volume} {84}},\ \bibinfo {pages} {054008}
  (\bibinfo {year} {2011})},\ \Eprint {http://arxiv.org/abs/1107.0602}
  {arXiv:1107.0602 [nucl-th]} \BibitemShut {NoStop}%
\bibitem [{\citenamefont {Piarulli}\ \emph {et~al.}(2013)\citenamefont
  {Piarulli}, \citenamefont {Girlanda}, \citenamefont {Marcucci}, \citenamefont
  {Pastore}, \citenamefont {Schiavilla},\ and\ \citenamefont
  {Viviani}}]{Piarulli:2012bn}%
  \BibitemOpen
  \bibfield  {author} {\bibinfo {author} {\bibfnamefont {M.}~\bibnamefont
  {Piarulli}}, \bibinfo {author} {\bibfnamefont {L.}~\bibnamefont {Girlanda}},
  \bibinfo {author} {\bibfnamefont {L.~E.}\ \bibnamefont {Marcucci}}, \bibinfo
  {author} {\bibfnamefont {S.}~\bibnamefont {Pastore}}, \bibinfo {author}
  {\bibfnamefont {R.}~\bibnamefont {Schiavilla}}, \ and\ \bibinfo {author}
  {\bibfnamefont {M.}~\bibnamefont {Viviani}},\ }\href {\doibase
  10.1103/PhysRevC.87.014006} {\bibfield  {journal} {\bibinfo  {journal} {Phys.
  Rev. C}\ }\textbf {\bibinfo {volume} {87}},\ \bibinfo {pages} {014006}
  (\bibinfo {year} {2013})},\ \Eprint {http://arxiv.org/abs/1212.1105}
  {arXiv:1212.1105 [nucl-th]} \BibitemShut {NoStop}%
\bibitem [{\citenamefont {Schiavilla}\ \emph {et~al.}(2019)\citenamefont
  {Schiavilla} \emph {et~al.}}]{Schiavilla:2018udt}%
  \BibitemOpen
  \bibfield  {author} {\bibinfo {author} {\bibfnamefont {R.}~\bibnamefont
  {Schiavilla}} \emph {et~al.},\ }\href {\doibase 10.1103/PhysRevC.99.034005}
  {\bibfield  {journal} {\bibinfo  {journal} {Phys. Rev. C}\ }\textbf {\bibinfo
  {volume} {99}},\ \bibinfo {pages} {034005} (\bibinfo {year} {2019})},\
  \Eprint {http://arxiv.org/abs/1809.10180} {arXiv:1809.10180 [nucl-th]}
  \BibitemShut {NoStop}%
\bibitem [{\citenamefont {Bacca}\ and\ \citenamefont
  {Pastore}(2014)}]{Bacca:2014tla}%
  \BibitemOpen
  \bibfield  {author} {\bibinfo {author} {\bibfnamefont {S.}~\bibnamefont
  {Bacca}}\ and\ \bibinfo {author} {\bibfnamefont {S.}~\bibnamefont
  {Pastore}},\ }\href {\doibase 10.1088/0954-3899/41/12/123002} {\bibfield
  {journal} {\bibinfo  {journal} {J. Phys. G}\ }\textbf {\bibinfo {volume}
  {41}},\ \bibinfo {pages} {123002} (\bibinfo {year} {2014})},\ \Eprint
  {http://arxiv.org/abs/1407.3490} {arXiv:1407.3490 [nucl-th]} \BibitemShut
  {NoStop}%
\bibitem [{\citenamefont {Carlson}\ \emph {et~al.}(2015)\citenamefont
  {Carlson}, \citenamefont {Gandolfi}, \citenamefont {Pederiva}, \citenamefont
  {Pieper}, \citenamefont {Schiavilla}, \citenamefont {Schmidt},\ and\
  \citenamefont {Wiringa}}]{Carlson:2014vla}%
  \BibitemOpen
  \bibfield  {author} {\bibinfo {author} {\bibfnamefont {J.}~\bibnamefont
  {Carlson}}, \bibinfo {author} {\bibfnamefont {S.}~\bibnamefont {Gandolfi}},
  \bibinfo {author} {\bibfnamefont {F.}~\bibnamefont {Pederiva}}, \bibinfo
  {author} {\bibfnamefont {S.~C.}\ \bibnamefont {Pieper}}, \bibinfo {author}
  {\bibfnamefont {R.}~\bibnamefont {Schiavilla}}, \bibinfo {author}
  {\bibfnamefont {K.~E.}\ \bibnamefont {Schmidt}}, \ and\ \bibinfo {author}
  {\bibfnamefont {R.~B.}\ \bibnamefont {Wiringa}},\ }\href {\doibase
  10.1103/RevModPhys.87.1067} {\bibfield  {journal} {\bibinfo  {journal} {Rev.
  Mod. Phys.}\ }\textbf {\bibinfo {volume} {87}},\ \bibinfo {pages} {1067}
  (\bibinfo {year} {2015})},\ \Eprint {http://arxiv.org/abs/1412.3081}
  {arXiv:1412.3081 [nucl-th]} \BibitemShut {NoStop}%
\bibitem [{\citenamefont {Marcucci}\ \emph {et~al.}(2016)\citenamefont
  {Marcucci}, \citenamefont {Gross}, \citenamefont {Pena}, \citenamefont
  {Piarulli}, \citenamefont {Schiavilla}, \citenamefont {Sick}, \citenamefont
  {Stadler}, \citenamefont {Van~Orden},\ and\ \citenamefont
  {Viviani}}]{Marcucci:2015rca}%
  \BibitemOpen
  \bibfield  {author} {\bibinfo {author} {\bibfnamefont {L.~E.}\ \bibnamefont
  {Marcucci}}, \bibinfo {author} {\bibfnamefont {F.}~\bibnamefont {Gross}},
  \bibinfo {author} {\bibfnamefont {M.~T.}\ \bibnamefont {Pena}}, \bibinfo
  {author} {\bibfnamefont {M.}~\bibnamefont {Piarulli}}, \bibinfo {author}
  {\bibfnamefont {R.}~\bibnamefont {Schiavilla}}, \bibinfo {author}
  {\bibfnamefont {I.}~\bibnamefont {Sick}}, \bibinfo {author} {\bibfnamefont
  {A.}~\bibnamefont {Stadler}}, \bibinfo {author} {\bibfnamefont {J.~W.}\
  \bibnamefont {Van~Orden}}, \ and\ \bibinfo {author} {\bibfnamefont
  {M.}~\bibnamefont {Viviani}},\ }\href {\doibase
  10.1088/0954-3899/43/2/023002} {\bibfield  {journal} {\bibinfo  {journal} {J.
  Phys. G}\ }\textbf {\bibinfo {volume} {43}},\ \bibinfo {pages} {023002}
  (\bibinfo {year} {2016})},\ \Eprint {http://arxiv.org/abs/1504.05063}
  {arXiv:1504.05063 [nucl-th]} \BibitemShut {NoStop}%
\bibitem [{\citenamefont {Perry}\ and\ \citenamefont
  {Bame}(1955)}]{Perry:1955ptc}%
  \BibitemOpen
  \bibfield  {author} {\bibinfo {author} {\bibfnamefont {J.~E.}\ \bibnamefont
  {Perry}}\ and\ \bibinfo {author} {\bibfnamefont {S.~J.}\ \bibnamefont
  {Bame}},\ }\href {\doibase 10.1103/PhysRev.99.1368} {\bibfield  {journal}
  {\bibinfo  {journal} {Phys. Rev.}\ }\textbf {\bibinfo {volume} {99}},\
  \bibinfo {pages} {1368} (\bibinfo {year} {1955})}\BibitemShut {NoStop}%
\bibitem [{\citenamefont {Hahn}\ \emph {et~al.}(1995)\citenamefont {Hahn},
  \citenamefont {Brune},\ and\ \citenamefont {Kavanagh}}]{Hahn:1995ptc}%
  \BibitemOpen
  \bibfield  {author} {\bibinfo {author} {\bibfnamefont {K.~I.}\ \bibnamefont
  {Hahn}}, \bibinfo {author} {\bibfnamefont {C.~R.}\ \bibnamefont {Brune}}, \
  and\ \bibinfo {author} {\bibfnamefont {R.~W.}\ \bibnamefont {Kavanagh}},\
  }\href {\doibase 10.1103/PhysRevC.51.1624} {\bibfield  {journal} {\bibinfo
  {journal} {Phys. Rev. C}\ }\textbf {\bibinfo {volume} {51}},\ \bibinfo
  {pages} {1624} (\bibinfo {year} {1995})}\BibitemShut {NoStop}%
\bibitem [{\citenamefont {Canon}\ \emph {et~al.}(2002)\citenamefont {Canon},
  \citenamefont {Nelson}, \citenamefont {Sabourov}, \citenamefont {Wulf},
  \citenamefont {Weller}, \citenamefont {Prior}, \citenamefont {Spraker},
  \citenamefont {Kelley},\ and\ \citenamefont {Tilley}}]{Canon:2002ds}%
  \BibitemOpen
  \bibfield  {author} {\bibinfo {author} {\bibfnamefont {R.~S.}\ \bibnamefont
  {Canon}}, \bibinfo {author} {\bibfnamefont {S.~O.}\ \bibnamefont {Nelson}},
  \bibinfo {author} {\bibfnamefont {K.}~\bibnamefont {Sabourov}}, \bibinfo
  {author} {\bibfnamefont {E.}~\bibnamefont {Wulf}}, \bibinfo {author}
  {\bibfnamefont {H.~R.}\ \bibnamefont {Weller}}, \bibinfo {author}
  {\bibfnamefont {R.~M.}\ \bibnamefont {Prior}}, \bibinfo {author}
  {\bibfnamefont {M.}~\bibnamefont {Spraker}}, \bibinfo {author} {\bibfnamefont
  {J.~H.}\ \bibnamefont {Kelley}}, \ and\ \bibinfo {author} {\bibfnamefont
  {D.~R.}\ \bibnamefont {Tilley}},\ }\href {\doibase
  10.1103/PhysRevC.65.044008} {\bibfield  {journal} {\bibinfo  {journal} {Phys.
  Rev. C}\ }\textbf {\bibinfo {volume} {65}},\ \bibinfo {pages} {044008}
  (\bibinfo {year} {2002})}\BibitemShut {NoStop}%
\bibitem [{\citenamefont {Komar}\ \emph {et~al.}(1993)\citenamefont {Komar},
  \citenamefont {Mak}, \citenamefont {Leslie}, \citenamefont {Evans},
  \citenamefont {Bonvin}, \citenamefont {Earle},\ and\ \citenamefont
  {Alexander}}]{Komar:1993nhc}%
  \BibitemOpen
  \bibfield  {author} {\bibinfo {author} {\bibfnamefont {R.~J.}\ \bibnamefont
  {Komar}}, \bibinfo {author} {\bibfnamefont {H.-B.}\ \bibnamefont {Mak}},
  \bibinfo {author} {\bibfnamefont {J.~R.}\ \bibnamefont {Leslie}}, \bibinfo
  {author} {\bibfnamefont {H.~C.}\ \bibnamefont {Evans}}, \bibinfo {author}
  {\bibfnamefont {E.}~\bibnamefont {Bonvin}}, \bibinfo {author} {\bibfnamefont
  {E.~D.}\ \bibnamefont {Earle}}, \ and\ \bibinfo {author} {\bibfnamefont
  {T.~K.}\ \bibnamefont {Alexander}},\ }\href {\doibase
  10.1103/PhysRevC.48.2375} {\bibfield  {journal} {\bibinfo  {journal} {Phys.
  Rev. C}\ }\textbf {\bibinfo {volume} {48}},\ \bibinfo {pages} {2375}
  (\bibinfo {year} {1993})}\BibitemShut {NoStop}%
\bibitem [{\citenamefont {Kievsky}\ \emph {et~al.}(2008)\citenamefont
  {Kievsky}, \citenamefont {Rosati}, \citenamefont {Viviani}, \citenamefont
  {Marcucci},\ and\ \citenamefont {Girlanda}}]{Kievsky:2008es}%
  \BibitemOpen
  \bibfield  {author} {\bibinfo {author} {\bibfnamefont {A.}~\bibnamefont
  {Kievsky}}, \bibinfo {author} {\bibfnamefont {S.}~\bibnamefont {Rosati}},
  \bibinfo {author} {\bibfnamefont {M.}~\bibnamefont {Viviani}}, \bibinfo
  {author} {\bibfnamefont {L.}~\bibnamefont {Marcucci}}, \ and\ \bibinfo
  {author} {\bibfnamefont {L.}~\bibnamefont {Girlanda}},\ }\href {\doibase
  10.1088/0954-3899/35/6/063101} {\bibfield  {journal} {\bibinfo  {journal} {J.
  Phys. G}\ }\textbf {\bibinfo {volume} {35}},\ \bibinfo {pages} {063101}
  (\bibinfo {year} {2008})},\ \Eprint {http://arxiv.org/abs/0805.4688}
  {arXiv:0805.4688 [nucl-th]} \BibitemShut {NoStop}%
\bibitem [{\citenamefont {Marcucci}\ \emph {et~al.}(2020)\citenamefont
  {Marcucci}, \citenamefont {Dohet-Eraly}, \citenamefont {Girlanda},
  \citenamefont {Gnech}, \citenamefont {Kievsky},\ and\ \citenamefont
  {Viviani}}]{Marcucci:2020fip}%
  \BibitemOpen
  \bibfield  {author} {\bibinfo {author} {\bibfnamefont {L.~E.}\ \bibnamefont
  {Marcucci}}, \bibinfo {author} {\bibfnamefont {J.}~\bibnamefont
  {Dohet-Eraly}}, \bibinfo {author} {\bibfnamefont {L.}~\bibnamefont
  {Girlanda}}, \bibinfo {author} {\bibfnamefont {A.}~\bibnamefont {Gnech}},
  \bibinfo {author} {\bibfnamefont {A.}~\bibnamefont {Kievsky}}, \ and\
  \bibinfo {author} {\bibfnamefont {M.}~\bibnamefont {Viviani}},\ }\href
  {\doibase 10.3389/fphy.2020.00069} {\bibfield  {journal} {\bibinfo  {journal}
  {Front. in Phys.}\ }\textbf {\bibinfo {volume} {8}},\ \bibinfo {pages} {69}
  (\bibinfo {year} {2020})},\ \Eprint {http://arxiv.org/abs/1912.09751}
  {arXiv:1912.09751 [nucl-th]} \BibitemShut {NoStop}%
\bibitem [{\citenamefont {Schiavilla}\ \emph {et~al.}(1989)\citenamefont
  {Schiavilla}, \citenamefont {Pandharipande},\ and\ \citenamefont
  {Riska}}]{Schiavilla:1989zz}%
  \BibitemOpen
  \bibfield  {author} {\bibinfo {author} {\bibfnamefont {R.}~\bibnamefont
  {Schiavilla}}, \bibinfo {author} {\bibfnamefont {V.~R.}\ \bibnamefont
  {Pandharipande}}, \ and\ \bibinfo {author} {\bibfnamefont {D.-O.}\
  \bibnamefont {Riska}},\ }\href {\doibase 10.1103/PhysRevC.40.2294} {\bibfield
   {journal} {\bibinfo  {journal} {Phys. Rev. C}\ }\textbf {\bibinfo {volume}
  {40}},\ \bibinfo {pages} {2294} (\bibinfo {year} {1989})}\BibitemShut
  {NoStop}%
\bibitem [{\citenamefont {Gustavino}\ \emph {et~al.}(2021)\citenamefont
  {Gustavino} \emph {et~al.}}]{Gustavino:2021abcd}%
  \BibitemOpen
  \bibfield  {author} {\bibinfo {author} {\bibfnamefont {C.}~\bibnamefont
  {Gustavino}} \emph {et~al.},\ }\href@noop {} {}\bibinfo {howpublished}
  {PRIN2020.0004062.26-01-2021, unpublished} (\bibinfo {year}
  {2021})\BibitemShut {NoStop}%
\bibitem [{\citenamefont {Sabat\'e-Gilarte}\ \emph {et~al.}(2017)\citenamefont
  {Sabat\'e-Gilarte} \emph {et~al.}}]{Sabate-Gilarte:2017biu}%
  \BibitemOpen
  \bibfield  {author} {\bibinfo {author} {\bibfnamefont {M.}~\bibnamefont
  {Sabat\'e-Gilarte}} \emph {et~al.},\ }\href {\doibase
  10.1140/epja/i2017-12392-4} {\bibfield  {journal} {\bibinfo  {journal} {Eur.
  Phys. J. A}\ }\textbf {\bibinfo {volume} {53}},\ \bibinfo {pages} {210}
  (\bibinfo {year} {2017})}\BibitemShut {NoStop}%
\bibitem [{\citenamefont {Baroni}\ \emph {et~al.}(2018)\citenamefont {Baroni}
  \emph {et~al.}}]{Baroni:2018fdn}%
  \BibitemOpen
  \bibfield  {author} {\bibinfo {author} {\bibfnamefont {A.}~\bibnamefont
  {Baroni}} \emph {et~al.},\ }\href {\doibase 10.1103/PhysRevC.98.044003}
  {\bibfield  {journal} {\bibinfo  {journal} {Phys. Rev. C}\ }\textbf {\bibinfo
  {volume} {98}},\ \bibinfo {pages} {044003} (\bibinfo {year} {2018})},\
  \Eprint {http://arxiv.org/abs/1806.10245} {arXiv:1806.10245 [nucl-th]}
  \BibitemShut {NoStop}%
\bibitem [{\citenamefont {Viviani}\ \emph {et~al.}(2011)\citenamefont
  {Viviani}, \citenamefont {Deltuva}, \citenamefont {Lazauskas}, \citenamefont
  {Carbonell}, \citenamefont {Fonseca}, \citenamefont {Kievsky}, \citenamefont
  {Marcucci},\ and\ \citenamefont {Rosati}}]{Viviani:2011ax}%
  \BibitemOpen
  \bibfield  {author} {\bibinfo {author} {\bibfnamefont {M.}~\bibnamefont
  {Viviani}}, \bibinfo {author} {\bibfnamefont {A.}~\bibnamefont {Deltuva}},
  \bibinfo {author} {\bibfnamefont {R.}~\bibnamefont {Lazauskas}}, \bibinfo
  {author} {\bibfnamefont {J.}~\bibnamefont {Carbonell}}, \bibinfo {author}
  {\bibfnamefont {A.}~\bibnamefont {Fonseca}}, \bibinfo {author} {\bibfnamefont
  {A.}~\bibnamefont {Kievsky}}, \bibinfo {author} {\bibfnamefont {L.~E.}\
  \bibnamefont {Marcucci}}, \ and\ \bibinfo {author} {\bibfnamefont
  {S.}~\bibnamefont {Rosati}},\ }\href {\doibase 10.1103/PhysRevC.84.054010}
  {\bibfield  {journal} {\bibinfo  {journal} {Phys. Rev. C}\ }\textbf {\bibinfo
  {volume} {84}},\ \bibinfo {pages} {054010} (\bibinfo {year} {2011})},\
  \Eprint {http://arxiv.org/abs/1109.3625} {arXiv:1109.3625 [nucl-th]}
  \BibitemShut {NoStop}%
\bibitem [{\citenamefont {Viviani}\ \emph {et~al.}(2017)\citenamefont
  {Viviani}, \citenamefont {Deltuva}, \citenamefont {Lazauskas}, \citenamefont
  {Fonseca}, \citenamefont {Kievsky},\ and\ \citenamefont
  {Marcucci}}]{Viviani:2016cww}%
  \BibitemOpen
  \bibfield  {author} {\bibinfo {author} {\bibfnamefont {M.}~\bibnamefont
  {Viviani}}, \bibinfo {author} {\bibfnamefont {A.}~\bibnamefont {Deltuva}},
  \bibinfo {author} {\bibfnamefont {R.}~\bibnamefont {Lazauskas}}, \bibinfo
  {author} {\bibfnamefont {A.}~\bibnamefont {Fonseca}}, \bibinfo {author}
  {\bibfnamefont {A.}~\bibnamefont {Kievsky}}, \ and\ \bibinfo {author}
  {\bibfnamefont {L.~E.}\ \bibnamefont {Marcucci}},\ }\href {\doibase
  10.1103/PhysRevC.95.034003} {\bibfield  {journal} {\bibinfo  {journal} {Phys.
  Rev. C}\ }\textbf {\bibinfo {volume} {95}},\ \bibinfo {pages} {034003}
  (\bibinfo {year} {2017})},\ \Eprint {http://arxiv.org/abs/1610.09140}
  {arXiv:1610.09140 [nucl-th]} \BibitemShut {NoStop}%
\bibitem [{\citenamefont {Kegel}\ \emph {et~al.}(2021)\citenamefont {Kegel}
  \emph {et~al.}}]{Kegel:2021abcd}%
  \BibitemOpen
  \bibfield  {author} {\bibinfo {author} {\bibfnamefont {S.}~\bibnamefont
  {Kegel}} \emph {et~al.},\ }\href@noop {} {}\bibinfo {howpublished} {in
  preparation} (\bibinfo {year} {2021})\BibitemShut {NoStop}%
\bibitem [{\citenamefont {Piarulli}\ \emph {et~al.}(2014)\citenamefont
  {Piarulli}, \citenamefont {Girlanda}, \citenamefont {Marcucci}, \citenamefont
  {Pastore}, \citenamefont {Schiavilla},\ and\ \citenamefont
  {Viviani}}]{Piarulli:2014fta}%
  \BibitemOpen
  \bibfield  {author} {\bibinfo {author} {\bibfnamefont {M.}~\bibnamefont
  {Piarulli}}, \bibinfo {author} {\bibfnamefont {L.}~\bibnamefont {Girlanda}},
  \bibinfo {author} {\bibfnamefont {L.}~\bibnamefont {Marcucci}}, \bibinfo
  {author} {\bibfnamefont {S.}~\bibnamefont {Pastore}}, \bibinfo {author}
  {\bibfnamefont {R.}~\bibnamefont {Schiavilla}}, \ and\ \bibinfo {author}
  {\bibfnamefont {M.}~\bibnamefont {Viviani}},\ }\href {\doibase
  10.22323/1.212.0040} {\bibfield  {journal} {\bibinfo  {journal} {PoS}\
  }\textbf {\bibinfo {volume} {Bormio2014}},\ \bibinfo {pages} {040} (\bibinfo
  {year} {2014})}\BibitemShut {NoStop}%
\bibitem [{\citenamefont {Walecka}(1995)}]{Walecka1995}%
  \BibitemOpen
  \bibfield  {author} {\bibinfo {author} {\bibfnamefont {J.~D.}\ \bibnamefont
  {Walecka}},\ }\href@noop {} {\emph {\bibinfo {title} {Theoretical Nuclear and
  Subnuclear Physics}}}\ (\bibinfo  {publisher} {Oxford University Press},\
  \bibinfo {address} {New York},\ \bibinfo {year} {1995})\BibitemShut {NoStop}%
\bibitem [{\citenamefont {Marcucci}\ \emph {et~al.}(2001)\citenamefont
  {Marcucci}, \citenamefont {Schiavilla}, \citenamefont {Viviani},
  \citenamefont {Kievsky}, \citenamefont {Rosati},\ and\ \citenamefont
  {Beacom}}]{Marcucci:2000xy}%
  \BibitemOpen
  \bibfield  {author} {\bibinfo {author} {\bibfnamefont {L.~E.}\ \bibnamefont
  {Marcucci}}, \bibinfo {author} {\bibfnamefont {R.}~\bibnamefont
  {Schiavilla}}, \bibinfo {author} {\bibfnamefont {M.}~\bibnamefont {Viviani}},
  \bibinfo {author} {\bibfnamefont {A.}~\bibnamefont {Kievsky}}, \bibinfo
  {author} {\bibfnamefont {S.}~\bibnamefont {Rosati}}, \ and\ \bibinfo {author}
  {\bibfnamefont {J.~F.}\ \bibnamefont {Beacom}},\ }\href {\doibase
  10.1103/PhysRevC.63.015801} {\bibfield  {journal} {\bibinfo  {journal} {Phys.
  Rev. C}\ }\textbf {\bibinfo {volume} {63}},\ \bibinfo {pages} {015801}
  (\bibinfo {year} {2001})},\ \Eprint {http://arxiv.org/abs/nucl-th/0006005}
  {arXiv:nucl-th/0006005} \BibitemShut {NoStop}%
\bibitem [{\citenamefont {Schiavilla}\ and\ \citenamefont
  {Wiringa}(2002)}]{Schiavilla:2002}%
  \BibitemOpen
  \bibfield  {author} {\bibinfo {author} {\bibfnamefont {R.}~\bibnamefont
  {Schiavilla}}\ and\ \bibinfo {author} {\bibfnamefont {R.~B.}\ \bibnamefont
  {Wiringa}},\ }\href {\doibase 10.1103/PhysRevC.65.054302} {\bibfield
  {journal} {\bibinfo  {journal} {Phys. Rev. C}\ }\textbf {\bibinfo {volume}
  {65}},\ \bibinfo {pages} {054302} (\bibinfo {year} {2002})}\BibitemShut
  {NoStop}%
\bibitem [{\citenamefont {Edmonds}(1957)}]{Edmond1957}%
  \BibitemOpen
  \bibfield  {author} {\bibinfo {author} {\bibfnamefont {A.}~\bibnamefont
  {Edmonds}},\ }\href@noop {} {\emph {\bibinfo {title} {Angular Momentum in
  Quantum Mechanics}}}\ (\bibinfo  {publisher} {Princeton University Press},\
  \bibinfo {address} {Princeton},\ \bibinfo {year} {1957})\BibitemShut
  {NoStop}%
\bibitem [{\citenamefont {Siegert}(1937)}]{Siegert1937}%
  \BibitemOpen
  \bibfield  {author} {\bibinfo {author} {\bibfnamefont {A.~J.~F.}\
  \bibnamefont {Siegert}},\ }\href@noop {} {\bibfield  {journal} {\bibinfo
  {journal} {Phys. Rev.}\ }\textbf {\bibinfo {volume} {52}},\ \bibinfo {pages}
  {787} (\bibinfo {year} {1937})}\BibitemShut {NoStop}%
\bibitem [{\citenamefont {Peskin}\ and\ \citenamefont
  {Schroeder}(1995)}]{Peskin:1995ev}%
  \BibitemOpen
  \bibfield  {author} {\bibinfo {author} {\bibfnamefont {M.~E.}\ \bibnamefont
  {Peskin}}\ and\ \bibinfo {author} {\bibfnamefont {D.~V.}\ \bibnamefont
  {Schroeder}},\ }\href@noop {} {\emph {\bibinfo {title} {{An Introduction to
  quantum field theory}}}}\ (\bibinfo  {publisher} {Addison-Wesley},\ \bibinfo
  {address} {Reading, USA},\ \bibinfo {year} {1995})\BibitemShut {NoStop}%
\bibitem [{\citenamefont {Scherer}\ and\ \citenamefont
  {Schindler}(2005)}]{Scherer:2005ri}%
  \BibitemOpen
  \bibfield  {author} {\bibinfo {author} {\bibfnamefont {S.}~\bibnamefont
  {Scherer}}\ and\ \bibinfo {author} {\bibfnamefont {M.~R.}\ \bibnamefont
  {Schindler}},\ }\href@noop {} {\  (\bibinfo {year} {2005})},\ \Eprint
  {http://arxiv.org/abs/hep-ph/0505265} {arXiv:hep-ph/0505265} \BibitemShut
  {NoStop}%
\bibitem [{\citenamefont {Green}\ \emph {et~al.}(2017)\citenamefont {Green},
  \citenamefont {Hasan}, \citenamefont {Meinel}, \citenamefont {Engelhardt},
  \citenamefont {Krieg}, \citenamefont {Laeuchli}, \citenamefont {Negele},
  \citenamefont {Orginos}, \citenamefont {Pochinsky},\ and\ \citenamefont
  {Syritsyn}}]{Green:2017keo}%
  \BibitemOpen
  \bibfield  {author} {\bibinfo {author} {\bibfnamefont {J.}~\bibnamefont
  {Green}}, \bibinfo {author} {\bibfnamefont {N.}~\bibnamefont {Hasan}},
  \bibinfo {author} {\bibfnamefont {S.}~\bibnamefont {Meinel}}, \bibinfo
  {author} {\bibfnamefont {M.}~\bibnamefont {Engelhardt}}, \bibinfo {author}
  {\bibfnamefont {S.}~\bibnamefont {Krieg}}, \bibinfo {author} {\bibfnamefont
  {J.}~\bibnamefont {Laeuchli}}, \bibinfo {author} {\bibfnamefont
  {J.}~\bibnamefont {Negele}}, \bibinfo {author} {\bibfnamefont
  {K.}~\bibnamefont {Orginos}}, \bibinfo {author} {\bibfnamefont
  {A.}~\bibnamefont {Pochinsky}}, \ and\ \bibinfo {author} {\bibfnamefont
  {S.}~\bibnamefont {Syritsyn}},\ }\href {\doibase 10.1103/PhysRevD.95.114502}
  {\bibfield  {journal} {\bibinfo  {journal} {Phys. Rev. D}\ }\textbf {\bibinfo
  {volume} {95}},\ \bibinfo {pages} {114502} (\bibinfo {year} {2017})},\
  \Eprint {http://arxiv.org/abs/1703.06703} {arXiv:1703.06703 [hep-lat]}
  \BibitemShut {NoStop}%
\bibitem [{\citenamefont {Weinberg}(1995)}]{Weinberg1995}%
  \BibitemOpen
  \bibfield  {author} {\bibinfo {author} {\bibfnamefont {S.}~\bibnamefont
  {Weinberg}},\ }\href@noop {} {\emph {\bibinfo {title} {The Quantum Theory of
  Fields}}}\ (\bibinfo  {publisher} {Cambridge University Press},\ \bibinfo
  {address} {New York},\ \bibinfo {year} {1995})\BibitemShut {NoStop}%
\bibitem [{\citenamefont {Hoferichter}\ \emph {et~al.}(2016)\citenamefont
  {Hoferichter}, \citenamefont {Ruiz~de Elvira}, \citenamefont {Kubis},\ and\
  \citenamefont {Mei{\ss}ner}}]{Hoferichter:2015hva}%
  \BibitemOpen
  \bibfield  {author} {\bibinfo {author} {\bibfnamefont {M.}~\bibnamefont
  {Hoferichter}}, \bibinfo {author} {\bibfnamefont {J.}~\bibnamefont {Ruiz~de
  Elvira}}, \bibinfo {author} {\bibfnamefont {B.}~\bibnamefont {Kubis}}, \ and\
  \bibinfo {author} {\bibfnamefont {U.-G.}\ \bibnamefont {Mei{\ss}ner}},\
  }\href {\doibase 10.1016/j.physrep.2016.02.002} {\bibfield  {journal}
  {\bibinfo  {journal} {Phys. Rept.}\ }\textbf {\bibinfo {volume} {625}},\
  \bibinfo {pages} {1} (\bibinfo {year} {2016})},\ \Eprint
  {http://arxiv.org/abs/1510.06039} {arXiv:1510.06039 [hep-ph]} \BibitemShut
  {NoStop}%
%%CITATION = ARXIV:1510.06039;%%
\bibitem [{\citenamefont {Bsaisou}\ \emph {et~al.}(2015)\citenamefont
  {Bsaisou}, \citenamefont {Mei{\ss}ner}, \citenamefont {Nogga},\ and\
  \citenamefont {Wirzba}}]{Bsaisou:2014oka}%
  \BibitemOpen
  \bibfield  {author} {\bibinfo {author} {\bibfnamefont {J.}~\bibnamefont
  {Bsaisou}}, \bibinfo {author} {\bibfnamefont {U.-G.}\ \bibnamefont
  {Mei{\ss}ner}}, \bibinfo {author} {\bibfnamefont {A.}~\bibnamefont {Nogga}},
  \ and\ \bibinfo {author} {\bibfnamefont {A.}~\bibnamefont {Wirzba}},\ }\href
  {\doibase 10.1016/j.aop.2015.04.031} {\bibfield  {journal} {\bibinfo
  {journal} {Annals Phys.}\ }\textbf {\bibinfo {volume} {359}},\ \bibinfo
  {pages} {317} (\bibinfo {year} {2015})},\ \Eprint
  {http://arxiv.org/abs/1412.5471} {arXiv:1412.5471 [hep-ph]} \BibitemShut
  {NoStop}%
%%CITATION = ARXIV:1412.5471;%%
\bibitem [{\citenamefont {Borsanyi}\ \emph {et~al.}(2015)\citenamefont
  {Borsanyi} \emph {et~al.}}]{Borsanyi:2014jba}%
  \BibitemOpen
  \bibfield  {author} {\bibinfo {author} {\bibfnamefont {S.}~\bibnamefont
  {Borsanyi}} \emph {et~al.},\ }\href {\doibase 10.1126/science.1257050}
  {\bibfield  {journal} {\bibinfo  {journal} {Science}\ }\textbf {\bibinfo
  {volume} {347}},\ \bibinfo {pages} {1452} (\bibinfo {year} {2015})},\ \Eprint
  {http://arxiv.org/abs/1406.4088} {arXiv:1406.4088 [hep-lat]} \BibitemShut
  {NoStop}%
%%CITATION = ARXIV:1406.4088;%%
\bibitem [{\citenamefont {Walker-Loud}(2014)}]{Walker-Loud:2014iea}%
  \BibitemOpen
  \bibfield  {author} {\bibinfo {author} {\bibfnamefont {A.}~\bibnamefont
  {Walker-Loud}},\ }\href {\doibase 10.22323/1.187.0013} {\bibfield  {journal}
  {\bibinfo  {journal} {PoS}\ }\textbf {\bibinfo {volume} {LATTICE2013}},\
  \bibinfo {pages} {013} (\bibinfo {year} {2014})},\ \Eprint
  {http://arxiv.org/abs/1401.8259} {arXiv:1401.8259 [hep-lat]} \BibitemShut
  {NoStop}%
\bibitem [{\citenamefont {Aoki}\ \emph {et~al.}(2019)\citenamefont {Aoki} \emph
  {et~al.}}]{Aoki:2019cca}%
  \BibitemOpen
  \bibfield  {author} {\bibinfo {author} {\bibfnamefont {S.}~\bibnamefont
  {Aoki}} \emph {et~al.} (\bibinfo {collaboration} {Flavour Lattice Averaging
  Group}),\ }\href@noop {} {\  (\bibinfo {year} {2019})},\ \Eprint
  {http://arxiv.org/abs/1902.08191} {arXiv:1902.08191 [hep-lat]} \BibitemShut
  {NoStop}%
%%CITATION = ARXIV:1902.08191;%%
\bibitem [{\citenamefont {Crivellin}\ \emph {et~al.}(2014)\citenamefont
  {Crivellin}, \citenamefont {Hoferichter},\ and\ \citenamefont
  {Procura}}]{Crivellin:2013ipa}%
  \BibitemOpen
  \bibfield  {author} {\bibinfo {author} {\bibfnamefont {A.}~\bibnamefont
  {Crivellin}}, \bibinfo {author} {\bibfnamefont {M.}~\bibnamefont
  {Hoferichter}}, \ and\ \bibinfo {author} {\bibfnamefont {M.}~\bibnamefont
  {Procura}},\ }\href {\doibase 10.1103/PhysRevD.89.054021} {\bibfield
  {journal} {\bibinfo  {journal} {Phys. Rev. D}\ }\textbf {\bibinfo {volume}
  {89}},\ \bibinfo {pages} {054021} (\bibinfo {year} {2014})},\ \Eprint
  {http://arxiv.org/abs/1312.4951} {arXiv:1312.4951 [hep-ph]} \BibitemShut
  {NoStop}%
\bibitem [{\citenamefont {Ruiz~de Elvira}\ \emph {et~al.}(2018)\citenamefont
  {Ruiz~de Elvira}, \citenamefont {Hoferichter}, \citenamefont {Kubis},\ and\
  \citenamefont {Mei\ss{}ner}}]{RuizdeElvira:2017stg}%
  \BibitemOpen
  \bibfield  {author} {\bibinfo {author} {\bibfnamefont {J.}~\bibnamefont
  {Ruiz~de Elvira}}, \bibinfo {author} {\bibfnamefont {M.}~\bibnamefont
  {Hoferichter}}, \bibinfo {author} {\bibfnamefont {B.}~\bibnamefont {Kubis}},
  \ and\ \bibinfo {author} {\bibfnamefont {U.-G.}\ \bibnamefont
  {Mei\ss{}ner}},\ }\href {\doibase 10.1088/1361-6471/aa9422} {\bibfield
  {journal} {\bibinfo  {journal} {J. Phys. G}\ }\textbf {\bibinfo {volume}
  {45}},\ \bibinfo {pages} {024001} (\bibinfo {year} {2018})},\ \Eprint
  {http://arxiv.org/abs/1706.01465} {arXiv:1706.01465 [hep-ph]} \BibitemShut
  {NoStop}%
\bibitem [{\citenamefont {Fettes}\ \emph {et~al.}(1998)\citenamefont {Fettes},
  \citenamefont {Meissner},\ and\ \citenamefont {Steininger}}]{Fettes:1998ud}%
  \BibitemOpen
  \bibfield  {author} {\bibinfo {author} {\bibfnamefont {N.}~\bibnamefont
  {Fettes}}, \bibinfo {author} {\bibfnamefont {U.-G.}\ \bibnamefont
  {Meissner}}, \ and\ \bibinfo {author} {\bibfnamefont {S.}~\bibnamefont
  {Steininger}},\ }\href {\doibase 10.1016/S0375-9474(98)00452-7} {\bibfield
  {journal} {\bibinfo  {journal} {Nucl. Phys. A}\ }\textbf {\bibinfo {volume}
  {640}},\ \bibinfo {pages} {199} (\bibinfo {year} {1998})},\ \Eprint
  {http://arxiv.org/abs/hep-ph/9803266} {arXiv:hep-ph/9803266} \BibitemShut
  {NoStop}%
\bibitem [{\citenamefont {Krasznahorkay}\ \emph
  {et~al.}(2021{\natexlab{b}})\citenamefont {Krasznahorkay} \emph
  {et~al.}}]{Krasznahorkay:2015iga:2021abcd}%
  \BibitemOpen
  \bibfield  {author} {\bibinfo {author} {\bibfnamefont {A.~J.}\ \bibnamefont
  {Krasznahorkay}} \emph {et~al.},\ }\href@noop {} {}\bibinfo {howpublished}
  {in preparation} (\bibinfo {year} {2021}{\natexlab{b}})\BibitemShut {NoStop}%
\bibitem [{\citenamefont {Gasser}\ and\ \citenamefont
  {Leutwyler}(1984)}]{Gasser:1983yg}%
  \BibitemOpen
  \bibfield  {author} {\bibinfo {author} {\bibfnamefont {J.}~\bibnamefont
  {Gasser}}\ and\ \bibinfo {author} {\bibfnamefont {H.}~\bibnamefont
  {Leutwyler}},\ }\href {\doibase 10.1016/0003-4916(84)90242-2} {\bibfield
  {journal} {\bibinfo  {journal} {Annals Phys.}\ }\textbf {\bibinfo {volume}
  {158}},\ \bibinfo {pages} {142} (\bibinfo {year} {1984})}\BibitemShut
  {NoStop}%
%%CITATION = APNYA,158,142;%%
\bibitem [{\citenamefont {Fettes}\ \emph {et~al.}(2000)\citenamefont {Fettes},
  \citenamefont {Mei{\ss}ner}, \citenamefont {Mojzis},\ and\ \citenamefont
  {Steininger}}]{Fettes:2000gb}%
  \BibitemOpen
  \bibfield  {author} {\bibinfo {author} {\bibfnamefont {N.}~\bibnamefont
  {Fettes}}, \bibinfo {author} {\bibfnamefont {U.-G.}\ \bibnamefont
  {Mei{\ss}ner}}, \bibinfo {author} {\bibfnamefont {M.}~\bibnamefont {Mojzis}},
  \ and\ \bibinfo {author} {\bibfnamefont {S.}~\bibnamefont {Steininger}},\
  }\href {\doibase 10.1006/aphy.2000.6059} {\bibfield  {journal} {\bibinfo
  {journal} {Annals Phys.}\ }\textbf {\bibinfo {volume} {283}},\ \bibinfo
  {pages} {273} (\bibinfo {year} {2000})},\ \bibinfo {note} {[Erratum: Annals
  Phys. 288, 249--250 (2001)]},\ \Eprint {http://arxiv.org/abs/hep-ph/0001308}
  {arXiv:hep-ph/0001308} \BibitemShut {NoStop}%
\end{thebibliography}%
\end{document}